\documentclass{article}

\newif\ifarxiv
\newcommand{\siref}[1]{\ifarxiv Appendix~\ref{#1}\else SM~\ref{SM-#1}\fi} 
\newcommand{\mainref}[1]{\ifarxiv\ref{#1}\else\ref{M-#1}\fi} 
\newcommand{\maineqref}[1]{\ifarxiv\eqref{#1}\else\eqref{M-#1}\fi}
\newcommand{\xref}[1]{\ifarxiv\ref{#1}\else\ref{SM-#1}\fi}
\newcommand{\sieqref}[1]{\ifarxiv\eqref{#1}\else\eqref{SM-#1}\fi}
\arxivtrue

\usepackage{lineno}

\usepackage[normalem]{ulem}       

\usepackage[verbose]{placeins} 		

\usepackage{xr-hyper} 
\usepackage{hyperref} 
\usepackage{bookmark} 
\hypersetup{
    colorlinks=false, 
    pdfborder={0 0 1}, 
    pdfborderstyle={/S /U /W 1}, 
    linktocpage, 
}

\usepackage[a4paper, left=2.5cm, right=2.5cm, top=3cm, bottom=4cm]{geometry}
\usepackage{needspace} 
\usepackage{fancyhdr} 

\usepackage{amssymb} 
\usepackage{amsmath} 

\usepackage{amssymb} 

\usepackage{lmodern} 

\allowdisplaybreaks	

\usepackage{bm}     
\usepackage{mathtools} 		

\usepackage{siunitx}
\DeclareSIUnit\bar{bar}     
\DeclareSIUnit\permille{\text{\textperthousand}}

\usepackage{float}	
\usepackage{graphicx}   
\usepackage[figuresright]{rotating}	
\usepackage{subcaption} 

\usepackage[labelfont=bf]{caption} 	

\usepackage{pdfpages} 
\usepackage{svg} 

\usepackage{pgfplots} 
\pgfplotsset{compat=newest}
\usepackage{tikz}
\pgfplotsset{plot coordinates/math parser=false}

\usepackage{changepage} 

\usepackage{longtable}	
\usepackage{supertabular}  
\usepackage{tabularx}  
\usepackage[para]{threeparttable}

\usepackage{multirow}   

\usepackage{enumitem}
\usepackage{makecell} 		
\usepackage{tablefootnote} 	
\usepackage{pdflscape} 		

\usepackage{booktabs}		
\usepackage{caption}        
\usepackage{boldline}       
\usepackage{arydshln} 		
\usepackage{dcolumn}
\newcolumntype{C}[1]{>{\centering\arraybackslash}p{#1}} 
\newcolumntype{Y}{>{\centering\arraybackslash}X} 
\newcolumntype{d}[1]{D{.}{.}{#1}} 
\newcommand*{\botrule}{\bottomrule}

\usepackage[bibstyle=ieee, citestyle=numeric-comp, sorting=none]{biblatex}
\usepackage{xspace} 	
\usepackage{ragged2e} 	

\newcommand{\cotwo}{CO\textsubscript{2}\xspace}		
\newcommand{\chfour}{CH\textsubscript{4}\xspace}	
\newcommand{\nhfour}{NH\textsubscript{4}\xspace}	
\newcommand{\RR}{R\textsuperscript{2}\xspace}	

\usepackage{xcolor} 

\usepackage{url}
\usepackage{xurl} 

\usepackage{lipsum} 

\usepackage{hyphenat} 
\begin{document}

\pagestyle{fancy}
\fancyhf{} 
\renewcommand{\headrulewidth}{0pt} 
\fancyhead[L]{Extended preprint of a research paper in \textit{IWA Water Research}} 
\fancyfoot[C]{\thepage} 
\thispagestyle{empty}   
\setcounter{page}{0}

\begin{center}
\LARGE Global Parameter Identification and Uncertainty Quantification of reduced-order ADM1 using Maximum Likelihood Estimation and Bayesian Inference \\ [5mm]
\end{center}

\begin{center}

\large
Simon Hellmann\textsuperscript{1,2}, Leander Lerch\textsuperscript{2}, Terrance Wilms\textsuperscript{3}, Moritz Reumschüssel\textsuperscript{4}, Alberto Meola\textsuperscript{1,5}, Félix Delory\textsuperscript{1,6}, Steffi Knorn\textsuperscript{3}, Stefan Streif\textsuperscript{2}, Sören Weinrich\textsuperscript{7,*} \\ [3mm]

\normalsize
\textsuperscript{1} DBFZ Deutsches Biomasseforschungszentrum gGmbH, Biochemical Conversion Department, Leipzig, Germany. simon.hellmann@dbfz.de\\ [3mm]

\textsuperscript{2} Chemnitz University of Technology, Professorship for Automatic Control and System Dynamics, Chemnitz, Germany. stefan.streif@etit.tu-chemnitz.de\\ [3mm]

\textsuperscript{3} Technische Universität Berlin, Chair of Control, Berlin, Germany. knorn@tu-berlin.de \\ [3mm]

\textsuperscript{4} Department of Engineering, University of Cambridge, Cambridge, United Kingdom \\ [3mm]

\textsuperscript{5} Faculty of Mathematics and Computer Science, Leipzig University, Leipzig, Germany \\ [3mm]

\textsuperscript{6} Institute of Biotechnology, Technische Universität Berlin, Berlin, Germany \\ [3mm]

\textsuperscript{7} FH Münster University of Applied Sciences, Faculty of Energy, Building Services, Environmental~Engineering, Steinfurt, Germany. weinrich@fh-muenster.de \\ [3mm]

\textsuperscript{*}\textit{corresponding author}
\end{center}

\vspace{1cm}

\setlength{\parindent}{0pt}
\Large
\copyright \, The authors 2026. This is an extended preprint of a research paper submitted to \textit{IWA Water Research}. This preprint is made available under the CC-BY-NC-ND 4.0 license \href{https://creativecommons.org/licenses/by-nc-nd/4.0/}{(https://creativecommons.org/licenses/by-nc-nd/4.0/)}. 


\vskip 1cm


\normalsize

\setlength{\parindent}{15pt} 
\clearpage

\phantomsection 
\addcontentsline{toc}{section}{Abstract}
\begin{abstract}
Anaerobic digestion (AD) models are challenging to calibrate against real data due to limited practical identifiability, data scarcity, and model complexity. Weighted least squares objective functions and gradient-based solvers using finite differences are often used in literature, but result in suboptimal parameter estimates and large uncertainty bands. 

This paper identifies the parameters of a reduced-order ADM1 by comparing the frameworks of maximum likelihood estimation (MLE) and Bayesian inference (BI), and quantifies uncertainties of parameters and outputs. Multirate experimental data are taken from a lab-scale AD reactor fed with agricultural substrates and operated at organic loading rates of 3\,kg VS $\mathrm{m^{-3}d^{-1}}$. 

Global optima are found via global and gradient-based local search algorithms using forward sensitivities. Enhanced scatter search provides the best global search results. These are refined by the Gauss-Newton algorithm as the best local search, which delivers \RR values of 0.95 for acetic acid and 0.77 for biogas production. BI yields substantially narrower uncertainty bands than MLE and avoids the need for a global search by incorporating prior knowledge about parameter distributions. The search space is captured efficiently by log10 scaling of parameters. Algorithmic details on its implications are provided. 

Despite a high-resolution lab-scale dataset, the ADM1-specific inhibition parameter $K_\mathrm{I,nh3}$ is practically not identifiable with MLE, and shows a low output sensitivity, whereas a parameter for the effective ion concentration shows superb practical identifiability at high confidence with both MLE and BI. 

The present work lays a scalable base for calibration of nonlinear AD models by providing reliable simulations. It thus advances model-based monitoring and control applications of multirate processes such as agricultural AD plants.
\end{abstract}

\section*{Graphical Abstract}
\begin{figure*}
\centering
\includegraphics[width=0.8\linewidth]{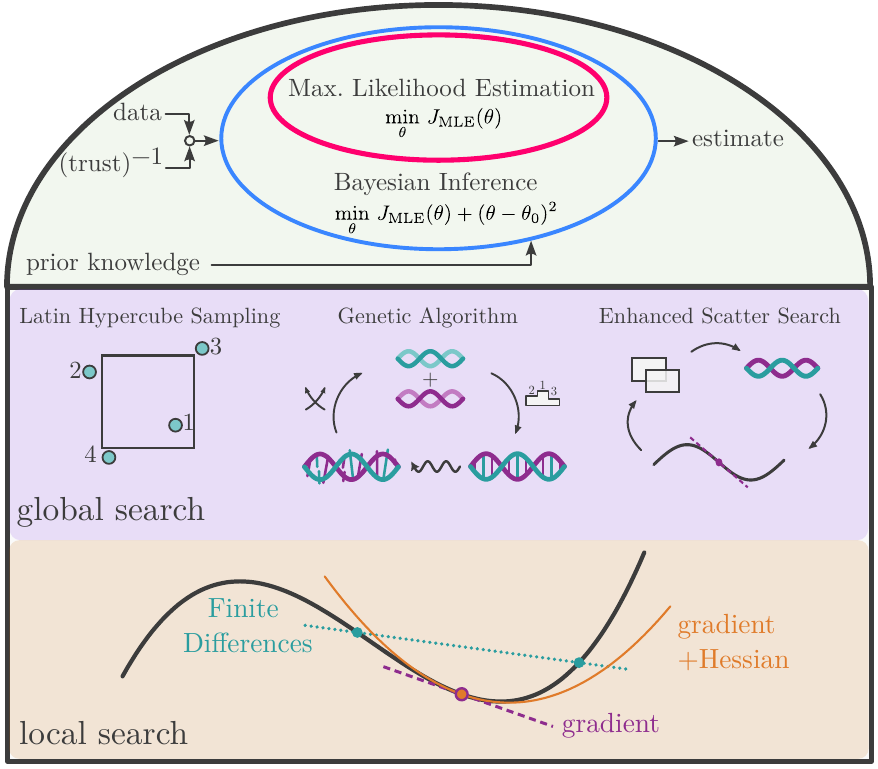}
\end{figure*}

\section*{Keywords}
Biogas technology, Gradient-based optimization, A-posteriori, Global optimization, Hessian, Sloppiness, Agricultural substrates
\newpage

\section{Introduction}
Anaerobic digestion (AD) is a biological process which degrades biogenic substrates to biogas, a mixture of mostly methane and carbon dioxide, via the formation of organic acids~\cite{nathia-neves2018,mo2023}. Agricultural AD plants 
provide a controllable source of renewable electricity and heat~\cite{grim2015}, and recover nutrients from agricultural residues~\cite{theuerl2019}. The methane produced can be converted to electricity in internal combustion engines~\cite{zhang2021} or fuel cells~\cite{tamburrano2024}, or upgraded and injected into the gas grid~\cite{steindl2025}. Moreover, intermediates such as organic acids can be separated and upgraded to products of high value to the pharmaceutical and chemical industry, placing AD as a central element in biorefineries~\cite{zhu2026}. 

Full-scale AD plants were conventionally operated at steady state~\cite{theuerl2019}, aiming at maximum methane production for base-load electricity production~\cite{mauky2017}. The uprise of other, volatile renewable sources such as wind and solar, however, requires AD plants to adopt more innovative and flexible operating strategies to sustain their economic viability~\cite{nikolausz2020}. 
These strategies include producing biogas on-demand~\cite{hahn2014}, substrate flexibility~\cite{theuerl2019}, or targeting of organic acids production~\cite{nikolausz2025}. However, to avoid process inhibition and failure, flexible process operation requires reliable monitoring and control~\cite{ahmed2020,mo2023,neubauer2025}. Due to the highly nonlinear nature of the AD process, these depend on reliable process models to deliver accurate process predictions~\cite{kegl2025}. 

The Anaerobic Digestion Model No 1 (ADM1)~\cite{batstone2002} is the most prominent AD model and describes all four steps of the AD process (hydrolysis, acidogenesis, acetogenesis, and methanogenesis)~\cite{mo2023}. However, the sparse and inaccurate measurements in full-scale AD operation contrast with the ADM1's high number of uncertain process parameters and states, which render it prohibitive for model-based control and monitoring applications in regular operation~\cite{ahmed2020}. Therefore, Weinrich and Nelles~\cite{weinrich2021} proposed a variety of ADM1 reductions, which maintain the essence of the original stoichiometry, and transform the ADM1 reference unit of chemical oxygen demand (COD) to mass basis. Among these model reductions, the ADM1-R3 describes AD as a two-step process (hydrolysis and methanogenesis), and captures inhibitions through pH, ammonia, and nitrogen limitation. The ADM1-R3 thus provides a promising alternative to the model of Bernard et al.~\cite{bernard2001}, which is frequently applied in modeling and control studies in the AD literature~\cite{ahmed2020,corbellini2019,garcia-sandoval2016}, but is formulated in COD and uses a hypothetical stoichiometry.

Despite the ADM1's level of detail, 
not all of its parameters are uniquely defined by first principles~\cite{kegl2025}. Instead, specific parameters such as hydrolysis rate and inhibition constants depend on operational process parameters and utilized uncertain substrates, and therefore require calibration~\cite{segura2025,batstone2009,tisocco2024}. This is even more the case for reduced-order AD models, which summarize nutrient fractions and cluster functional microbial groups, resulting in lumped parameters~\cite{carecci2026}. 

Parameter identification (PI) of biological processes, 
and specifically AD, is challenging due to practical identifiability issues~\cite{villaverde2019, brun2001, lopezc.2015} caused by 
uncertain measurements (such as uncertain influent concentrations~\cite{tisocco2024, segura2025} and noisy multirate measurements~\cite{durr2021, jimenez2015}), plant-model mismatch~\cite{wilmsterrance2026}, 
computational complexity (including stiff, nonlinear process models~\cite{donoso-bravo2013, weinrich2021a}, and large numbers of model parameters and states~\cite{donoso-bravo2011, villaverde2019}), 
and the existence of multiple local optima~\cite{donoso-bravo2013, villaverde2019, frohlich2017a}.

To identify model parameters of biological and specifically AD processes, the framework of maximum likelihood estimation (MLE) is frequently applied~\cite{barz2018, lopezc.2015, frohlich2017a, donoso-bravo2013, delory2025}. 
MLE seeks the parameters under which the observed data are most probable. In practice, MLE leads to a minimization problem of a weighted sum of squares of residuals. 
The associated estimation uncertainty can be found via the Cramér-Rao bound, which uses the Hessian of the MLE objective function (OF)~\cite{lopezc.2015}.

Many uncertain AD model parameters are not entirely unknown but well documented in literature~\cite{kegl2025,weinrich2021,donoso-bravo2011}. A PI framework that leverages this prior knowledge is Bayesian inference (BI). BI can be thought of as an extension to MLE, as it takes into account not only residuals, 
but also considers prior knowledge about model parameters~\cite{walter1997}. However, BI is rarely pursued in AD~\cite{liu2026, picard-weibel2024, donoso-bravo2013}. 
An efficient method to quantify estimation uncertainty in BI is Laplace's method, which also uses the Hessian of the OF, and is thus highly similar to MLE~\cite{yoko2026}.



Both MLE and BI cast PI into nonlinear optimization problems. Various optimization methods are available in the literature to solve these~\cite{lopezc.2015,frohlich2017,egea2014}. 
AD-specific studies 
frequently apply 
simple gradient-free methods such as plain simplex algorithms~\cite{donoso-bravo2011} or use finite difference (FD) approximations in gradient-based approaches~\cite{donoso-bravo2013,frohlich2017a}. 
However, significant potential for improved search efficiency and higher precision lies in the use of gradients obtained via numerically elaborate methods or analytical derivations~\cite{dierkes2013}. Even larger improvements can be found when using curvature information from second-order derivatives. 
Analytical gradients, for instance, improve the search efficiency substantially over FD~\cite{mitra2019,almquist2015}, and even more so the use of second-order curvature information~\cite{stapor2018}. 

Additionally, many applications do not consider global optimization strategies, and thus risk to identify parameter values based on suboptimal, local minima~\cite{tisocco2024,kegl2022,batstone2009}. Some exceptions can be found in~\cite{donoso-bravo2011} and~\cite{denobel2024}, which use multistart strategies to address this~\cite{weinrich2021a}. 
Alternative global search approaches were shown to be valuable in 
biological modeling~\cite{frohlich2017a}. These approaches include evolutionary algorithms such as genetic algorithm~\cite{wichern2009}, particle swarm optimization~\cite{wolf2009}, or enhanced scatter search (ESS)~\cite{egea2014}.  


Best practices for PI of biological processes have been proposed by Villaverde et al.~\cite{villaverde2019}. The guidelines include maximum likelihood estimation (MLE), analytical gradient information for gradient-based solvers, log10-scaling of parameters, and ESS as global search. 

This paper applies the essence of these best practices to an experimental multirate dataset of a lab-scale AD reactor utilizing agricultural residues and energy crops. In addition, the best practices are complemented by improving gradient-based search with the Gauss-Newton Hessian approximation, and parallel computing for numerical efficiency. 
Furthermore, prior knowledge about parameters is utilized by applying BI, which is demonstrated as a viable extension to MLE. 
Moreover, both frameworks are highlighted to deliver approximately Gaussian parameter estimates, resulting from second-order approximations.   
Finally, the paper provides crucial implementation guidance relevant to practitioners aiming for efficient and accurate parameter estimation of biological models, and in particular AD models. 


\section{Methods}
\subsection{Anaerobic digestion process}

\subsubsection{Experimental setup} 
\begin{figure}
\centering
\includegraphics[width=0.4\linewidth]{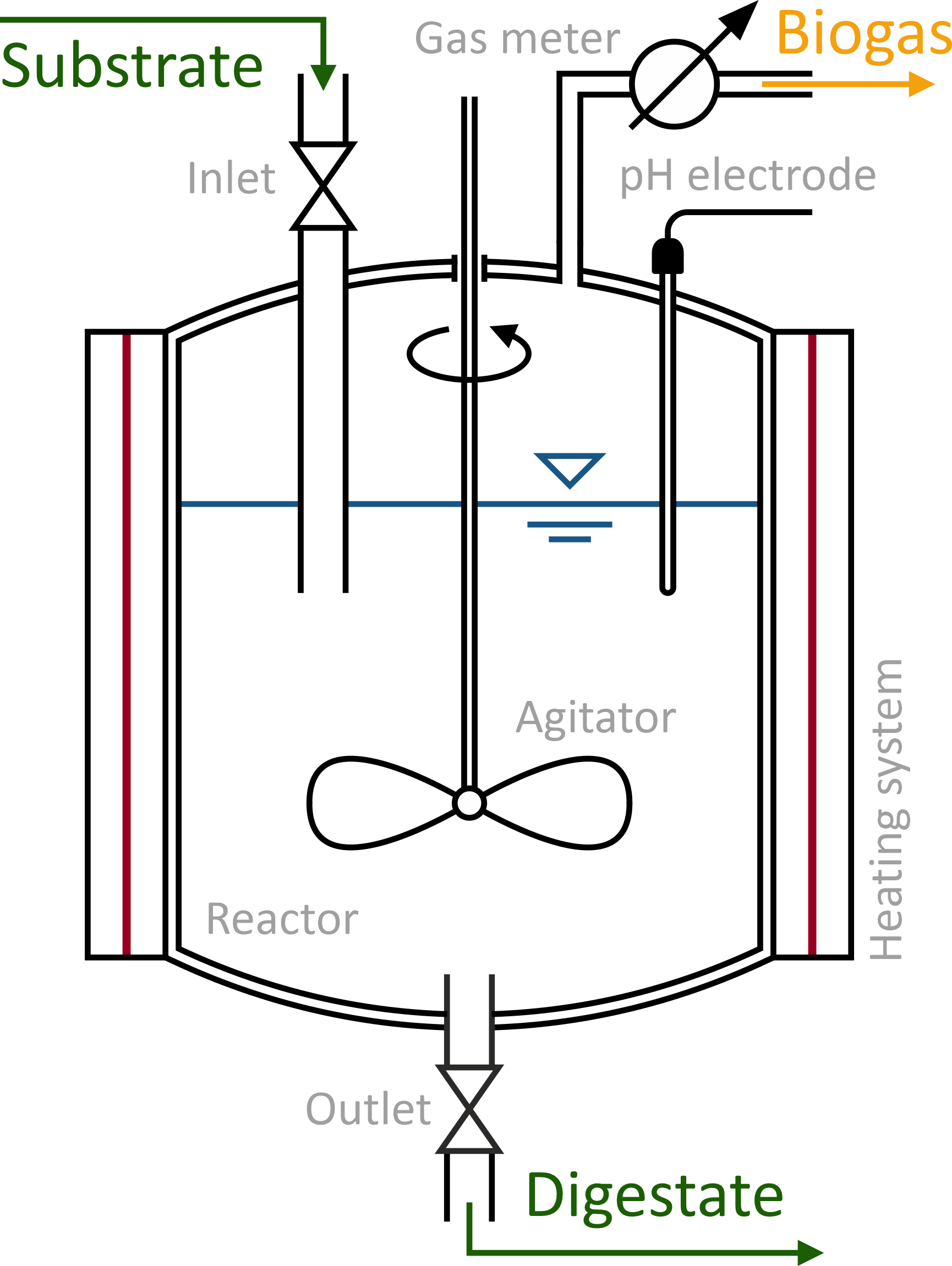}
\caption{Lab-scale anaerobic digestion reactor with substrate feeding, online measurements and offline digestate sampling.}
\label{fig:meth:reactor}
\end{figure}

\begin{figure}
\centering
\includegraphics[width=\linewidth]{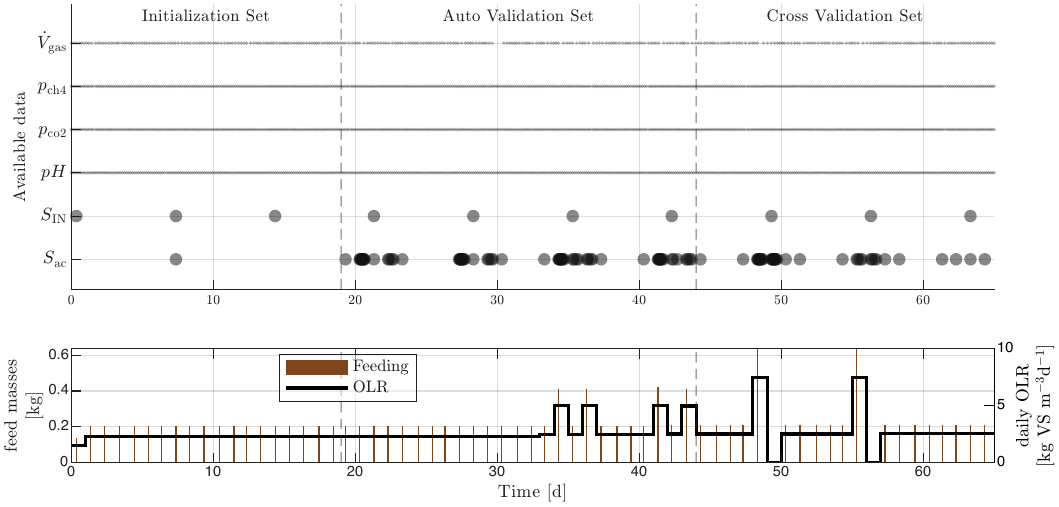}
\caption{Data partitioning, density of measurement samples, and substrate feedings by means of feed masses and daily organic loading rate (OLR) of the complete dataset. The data was partitioned into initialization, auto validation and cross validation datasets. Substrates were fed once a day except days 49 and 56. Only \SI{5}{\permille} of online samples are shown for better visibility. Offline samples are shown as semi-transparent grey dots, hence increasingly dark shades represent sampling at elevated frequency.}
\label{fig:meth:feeding}
\end{figure}

A lab-scale AD reactor with a total volume of \SI{15}{\liter} (\SI{3}{\liter} headspace) digested maize silage (MS) and cattle manure (CM) at a mesophilic temperature of \SI{40}{\celsius}, Fig.~\ref{fig:meth:reactor}. The reactor was semi-continuously fed once per day, and maintained an average organic loading rate (OLR) of 3\,$\mathrm{kg\,VS \,m}^{-3} \,\mathrm{d}^{-1}$ 
throughout the experiment (Fig.~\ref{fig:meth:feeding}). VS denotes volatile solids. MS and CM contributed to the OLR by 80/\SI{20}{\percent} based on VS. Digestate from a full-scale AD plant inoculated the reactor. 

Biogas production $\dot V_\mathrm{gas}$ was measured by volumetric displacement using a magnetic piston (BlueVCount, BlueSens GmbH, Germany). Partial pressures $p_\mathrm{ch4}$ and $p_\mathrm{co2}$ of methane (\chfour) and carbon dioxide (\cotwo) were measured using infrared sensors (BlueVary, BlueSens GmbH, Germany). Furthermore, a polymer electrode recorded pH data (SE 554 \textsc{Memosens}, \textsc{Knick} GmbH, Germany). All online measurements ($\dot V_\mathrm{gas}$, $p_\mathrm{ch4}$, $p_\mathrm{co2}$, and pH) were recorded at a \SI{1}{\minute} sample time.

Digestate samples were taken manually, centrifuged, and analyzed for total inorganic nitrogen (IN) using a photometer (\textsc{Hach} GmbH, Austria)~\cite{strach2020} and for acetic acid (AC) concentrations using a headspace gas chromatograph (\textsc{Agilent Technologies}, USA)~\cite{apelt2020}. Table~\ref{tab:meth:measurements} lists standard deviations (SD) and sample times of measurements. 
Delory et al.~\cite{delory2025} reported details on the experimental conditions.
\begin{table}
\begin{threeparttable}
\centering
\renewcommand{\arraystretch}{1.3}
\caption{Measurement signals $y$ with standard deviations $\sigma_\mathrm{y}$ and sample times.}
\label{tab:meth:measurements}
\begin{tabularx}{\linewidth}{lYYY}
    \toprule
    Symbol$^{\mathrm{a,d}}$ & $\sigma_\mathrm{y}^{\mathrm{b}}$ & Unit & Sample time$^{\mathrm{e}}$ \\
    \midrule
    $\dot V_\mathrm{gas}$ & 0.0004 & \si{\cubic\meter\per\day} & \SI{1}{\minute} \\
    $p_\mathrm{ch4}$ & 0.0178 & \si{\bar} & \SI{1}{\minute} \\
    $p_\mathrm{co2}$ & 0.0268 & \si{\bar} & \SI{1}{\minute} \\
    $pH$ & 0.02 & $-$ & \SI{1}{\minute} \\
    \cdashline{1-4}
    $S_\mathrm{IN}$ & 0.12$^{\mathrm{c}}$& \si{\kilogram\per\cubic\meter} & \SI{7}{\day} \\
    $S_\mathrm{ac}$ & 0.05$^{\mathrm{c}}$ & \si{\kilogram\per\cubic\meter} & \SI{1}{\hour} -- \SI{12}{\day}  \\
    \bottomrule
\end{tabularx}
\begin{tablenotes}
\footnotesize
\item[a] In row order: biogas production $\dot V_\mathrm{gas}$, partial pressure of methane/carbon dioxide $p_\mathrm{ch4}$, $p_\mathrm{co2}$, pH value, concentration of total inorganic nitrogen $S_\mathrm{IN}$, effective acetic acid concentration $S_\mathrm{ac}$ (mole-based sum of acetic through valeric acid, cf. \siref{sec:si:influent}). Symbols according to Weinrich and Nelles~\cite{weinrich2021}.

\item[b] Source: Online sensor data sheets (BlueSens BlueVCount (biogas production), BlueSens BlueVary (\chfour), and \textsc{Knick} \mbox{SE 554} \textsc{Memosens} (pH)), assuming mean biogas production of \SI{20}{\litre\per\day} with 52/\SI{48}{\percent} \chfour/\cotwo.

\item[c] Standard deviations of offline measurements were determined based on triplicates drawn throughout a 6-week steady-state operation 
with sample sizes of 9 ($S_\mathrm{IN}$) and 14 ($S_\mathrm{ac}$).

\item[d] Online measurements of 
$p_\mathrm{co2}$ were available but disregarded except in \siref{sec:res:ch_fractionation}. This is because they overdetermine the output equations of the gas phase in the case that $\dot V_\mathrm{gas}$ and $p_\mathrm{ch4}$ are available yet, cf. Eqs.~\sieqref{eq:si:r3-core:y} in \siref{sec:si:model_equations}. Two out of three gas phase outputs determine the gas phase's states uniquely.

\item[e] AC sampling frequency varied widely throughout the experiment, as shown in Fig.~\ref{fig:meth:feeding}, with a mean of \SI{13}{\hour} and a median of \SI{2}{\hour}.
\end{tablenotes}
\end{threeparttable}
\end{table}
\subsubsection{Anaerobic digestion models}
\label{sec:meth:ad_models}
%

The ADM1-R3 describes the AD process in three stages. Macronutrients (carbohydrates (CH), proteins, and lipids) are hydrolyzed, thus decomposed to monomers, and further degraded to organic acids. These further degrade to acetic acid. Finally, acetoclastic methanogenesis produces biogas, i.e., methane (\chfour) and carbon dioxide (\cotwo)~\cite{weinrich2021}. 

Two variants of the ADM1-R3 were considered. First, the ADM1-R3-core neglects water and ions as state variables~\cite{hellmann2026}. Second, the ADM1-R3-core-frac additionally splits CH into a rapidly and a slowly degradable fraction~\cite{delory2025,waszkielis2022}. A set of continuous-time ordinary differential equations (ODEs) and 
discrete-time output equations 
describes each model in the form
\begin{subequations}
\begin{align}
    \dot{x} &= f(x, \theta, u, \xi), \quad x(t_{0}) = x_{0}, \label{eq:meth:ode_x} \\
    y_k &= g(x(t_k), \theta) + \epsilon_k . \label{eq:meth:output}
\end{align}
\end{subequations}
%
$u \in \mathbb{R}_{\ge 0}$ denotes feed volume flow with associated influent concentrations $\xi\in \mathbb{R}^{n_\mathrm{x}}_{\ge 0}$. Furthermore, \mbox{$x\in \mathbb{R}^{n_\mathrm{x}}_{\ge 0}$}, $y \in \mathbb{R}^{n_\mathrm{y}}_{\ge 0}$, and $\theta\in \mathbb{R}^{n_\mathrm{\theta}}$ denote states, outputs, and tunable model parameters. 
$\epsilon_k \sim \mathcal N (0, C_{\mathrm{y},k})$ denotes additive Gaussian noise with diagonal covariance $C_{\mathrm{y},k}$~\cite{martin2010} (Table~\ref{tab:meth:measurements}). 
The full sets of model equations are provided in {} \siref{sec:si:model_equations}. Tunable model parameters $\theta$ are listed in Table~\ref{tab:meth:parameters}.  
In-house substrate characterization delivered influent concentrations $\xi$, which were assumed as known and fixed, cf. Table~\xref{tab:si:influent_init} in \siref{sec:si:data_preprocessing}. Computations for AC, IN, and macronutrients are detailed in \siref{sec:si:influent}. 
All other components were neglected in the influent.
\begin{table}
\begin{threeparttable}
\centering
\renewcommand{\arraystretch}{1.1} 
\caption{Model parameters $\theta$ subject to calibration, along with bounds and prior mean and standard deviation.}
\label{tab:meth:parameters}
\begin{tabularx}{\linewidth}{lccYYd{2}d{4}d{3}d{1}}
\toprule
& \multicolumn{2}{c}{Index} & & & \multicolumn{2}{c}{Prior$^{\mathrm{c,d}}$} & \multicolumn{2}{c}{Bounds$^{\mathrm{e}}$} \\
\cmidrule(lr){2-3} \cmidrule(lr){6-7} \cmidrule(lr){8-9}
Symbol$^{\mathrm{a}}$ & R3-core$^{\mathrm{b}}$ & R3-frac$^{\mathrm{b}}$ & Scale & Unit & \multicolumn{1}{c}{$\mu$} & \multicolumn{1}{c}{$\sigma$} & \multicolumn{1}{c}{min} & \multicolumn{1}{c}{max} \\
\midrule
$k_\mathrm{ch}$ & 1 & & log10 & \si{\per\day} & 0.6 & 0.6 & 0.001 & 100 \\
$k_\mathrm{ch,f}$ & $-$ & 1 & log10 & \si{\per\day} & 67.57 & 3.3785 & 0.01 & 100 \\
$k_\mathrm{ch,s}$ & $-$ & 2 & log10 & \si{\per\day} & 0.81 & 0.4374 & 0.001 & 10 \\
$k_\mathrm{pr}$ & 2 & 3 & log10 & \si{\per\day} & 0.12 & 0.1344 & 0.001 & 20 \\
$k_\mathrm{li}$ & 3 & 4 & log10 & \si{\per\day} & 0.63 & 0.3024 & 0.0001 & 10 \\
$k_\mathrm{dec}$ & 4 & 5 & log10 & \si{\per\day} & 0.02 & 0.0292 & 0.0001 & 5 \\
$\mu_\mathrm{m,ac}$ & 5 & 6 & log10 & \si{\per\day} & 0.71 & 0.5183 & 0.01 & 5 \\
$K_{\mathrm{S,ac}}$ & 6 & 7 & log10 & \si{\kilogram\per\cubic\meter} & 2.18 & 0.5014 & 0.001 & 10 \\
$K_{\mathrm{I,nh3}}$ & 7 & 8 & log10 & \si{\kilogram\per\cubic\meter} & 0.06 & 0.0312 & 0.001 & 5 \\
$\Delta S_\mathrm{ion,eff}$ & 8 & 9 & abs. & \si{\kilo\mole\per\cubic\meter} & 0.13 & 0.0875 & -0.05 & 0.3 \\
$\varphi_\mathrm{IN,in}$ & 9 & 10 & abs. & $-$ & 1 & 0.125 & 0.75 & 1.25 \\
$\varphi_\mathrm{ch,in,f}$ & $-$ & 11 & abs. & $-$ & 0.32 & 0.0192 & 0 & 1 \\
\botrule
\end{tabularx}
\begin{tablenotes}
\footnotesize
\item[a] In row order: $k_\mathrm{ch}$, $k_\mathrm{pr}$, $k_\mathrm{li}$: hydrolysis constants of carbohydrates (CH), proteins, lipids. $k_\mathrm{ch,f}$, $k_\mathrm{ch,s}$: hydrolysis constants of fast and slow CH fraction. $\varphi_\mathrm{ch,in,f}$: share of total influent CH allocated to the fast CH fraction. $k_\mathrm{dec}$: first-order decay constant of microbial biomass. $\mu_\mathrm{m,ac}$ and $K_{\mathrm{S,ac}}$: max. growth rate and half saturation constant of acetoclastic methanogens. $K_{\mathrm{I,nh3}}$: free ammonia inhibition constant. $\Delta S_\mathrm{ion,eff}$: effective cation concentration. $\varphi_\mathrm{IN,in}$: amplification factor of influent inorganic nitrogen. Details are provided in \siref{sec:si:model_equations}. Symbols according to \cite{weinrich2021} and \cite{hellmann2026}. 

\item[b] R3-core: ADM1-R3-core, R3-frac: ADM1-R3-core-frac.

\item[c] The prior denotes a typical literature value. In the BI framework, it was assumed to follow a Gaussian distribution. Mean $\mu$ and standard deviation $\sigma$ refer to linear-scale values of the prior distribution. Squared standard deviations give variances, i.e., the diagonal entries of $C_{\theta \theta,0}$, cf. Sec.~\ref{sec:meth:BI}.

\item[d] Values of $k_\mathrm{ch,f}, k_\mathrm{ch,s}, k_\mathrm{pr}, k_\mathrm{li}, k_\mathrm{dec}, \mu_\mathrm{m,ac}, K_{\mathrm{S,ac}}, K_{\mathrm{I,nh3}}$, and $\varphi_\mathrm{ch,in,f}$ taken from Delory et al. \cite[Table~4]{delory2025}: median fit interpreted as $\mu$; $\sigma$ derived from coefficients of variation and median fit. Values of $k_\mathrm{ch}$ taken from Tisocco et al. \cite[Table~4]{tisocco2024}. Prior values and bounds of $\Delta S_\mathrm{ion,eff}$ and $\varphi_\mathrm{IN,in}$ chosen by the authors. 

\item[e] Authors' choice, inspired by Weinrich and Nelles~\cite{weinrich2021} and Delory et al.~\cite{delory2025}.
\end{tablenotes}
\end{threeparttable}
\end{table}

Provided all online and offline measurements are available, the ORC-DF algorithm of the STRIKE-GOLDD toolbox~\cite{martinez2020} showed that both model variants are locally observable and structurally identifiable~\cite{hellmann2023}. 
\subsubsection{Data preprocessing and partitioning}
\label{sec:meth:data_preprocessing}
Raw experimental measurement data was processed via the first step of the pipeline of Meola et al.~\cite{meola2023}. Particularly, this covered removal of outliers 
and missing data handling. Additional preprocessing steps involve the conversion of raw substrate characterization data to influent concentrations $\xi$, cf. \siref{sec:si:data_preprocessing}.  

 
The excerpt of the total experimental dataset considered spans a period of \SI{65}{\day}, owing to its high density of offline digestate analyses and consecutive online measurements, Fig.~\ref{fig:meth:feeding}. Thereof, a subset of \SI{25}{d} was selected as auto validation (training), and a subset of \SI{21}{\day} as cross validation (test) data. An initialization set of \SI{19}{d} preceds the training data. 

\subsection{Algorithms}
\label{sec:meth:algorithms}
%
Parameter identification (PI) was pursued using two opposing frameworks: maximum likelihood estimation (MLE) and Bayesian inference (BI). Both 
%
BI and MLE 
can be illustrated by means of Bayes' theorem which describes the \textit{a-posteriori} probability density function (pdf) $p(\theta|\mathcal Y)$ of parameters given the observed data\ifarxiv\footnote{An additional conditioning on the hypothesized model $\mathcal M$ is suppressed for simplicity.}\fi{} $\mathcal Y$ (posterior), as a consequence of the likelihood pdf $p(\mathcal Y| \theta)$ of the data, the \textit{a-priori} assumption on the parameter pdf $p(\theta)$ (prior), and the model evidence $p(\mathcal Y)$
\begin{equation}
    p(\theta|\mathcal Y) = \frac{p(\mathcal Y|\theta) p(\theta)}{p(\mathcal Y)}.
\label{eq:meth:bayes}
\end{equation}
While MLE considers parameters as deterministic, BI perceives them as random variables~\cite{donoso-bravo2011}. Specifically, BI maximizes the \textit{a-posteriori} pdf of Eq.~\eqref{eq:meth:bayes}~\cite{garnett2023}, delivering a maximum \textit{a-posteriori} estimate (MAP). By contrast, MLE is agnostic of (prior) knowledge about parameter distributions, and instead maximizes the likelihood $p(\mathcal Y|\theta)$, giving a maximum likelihood estimate~\cite{villaverde2019,frohlich2017a}. 

In both MLE and BI, Gaussian noise is frequently assumed~\cite{martin2010,donoso-bravo2011,ninness2010}, Eq.~\eqref{eq:meth:output}. In BI, parameters are commonly assumed to be distributed uniformly~\cite{martin2010} or Gaussian~\cite{yoko2026,picard-weibel2024,carecci2025}. The latter is considered in this work, cf. Sec.~\ref{sec:meth:log10}.
%
\subsubsection{Maximum likelihood estimation}
\label{sec:meth:MLE}
MLE seeks those parameters $\theta_\mathrm{MLE}$ that maximize the likelihood $p(\mathcal Y|\theta)$ of observed measurements
\begin{equation}
    \theta_\mathrm{MLE} = \arg \max_{\theta} \, p(\mathcal Y|\theta). \label{eq:meth:mle_orig}
\end{equation}
This equates to minimizing the negative log-likelihood~\cite{frohlich2017a}. 
The measurement data $\mathcal Y$ is assumed to originate from a stochastic process whose deterministic part is given by the model, Eq.~\eqref{eq:meth:output}. 
Then for $N$ independent samples, Eq.~\eqref{eq:meth:mle_orig} can be expressed as a weighted least squares problem 
\begin{subequations}
\label{eq:meth:mle_problem}
\begin{align} 
    \theta_\mathrm{MLE} &= \arg \min_{\theta} \, J_\mathrm{MLE} (\theta), \, s.t. \label{eq:meth:mle_problem:theta}
    \\
    J_\mathrm{MLE}(\theta) &= - \ln (p(\mathcal Y|\theta)) = \frac{1}{2} \sum_{k=1}^{N} \left(y_k - g_k\right)^T \, \left( W_k \, C_{\mathrm{y},k}\right)^{-1} \, \left(y_k - g_k \right), \label{eq:meth:mle_obj_fun} \\
    \theta_\mathrm{min} &\le \theta \le \theta_\mathrm{max}.
\end{align}
\end{subequations}
%
$W_k$ accounts for fair contribution of online and offline measurements by holding on its diagonal the total number of samples $N_l$ per model output $l=1,\ldots, n_\mathrm{y}$. This is combined with the noise covariance as per Table~\ref{tab:meth:measurements} to form a weighting matrix $\tilde C_k$\ifarxiv{}, which simplifies the notation\fi
\begin{equation}
    \tilde C_k = W_k \, C_{\mathrm{y},k}. \label{eq:meth:weighting_matrix}
\end{equation}
%
Output-dependent variables $y_k, g_k$, and $\tilde C_k$ are masked for those 
entries available at time $t_k$, formulating a variable structure multirate output~\cite{kramer2005}. Eq.~\sieqref{eq:si:obj_fun_mle_mr} in \siref{sec:si_meth:algorithms} shows a dedicated multirate notation of the objective function (OF). $\theta_\mathrm{min}$ and $\theta_\mathrm{max}$ denote lower and upper bounds of the parameters (Table~\ref{tab:meth:parameters}). 

Derivations of analytical gradient $\nabla_{\theta} J_\mathrm{MLE}$ and Hessian approximation $\nabla^2_{\theta} J_\mathrm{MLE}$ according to the Gauss-Newton algorithm involve output sensitivities $S_\mathrm{y}$ and are given in \siref{sec:meth_si:grad_hess_derivation}. 

\subsubsection{Bayesian inference} 
\label{sec:meth:BI}
With Bayes' theorem in Eq.~\eqref{eq:meth:bayes} and a Gaussian prior of mean $\theta_0$ and covariance $C_{\theta\theta,0}$, \mbox{$p(\theta) \sim \mathcal{N}(\theta_0, C_{\theta\theta,0})$}, a MAP estimate can be found by minimizing the negative log posterior pdf 
%
\begin{subequations} \label{eq:meth:map_problem}
\begin{align}
    \theta_\mathrm{MAP} &= \arg \min_{\theta} J_\mathrm{MAP} (\theta), \, s.t.
    \\
    J_\mathrm{MAP}(\theta) &= - \ln \left(p(\theta|\mathcal Y) \right) = \frac{1}{2} \sum_{k=1}^{N} \left(y_k - g_k\right)^T \, \tilde C_k^{-1} \, \left(y_k - g_k \right) + \frac{1}{2} (\theta - \theta_0)^T C_{\theta\theta,0}^{-1} (\theta - \theta_0), \label{eq:meth:map_obj_fun} \\
    \theta_\mathrm{min} &\le \theta \le \theta_\mathrm{max}.
\end{align}
\end{subequations}
Eq.~\eqref{eq:meth:map_obj_fun} can be viewed as an extension of the MLE OF (Eq.~\eqref{eq:meth:mle_obj_fun}), which translates into analogous expressions of the analytical gradient $\nabla_{\theta} J_\mathrm{MAP}$ and the Hessian approximation $\nabla^2_{\theta} J_\mathrm{MAP}$ described in \siref{sec:meth_si:grad_hess_derivation}. 

\subsubsection{Uncertainty quantification and propagation}
\label{sec:meth:uncertainty_propagation}
The Gauss-Newton Hessian approximation of the MLE OF (cf. Eq.~\sieqref{eq:meth:hessian_mle} in \siref{sec:si_meth:algorithms}) is equivalent to the common definition of the Fisher information matrix $C_\mathrm{FIM}$~\cite{heine2008,lopezc.2015}. Its inverse delivers a proxy for the MLE parameter covariance matrix, which is known as the Cramér-Rao bound~\cite{lopezc.2015,ljung1979} 
\begin{equation}
    C_{\theta\theta, \mathrm{MLE}} \approx C_\mathrm{FIM}^{-1} = (\nabla^2_{\theta} J_\mathrm{MLE})^{-1}.
    \label{eq:meth:cramer_rao}
\end{equation}
Although MLE considers deterministic parameters, its estimates can be shown to be asymptotically Gaussian~\cite{ljung1979}. This renders $C_{\theta\theta, \mathrm{MLE}}$ the covariance of a multivariate Gaussian centered at $\theta_\mathrm{MLE}$~\cite{lopezc.2015}.

Around the MAP estimate $\theta_\mathrm{MAP}$ of BI, the posterior distribution $p(\theta|\mathcal Y)$ can be approximated by a Gaussian, the covariance of which is given by the inverse of the associated Hessian~\cite{yoko2026} 
\begin{equation}
    C_{\theta\theta, \mathrm{MAP}} \approx (\nabla^2_{\theta} J_\mathrm{MAP})^{-1}.
    \label{eq:meth:laplace}
\end{equation}
This follows from expanding the negative log posterior, $J_\mathrm{MAP}$, around $\theta_\mathrm{MAP}$ up to second order and considering that first-order derivatives vanish in the optimum \cite{yoko2026}. 
This is referred to as Laplace's method, and provides a direct analogy to the Cramér-Rao bound of the MLE covariance (Eq.~\eqref{eq:meth:cramer_rao}). 

\paragraph{Uncertainty propagation}
A first-order Taylor approximation of the model output $y$ (Eq.~\eqref{eq:meth:output}) delivers the time-variant output covariance $C_{\mathrm{yy},k}$ at time $t_k$ with  
\begin{equation}
    C_{\mathrm{yy},k} = S_{\mathrm{y},k} \, C_{\theta \theta} \, S_{\mathrm{y},k}^T.
    \label{eq:meth:uncertainty_propagation}
\end{equation}
It describes the output uncertainty resulting from output sensitivities $S_\mathrm{y}$ and parameter uncertainty $C_\mathrm{\theta\theta}$, for which the covariance matrix of MLE or MAP estimates must be inserted (Eqs.~\eqref{eq:meth:cramer_rao} and \eqref{eq:meth:laplace}). 

\subsubsection{Global optimization}
The availability of gradient and Hessian information improves the use of gradient-based optimization methods for the optimization problems in Eqs.~\eqref{eq:meth:mle_problem} and \eqref{eq:meth:map_problem}~\cite{mitra2019,stapor2018}. 
Both OFs, $J_\mathrm{MLE}$ and $J_\mathrm{MAP}$, can show multiple local minima~\cite{donoso-bravo2013,frohlich2017a}, which entails the risk of premature termination. To prevent this, 
three global optimization algorithms were considered: Latin hypercube sampling (LHS), genetic algorithm (GA), and ESS~\cite{egea2014}. All three have been reported as powerful tools in global calibration of biological models~\cite{wichern2009, villaverde2019, frohlich2017a,wilmsterrance2026}.  

LHS assigns a predefined number of samples to a multidimensional search space by dividing it into evenly spaced intervals along each dimension, from which random numbers are selected and then permuted randomly~\cite{mckay1979}. LHS samples the search space uniformly, and thus defines unbiased multistart locations without requiring prior knowledge beyond parameter bounds.  

GA is an evolutionary approach inspired by natural selection, which iterates across the search space via populations of candidate solutions~\cite{katoch2021,el-mihoub2006}. Individual candidates are evaluated by their OF value (\textit{fitness score}), combined to form new candidates (\textit{offspring}), and undergo random perturbations (\textit{mutation}) until a satisfactory solution is found or a fixed budget is exceeded. Like LHS, GA does not require prior knowledge about the search space except its bounds. 

Lastly, ESS combines the concepts of gradient-free, evolutionary search with gradient-based local search~\cite{egea2014}. An initial population (\textit{reference set}) is established via LHS, and propagated via alternating iterations of a diversification (global search) and an improvement method (local search). As for GA, iterations progress up to an accuracy or budget threshold.


\subsection{Implementation}
\label{sec:meth:implementation}
\subsubsection{Log10 scaling}
\label{sec:meth:log10}
When calibrating biological models over a large, multidimensional search space, 
converting parameters to log10 scale is advised~\cite{villaverde2019,raue2013}. This improves the numerical conditioning of the problem by equalizing effective solver step sizes in all dimensions of the parameter search space. 
However, when considering Gaussian approximations of parameter uncertainties as per Eqs.~\eqref{eq:meth:cramer_rao} and \eqref{eq:meth:laplace}, and Gaussian priors (in BI), log10 scaling implicitly changes the underlying parameter distribution from normal to log10-normal. 
This affects how priors need to be parameterized and how resulting parameter estimates and (co)variances are interpreted, i.e., retransformed to linear scale. 
These aspects are detailed in \siref{sec:si:log10_conversion}. 
\subsubsection{Global and local searches}
\begin{figure}
    \centering
    \begin{subfigure}[c]{0.50\linewidth}
        \includegraphics[width=\linewidth]{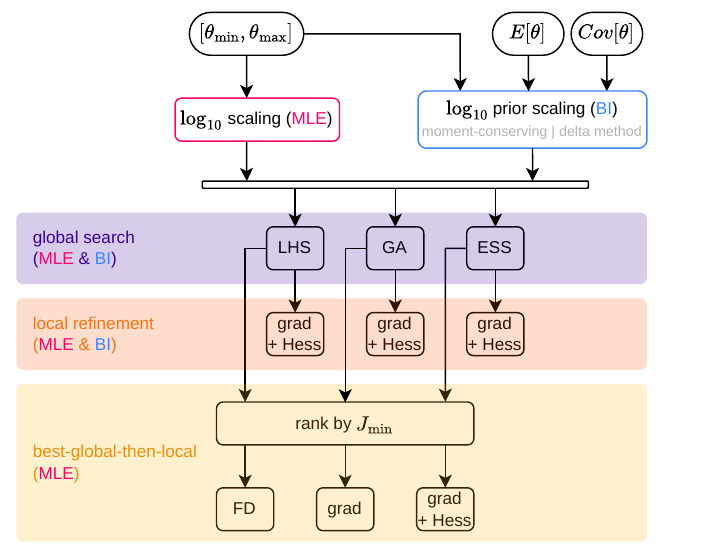}
        \caption{Global search, local refinement, and best-global-then-local search.}
        \label{fig:meth:metaheuristics_global_local}
    \end{subfigure}
    \begin{subfigure}[c]{0.49\linewidth}
        \includegraphics[width=\linewidth]{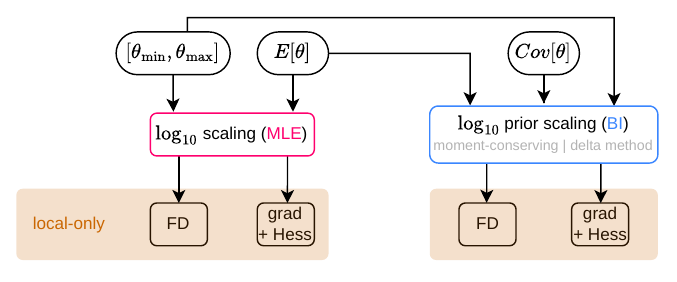}
        \caption{Local-only searches.}
        \label{fig:meth:metaheuristics_local_only}
    \end{subfigure}
    \caption{Blockflow diagrams of search configurations applied for MLE (magenta) and BI (blue): global search (purple), local refinement (orange), best-global-then-local search (yellow), and local-only search (brown). \newline 
    \footnotesize
    Abbreviations: MLE: maximum likelihood estimation, BI: Bayesian inference, LHS: Latin hypercube sampling, GA: genetic algorithm, ESS: enhanced scatter search, FD: finite differences, grad: gradient, Hess: Hessian. $E[\theta]$ and $Cov[\theta]$ denote expected value (mean) and covariance of uncertain model parameters $\theta$.} 
    \label{fig:meth:metaheuristics}
\end{figure}

Both MLE and BI were investigated by means of global and local search algorithms, 
Fig.~\ref{fig:meth:metaheuristics}. 
%
All schemes considered kinetic parameters in log10 scale, and remaining parameters in linear scale due to their smaller value ranges (Table~\ref{tab:meth:parameters}). For BI, the prior was transformed to log10 scale via conservation of moments (ADM1-R3-core) and via the delta method (ADM1-R3-core-frac), cf. \siref{sec:si:log10_conversion}. Both methods consider the prior means and covariances (Table~\ref{tab:meth:parameters}). 
MLE, by contrast, ignored distributions of the prior and instead directly applied log10 to parameters. 
\subsubsection{Global and local search}
\label{sec:meth_si:global_local_search_impl}
\paragraph{Global searches}
%
Three global search algorithms (LHS, GA, and ESS) were compared 
(purple box in Fig.~\ref{fig:meth:metaheuristics_global_local}). All global solvers considered the prior, i.e., as a distinct sample point (LHS), or as part of the initial population (GA, ESS). Moreover, all global solvers considered parameter bounds as per Table~\ref{tab:meth:parameters}. 

The number of OF evaluations was capped at 5000, which initial trials revealed as sufficient. LHS samples respected the log10 or linear scales of individual parameters. A GA population size of 100 ensured sufficient generations and marked a tradeoff between the default values of 50 or 200, which depend on the problem size~\cite{matlab.ga}. 
LHS and GA were implemented with \textsc{Matlab}'s functions \texttt{lhsdesign} and \texttt{ga}~\cite{matlab2024a}. The \textsc{MEIGO} toolbox provided a full \textsc{Matlab} implementation of ESS~\cite{egea2014}. Furthermore, ESS respected a computation time cap of \SI{250}{\minute}, found as suitable during initial trials. Other ESS hyperparameters remained at default values~\cite{egea2014}. 

%
\paragraph{Local searches}
To find a global optimum from within its proximity, local searches refined all global search results (orange box in Fig.~\ref{fig:meth:metaheuristics_global_local}). These local searches considered analytical gradient and Hessian information (Sec.~\ref{sec:meth:algorithms}). A maximum of 500 OF evaluations was found sufficient. \textsc{Matlab}'s \texttt{fmincon} function solved local searches using the interior point algorithm~\cite{byrd2000}.

Next, three configurations of approximating the OF's topology were compared (yellow box in Fig.~\ref{fig:meth:metaheuristics_global_local}): finite differences (FD), analytical gradients (grad) via forward sensitivities, and analytical gradients and Hessian (grad + Hess) via the Gauss-Newton approximation. All three started at the global search delivering the lowest OF value.

Local-only searches investigated the added value of preceding global searches (Fig.~\ref{fig:meth:metaheuristics_local_only}). Furthermore, the two approximations of OF curvature grad + Hess and FD were compared (brown box in Fig.~\ref{fig:meth:metaheuristics_local_only}). FD is often pursued in PI of AD models~\cite{donoso-bravo2013,frohlich2017a} for its low implementation effort compared to sensitivity-dependent analytical gradient and/or Hessian approximations. All local-only searches started at literature values given in Table~\ref{tab:meth:parameters} (prior column), where log10-scaling and prior conversion remained as for global searches. 

\subsubsection{Soft- and hardware}
The PI code was implemented in \textsc{Matlab} R2024a (The Mathworks, Inc.) using the global optimization, parallel computing, and symbolic math toolboxes~\cite{matlab2024a}. 

The data preprocessing pipeline was implemented \ifarxiv{}as a \textsc{Jupyter} notebook \fi{}in \textsc{Python} 3.13\ifarxiv{} using as its main packages pandas, numpy, scipy, sklearn, and matplotlib\fi{}~\cite{meola2023}. All code ran on a standard laptop under \textsc{Windows} 11 (\textsc{Intel} core i5-1335U processor, 10 cores, 32GB RAM).


\section{Results and Discussion}
\label{sec:results}

Model fits obtained using the two PI frameworks, MLE and BI, are compared based on posterior means, while posterior uncertainties are examined with respect to (w.r.t.) their implications for model predictions. The analysis further addresses convergence and the associated implementation effort. \siref{sec:res:aspects} explores model and data extensions to gain additional insights into model applications.

\subsection{Global parameter estimation}
%
\subsubsection{Model performance} 
\begin{figure}
\centering
\includegraphics[width=\linewidth]{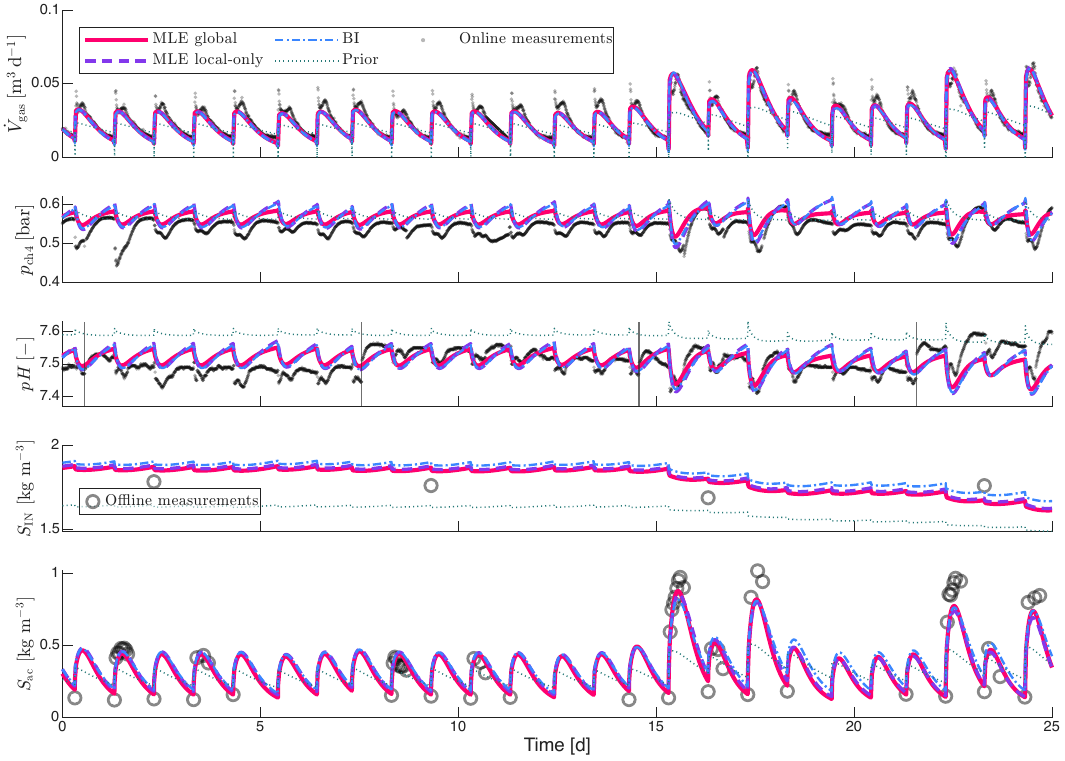}
\caption{Measurement data (auto validation dataset) and dynamic model ADM1-R3-core. Fits are shown for MLE and BI optima after global search with local refinement, after local-only search, and for literature prior. For BI, only one set of graphs is shown because resulting graphs overlap for global + local search and local-only search. Vertical grey lines in the pH tile represent re-calibration times of the pH electrode.}
\label{fig:res:fit_mle_bi}
\end{figure}

Both MLE and BI delivered meaningful parameter estimates for the ADM1-R3-core, as both captured the overall dynamics of the measurements well, Fig.~\ref{fig:res:fit_mle_bi}. 

BI yielded the same global optimum both for global search with local refinement (denoted \textit{global}) and local-only search (Table~\ref{tab:res:kpi_mle_vs_bi}). Consequently, only one graph is shown in blue in Fig.~\ref{fig:res:fit_mle_bi}. For MLE, by contrast, global search (magenta) gave a better fit than local-only search (purple), Fig.~\ref{fig:res:fit_mle_bi} and Table~\ref{tab:res:kpi_mle_global_local}. 

MLE and BI graphs for globally optimal estimates overlap for the most part, but notable differences exist for $p_\mathrm{ch4}$ and pH. While BI better maps the drops in both measurements following strong feeding events (days 15 and 17 in Fig.~\ref{fig:res:fit_mle_bi}), MLE graphs plateau faster in the aftermath of feedings, 
cf. days 19 and 24. A permanent bias remained for both fits of $p_\mathrm{ch4}$. 

For the pH, both fits showed alternating biases relative to the measurements at a weekly frequency. This coincides with weekly re-calibration of the pH electrode, indicated by vertical lines in the pH tile of Fig.~\ref{fig:res:fit_mle_bi}. These sudden changes in pH measurements are artefacts, which neither fit captured. This indicates overfitting is not present w.r.t. pH. 

These observations hold for simulations against the cross validation dataset, too, cf. Fig.~\xref{fig:si:fit_mle_bi_cross} in \siref{sec:si:res:plots_tables}. Here, the pH reached a maximum bias of 0.2 pH units at day 41. Step changes due to re-calibration remained visible (day 42), and the better $p_\mathrm{ch4}$ fit of MLE was confirmed (days 31, 38).

Short-term dynamics of biogas production were neither captured adequately by MLE nor BI, though residuals converged to zero before subsequent feedings. 

Intensive offline sampling for AC just after strong feeding events (days 15, 17, 23) revealed AC peaks and subsequent decay, which are described well by both fits, although the highest values at days 17 and 23 were not reached. AC peaks at ordinary feeding events (days 1, 3, and 8) were captured well by both MLE and BI. 

No substantial degradation dynamics were observed for IN. Instead, step-wise changes upon feedings indicate that dynamics were dominated by convection. Residuals at the few sampling points remained well visible. 

Overall, the ADM1-R3-core captured the dynamics observed in the measurements without overfitting, where MLE delivers slightly better fits. This is reflected in higher coefficients of determination (\RR) for MLE than BI, reaching values beyond 0.95 for AC during cross validation data, and 0.77 for biogas production (versus 0.89 and 0.76 for BI), Table~\ref{tab:res:kpi_mle_vs_bi} (global columns). 

Even with literature values (prior according to Table~\ref{tab:meth:parameters}), the model captured the measurement data reasonably well, except for an offset in pH (green dotted lines in Fig.~\ref{fig:res:fit_mle_bi}). 
For IN, improvements of the optimized fits over the prior are not obvious due to the low sample density, but a diminished sum of residuals can be inferred. 
Moreover, the prior neither reached the AC peaks nor showed a timely decay after feedings. In summary, the performance of the optimized variants was clearly superior to the prior's. 

%
\subsubsection{Parameter estimates}
\label{sec:res:param_estimates}
%
\begin{table}
\begin{threeparttable}
\centering
\renewcommand{\arraystretch}{1.1} 
\caption{Estimated parameters and corresponding standard deviations of ADM1-R3-core model according to MLE and BI.$^{\mathrm{a}}$}
\label{tab:res:mu_sigma_r3_core}
\begin{tabularx}{\linewidth}{clYYd{3}ld{3}d{2}d{3}d{2}}
\toprule
& & & & \multicolumn{2}{c}{MLE$^{\mathrm{b,c}}$} & \multicolumn{2}{c}{BI$^{\mathrm{b,c}}$} & \multicolumn{2}{c}{Prior$^{\mathrm{b}}$} \\
\cmidrule(lr){5-6} \cmidrule(lr){7-8} \cmidrule(lr){9-10}
Index & Symbol & Scale & Unit & \multicolumn{1}{c}{$\mu$} & \multicolumn{1}{c}{$\sigma$} & \multicolumn{1}{c}{$\mu$} & \multicolumn{1}{c}{$\sigma$} & \multicolumn{1}{c}{$\mu$} & \multicolumn{1}{c}{$\sigma$} \\
\midrule
1 & $k_\mathrm{ch}$ & log10 & \si{\per\day} & 1.566 & 1.12 & 2.206 & 1.07 & 0.424 & 2.30 \\
2 & $k_\mathrm{pr}$ & log10 & \si{\per\day} & 0.042 & 1.89 & 0.067 & 1.60 & 0.080 & 2.46 \\
3 & $k_\mathrm{li}$ & log10 & \si{\per\day} & 4.491 & 1.44 & 0.565 & 1.52 & 0.568 & 1.58 \\
4 & $k_\mathrm{dec}$ & log10 & \si{\per\day} & 0.139 & 8.56 & 0.029 & 1.91 & 0.011 & 2.91 \\
5 & $\mu_\mathrm{m,ac}$ & log10 & \si{\per\day} & 1.083 & 5.26 & 0.728 & 1.52 & 0.573 & 1.92 \\
6 & $K_{\mathrm{S,ac}}$ & log10 & \si{\kilogram\per\cubic\meter} & 1.744 & 6.58 & 2.044 & 1.25 & 2.125 & 1.25 \\
7 & $K_{\mathrm{I,nh3}}$ & log10 & \si{\kilogram\per\cubic\meter} & 2.009 & \num{3.87e21} & 0.055 & 1.59 & 0.053 & 1.63 \\
8 & $\Delta S_\mathrm{ion,eff}$ & abs. & \si{\kilo\mole\per\cubic\meter} & 0.086 & 0.01 & 0.083 & 0.01 & 0.130 & 0.09 \\
9 & $\varphi_\mathrm{IN,in}$ & abs. & $-$ & 0.751 & 0.47 & 0.911 & 0.12 & 1.000 & 0.13 \\
\botrule
\end{tabularx}
\begin{tablenotes}
\footnotesize
\item [a] Best final estimates of BI and MLE frameworks according to objective function value, obtained after global search with local refinement.

\item[b] For log10-scaled parameters, values $\mu$ refer to the geometric mean (median), and $\sigma$ denote geometric (multiplicative) standard deviations, cf. \siref{sec:si:log10_conversion}.  

\item[c] Values resulted from global presearch with ESS and local refinement using analytical gradient and Hessian approximation.
%
\end{tablenotes}
\end{threeparttable}
\end{table}

Optimal parameters of MLE and BI are summarized in Table~\ref{tab:res:mu_sigma_r3_core}. Parameters were similar, especially for $\Delta S_\mathrm{ion,eff}$, $\mu_\mathrm{m,ac}$ and $K_\mathrm{S,ac}$, and were of the same order of magnitude for $k_\mathrm{ch}$, $k_\mathrm{pr}$, $\varphi_\mathrm{IN,in}$. Different orders of magnitude were observed for $k_\mathrm{li}$, $k_\mathrm{dec}$, and $K_\mathrm{I,nh3}$. 


While observed hydrolysis constants lay far from default ADM1 values~\cite{batstone2002}, they remained in the range of reported literature values~\cite{pavlostathis1991,mata-alvarez2005}. 
The BI estimates of $k_\mathrm{li}$ agreed well with Delory et al.~\cite{delory2025}. However, lipids represented the smallest macronutrient fraction among the substrate mix applied~\cite{tisocco2024,carecci2024}, cf. Table~\xref{tab:si:influent_init} in \siref{sec:si:data_preprocessing}. This suggests a small influence of $k_\mathrm{li}$ on model predictions~\cite{delory2025,segura2025}. 

$\mu_\mathrm{m,ac}$ ranged higher than in literature~\cite{pavlostathis1991}, but at the same time, $K_\mathrm{S,ac}$ ranged much higher, too. The two Monod parameters $\mu_\mathrm{m,ac}$ and $K_\mathrm{S,ac}$ describing the growth of acetoclastic methanogens are known to be collinear~\cite{carecci2025, catenacci2024}, and thus difficult to identify simultaneously~\cite{weinrich2021a}. 
Thus, an increased max. growth rate $\mu_\mathrm{m,ac}$ can be compensated by an elevated half-saturation constant $K_\mathrm{S,ac}$. 

The BI estimate of $K_\mathrm{I,nh3}$ was in a plausible range~\cite{segura2025,delory2025,koch2010}, considering a base unit of \si{\kilogram\per\cubic\meter}. MLE, by contrast, gave a much higher median value, Table~\ref{tab:res:mu_sigma_r3_core}. However, its huge associated uncertainty indicates that this parameter was practically non-identifiable with MLE for the given dataset. While $K_\mathrm{I,nh3}$ was reported to show high sensitivity at low estimates~\cite{segura2025,koch2010}, it is also frequently kept fixed at default values~\cite{catenacci2021,weinrich2021a}, which appears to be prudent here, cf. also Sec.~\ref{sec:res:param_uncertainty}. 

Although the estimates of $k_\mathrm{dec}$ lay within the range reported in literature (0.004 -- \SI{0.3}{\per\day}) \cite{pavlostathis1991,mata-alvarez2005}, the MLE uncertainty was the second highest of all parameters, reducing the significance of the estimate (Table~\ref{tab:res:mu_sigma_r3_core}).

A high degree of correlation between parameters (such as for $\mu_\mathrm{m,ac}$ and $K_\mathrm{S,ac}$) and low output sensitivity are closely related to the concept of sloppiness~\cite{chis2016}. Large biological models were frequently reported to be sloppy~\cite{mannakee2016}, and the ADM1-R3 appears to involve sloppy parameters, too. Sloppiness is not equivalent to non-identifiability, though similar~\cite{white2016}, and thus the concept of (practical) identifiability is deemed superior w.r.t. improving the effectiveness and reliability of model calibration~\cite{chis2016}. 

%
\subsubsection{Value in global search}  
MLE clearly benefitted from a global presearch, which reflects in better fits shown in Fig.~\ref{fig:res:fit_mle_bi}, and an improved OF value of 59.7 over 60.3 for local-only search (Table~\ref{tab:res:kpi_mle_vs_bi}). For BI, by contrast, the global presearches did not outperform the local-only search, 
both delivering terminal log likelihoods\ifarxiv\footnote{MLE and BI apply different OFs, but both have the log likelihood in common (Eqs.~\eqref{eq:meth:mle_obj_fun} and \eqref{eq:meth:map_obj_fun}). They are thus compared by means of their log likelihoods in the following.}\fi{} of 62.2. 

\begin{table}
\begin{threeparttable}
\centering
\renewcommand{\arraystretch}{1.1} 
\caption{Comparison of MLE and BI optima. Global refers to global search with local refinement, local-only to local search only, initialized at the literature prior.}
\label{tab:res:kpi_mle_vs_bi}
\begin{tabularx}{\linewidth}{X*{4}{l}}
\toprule
 & \multicolumn{2}{c}{MLE} & \multicolumn{2}{c}{BI} \\
\cmidrule(lr){2-3} \cmidrule(lr){4-5} 
Criterion & global & local-only & global & local-only \\ 
\midrule
Min. obj. function value & 59.682 & 60.329 & 65.000 & 65.000 \\
Min. log-likelihood & 59.682 & 60.329 & 62.154 & 62.156 \\
Av.$^{\mathrm{a,c}}$ log10-scaled SD $\bar \sigma_\mathrm{z}$ & 0.198 & 0.171 & 0.080 & 0.080 \\
Av.$^{\mathrm{b,c}}$ linear-scaled geometric SD $\bar \sigma_\mathrm{x}$ & 1.578 & 1.483 & 1.202 & 1.202 \\
\RR (cross): Biogas production & 0.770 & 0.765 & 0.761 & 0.761 \\
\RR (cross): Acetic acid (AC) & 0.956 & 0.928 & 0.893 & 0.893 \\
Mean NRMSE (cross) & 0.179 & 0.194 & 0.200 & 0.200 \\
Global search method$^{\mathrm{c}}$ & ESS & $-$ & ESS & $-$ \\
Local search method$^{\mathrm{c}}$ & grad + Hess & grad + Hess & grad + Hess & grad + Hess \\
\botrule
\end{tabularx}
\begin{tablenotes}
\footnotesize
\item[a] Geometric average computed as $\bar \sigma_\mathrm{z} = (|C_{\mathrm{\theta_\mathrm{z} \theta_\mathrm{z}}}|)^{1/(2n_\mathrm{\theta})}$, where $|\cdot|$ denotes the determinant, and $\theta_\mathrm{z} \in \mathbb{R}^{n_\mathrm{\theta}}$ denotes log10-scaled parameters. Covariance entries of linear-scaled parameters were first converted to log10-scale via the delta method.

\item[b] Computed from log10-scale SD as $\bar \sigma_\mathrm{x} = 10^{\bar \sigma_\mathrm{z}}$

\item[c] Av.: average, SD: standard deviation, ESS: enhanced scatter search, GA: genetic algorithm, grad + Hess: gradient and Hessian.
\end{tablenotes}
\end{threeparttable}
\end{table}

Among BI posterior estimates, only $k_\mathrm{ch}$ and $\Delta S_\mathrm{ion,eff}$ clearly parted from their prior. The others stayed in the prior's proximity, Table~\ref{tab:res:mu_sigma_r3_core}. It is the regularization term of the BI OF that scales with the confidence in the prior, and thus provides a guide to the solver without enforcing a hard constraint (Eq.~\eqref{eq:meth:map_obj_fun}). This emphasizes the relevance of choosing appropriate prior mean and covariance in BI~\cite{ninness2010,yoko2026}: For a plausible and confident prior, the OF topology is re-shaped such that the solver is guided well towards a useful estimate. 
Evidently, this guidance was sufficiently strong in the present scenario, retaining the estimate close to the prior, which is why a global presearch showed no benefit in terms of an OF value, and a local-only search was sufficient (Table~\ref{tab:res:kpi_mle_vs_bi}).   
Contrarily, the solver may get astray for poor but overconfident priors~\cite{ninness2010}. 
In this case, an unbiased MLE may be preferred, as it is free to explore a wider range of the parameter space. 

All three MLE global searches (LHS, GA, and ESS) resulted in widely different parameter values, cf. Table~\xref{tab:si:mu_sigma_r3_core} in \siref{sec:si:res:plots_tables}. 
Model simulations (Fig.~\xref{fig:si:fits_global_only_cross} in \siref{sec:si:res:plots_tables}) show that GA and ESS found similarly suitable parameters, which agrees with their similar log likelihoods (Table~\ref{tab:res:kpi_mle_global_local}). 
ESS delivered the best estimate, LHS clearly the poorest, cf. Fig.~\xref{fig:si:fits_global_only_cross} in \siref{sec:si:res:plots_tables}. 
The ranking after subsequent local refinements remained the same (ESS best, LHS worst), showing that global searches determined different regions of attraction, cf. Fig.~\xref{fig:si:fits_mle_3global_then_local} and Table~\xref{tab:si:mu_sigma_r3_core} in \siref{sec:si:res:plots_tables}. 
Moreover, model simulations after local refinement only differed much for $p_\mathrm{ch4}$, pH and IN, while they almost overlapped for biogas production and AC, cf. Fig.~\xref{fig:si:fits_mle_3global_then_local} in \siref{sec:si:res:plots_tables}. This indicates a selective output sensitivity, which is expanded on in Secs.~\ref{sec:res:convergence} and \siref{sec:res:aspects}.

Among global BI presearches, ESS delivered the lowest log likelihood (62.19), closely followed by GA (62.6). LHS reached far worse estimates (log likelihood of 152.3). However, local refinements of all three global BI presearches using gradient and Hessian approximations delivered the same terminal estimate with a log likelihood of 62.15, cf. Table~\xref{tab:si:kpi_bi_global_local} in \siref{sec:si:res:plots_tables}. 

Yet, the BI estimates after global-only searches were nearly identical to those of local-only search, cf. Table~\xref{tab:si:mu_sigma_r3_core} in \siref{sec:si:res:plots_tables} and Table~\ref{tab:res:mu_sigma_r3_core}. This indicates a strong bias introduced by the prior, and suggests that indeed the global optimum of BI was found via both global and local-only search. 



\subsection{Uncertainty quantification}
\label{sec:res:uncertainty_quantification}
%

\subsubsection{Parameter uncertainty}
\label{sec:res:param_uncertainty}
MLE and BI apply different uncertainty measures, but both derive them from a second-order approximation of the inverse Hessian (Eqs.~\eqref{eq:meth:cramer_rao} and \eqref{eq:meth:laplace}). All BI estimates showed a lower uncertainty than for MLE except for $\Delta S_\mathrm{ion,eff}$, which showed equal values (Table~\ref{tab:res:mu_sigma_r3_core}). Furthermore, 
uncertainty bands of all BI posteriors fell below prior ranges except for $K_\mathrm{S,ac}$, which retained equal range. \ifarxiv{}Note that standard deviation (SD) of log10-scaled parameters in Table~\ref{tab:res:mu_sigma_r3_core} describe multiplicative (geometric) SD, cf. \siref{sec:si:log10_params_interpretation}.\fi 

SD differed by orders of magnitude for $k_\mathrm{dec}$, $\mu_\mathrm{m,ac}$, $K_\mathrm{S,ac}$, and $K_\mathrm{I,nh3}$. $K_\mathrm{I,nh3}$ appears to be practically non-identifiable with MLE despite the dynamic dataset with high online resolution, reflected by a huge SD, cf. Table~\ref{tab:res:mu_sigma_r3_core} and Sec.~\ref{sec:res:param_estimates}. This was likely caused by a (i) low sampling frequency of IN measurement data, prohibiting to draw conclusions on high-resolution dynamics of IN, and (ii) a high estimated value of $K_\mathrm{I,nh3}$, contrasting with low simulated values of free ammonia concentrations of $S_\mathrm{nh3} \approx \SI{0.06}{\kilogram\per\cubic\meter}$. This resulted in a low sensitivity of the ammonia inhibition factor $I_\mathrm{nh3}$ w.r.t. $K_\mathrm{I,nh3}$ (cf. Eq.~\sieqref{eq:si:R3-inhibit} in \siref{sec:si:model_equations}). Instead, $I_\mathrm{nh3}$ only shows high sensitivity for low values of $K_\mathrm{I,nh3}$~\cite{segura2025,koch2010}. 

High multiplicative SD of 5--7 for $\mu_\mathrm{m,ac}$ and $K_\mathrm{S,ac}$ reinforce that these coupled parameters of acetoclastic methanogenesis are difficult to identify jointly with least squares OFs~\cite{weinrich2021a, carecci2025, catenacci2024}. 

All in all, average log10-scaled SD $\bar \sigma_\mathrm{z}$ of BI (0.08) were less than half of those of MLE (0.2), cf. Table~\ref{tab:res:kpi_mle_vs_bi} (global columns). 

\begin{figure}
    \centering
    \begin{subfigure}[b]{0.43\linewidth}
        \includegraphics[width=\linewidth]{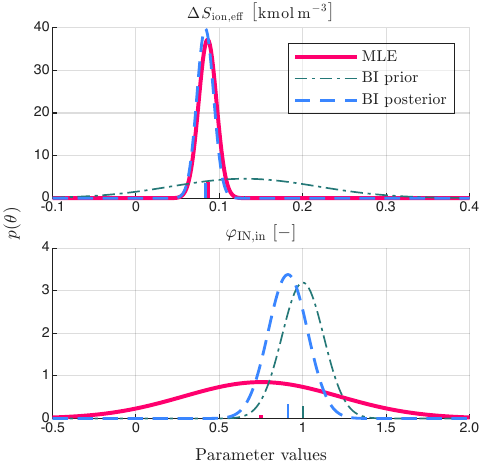}
        \caption{Linear-scaled parameters $\Delta S_\mathrm{ion,eff}$ (top) and $\varphi_\mathrm{IN,in}$ (bottom)}
        \label{fig:res:gaussians_abs}
    \end{subfigure}
    \par\vspace{0.5cm} 
    \begin{subfigure}[b]{0.43\linewidth}
        \includegraphics[width=\linewidth]{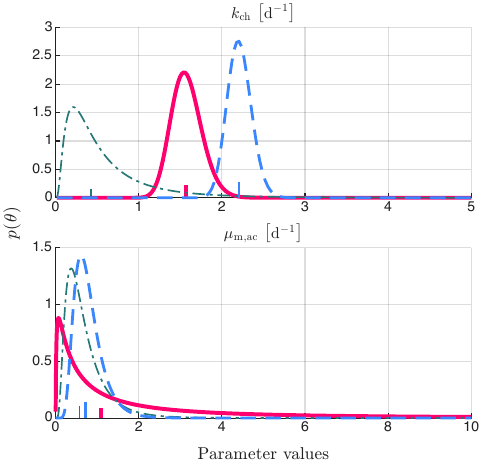}
        \caption{Log10-scaled parameters $k_\mathrm{ch}$ (top) and $\mu_\mathrm{m,ac}$ (bottom) on linear x-axis}
        \label{fig:res:gaussians_log_abs}
    \end{subfigure}
    \par\vspace{0.5cm}
    \begin{subfigure}[b]{0.43\linewidth}
        \includegraphics[width=\linewidth]{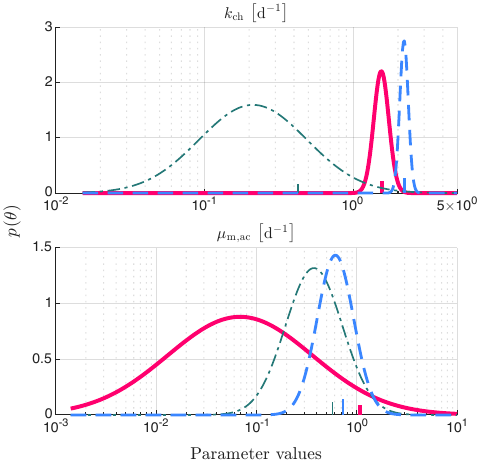}
        \caption{Log10-scaled parameters $k_\mathrm{ch}$ (top) and $\mu_\mathrm{m,ac}$ (bottom) on log10 x-axis}
        \label{fig:res:gaussians_log_log}
    \end{subfigure}
    \caption{Approximated prior and posterior distributions of MLE and BI estimates after global search and local refinement for selected linear- and log10-scaled model parameters.}
    \label{fig:res:gaussians}
\end{figure}
%


Estimated probability distributions 
illustrate the large bandwidths of uncertainty spreads particularly for MLE, Fig.~\ref{fig:res:gaussians}. Gaussians of the linear-scaled $\Delta S_\mathrm{ion,eff}$ are shown in Fig.~\ref{fig:res:gaussians_abs}. They agreed well between MLE (magenta) and BI (blue), indicating that an effective ion concentration inside the reactor 
is well identifiable at semi-continuous feeding. It may therefore pose a viable alternative to the conventional approach which (i) ties pH calibration to the influent concentrations, thereby taking effect only during feeding events, (ii) separates cations from anions~\cite{rosen_jeppsson.2006}, and (iii) considers ions as non-measurable states, impairing observability~\cite{hellmann2023}. 

\begin{figure}
\centering
\includegraphics[width=\linewidth]{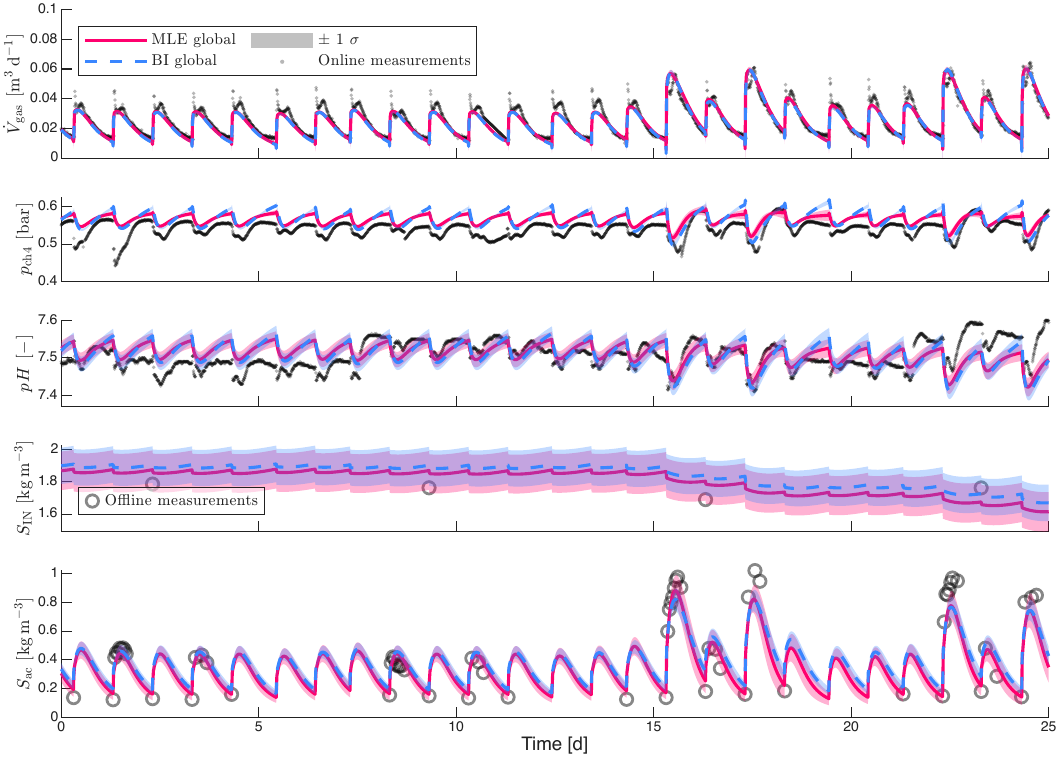}
\caption{Confidence intervals of model outputs (auto validation dataset) based on propagated parameter uncertainties ($\pm 1 \sigma$) of MLE and BI. Parameters were obtained via global search and subsequent local refinement. Log10-scaled parameters apply 1 geometric SD, linear-scaled parameters 1 additive SD (Sec.~\ref{sec:meth:log10}).}
\label{fig:res:conf_intervals}
\end{figure}

By contrast, estimates of 
$\varphi_\mathrm{IN,in}$ were smaller and less confident for MLE than BI, Fig.~\ref{fig:res:gaussians_abs}. Moreover, the BI posterior distribution of $\varphi_\mathrm{IN,in}$ did not move much from the prior's. This parameter is not commonly tuned in literature, and did not improve the fit of IN either, as shown in Fig.~\ref{fig:res:conf_intervals}. Therefore, it should rather be kept fixed at its prior value of 1 (Table~\ref{tab:meth:parameters}) and excluded from PI for the ADM1-R3.

Larger differences were observed for uncertainties associated with MLE and BI estimates of log10-scaled parameters such as $k_\mathrm{ch}$, and particularly $\mu_\mathrm{m,ac}$ (Fig.~\ref{fig:res:gaussians}). Distributions are shown for log10- (Fig.~\ref{fig:res:gaussians_log_log}) and linear-scaled (Fig.~\ref{fig:res:gaussians_log_abs}) x-axes. \ifarxiv{}Naturally, log10-normal distributions are skewed on a linear-scaled x-axis, and the median (small vertical line on x-axis) is shifted to the right of the mode, cf. \siref{sec:si:log10_conversion}.\fi

A less confident MLE estimate of $k_\mathrm{ch}$ as per Table~\ref{tab:res:mu_sigma_r3_core} corresponds with a wider distribution plot in Fig.~\ref{fig:res:gaussians_log_log}. 
Large uncertainties of log10-scaled parameters were not directly apparent on an absolute x-axis of Fig.~\ref{fig:res:gaussians_log_abs} (magenta), but were revealed on a log10 x-axis in Fig.~\ref{fig:res:gaussians_log_log} (bottom), particularly for $\mu_\mathrm{m,ac}$. This was even aggravated for distributions of $k_\mathrm{dec}$ and $K_\mathrm{S,ac}$, which spanned multiple orders of magnitude on a log10 x-axis, cf. Fig.~\xref{fig:si:gaussians_poor} in \siref{sec:si:res:plots_tables}. 

Smaller uncertainty bands of BI were a direct consequence of the non-negative regularization term of the prior, 
cf. Eq.~\sieqref{eq:meth:hessian_map} in \siref{sec:si_meth:algorithms}. 

\subsubsection{Output uncertainty}
\label{sec:res:output_uncertainty}
%
Parameter uncertainties translated into output uncertainties via linear uncertainty propagation, Eq.~\eqref{eq:meth:uncertainty_propagation}. Shaded areas in Fig.~\ref{fig:res:conf_intervals} show confidence intervals (CI) of $\pm 1$ SD of output predictions for the same dataset as in Fig.~\ref{fig:res:fit_mle_bi}. 

Biogas production showed the smallest CI, and IN the largest, cf. Fig.~\ref{fig:res:conf_intervals}. Those of pH and IN remained about constant over time for both MLE (magenta) and BI (blue). Conversely, they widened in response to feeding events for $p_\mathrm{ch4}$ (for MLE, magenta) and AC (for BI, blue). \ifarxiv{}$p_\mathrm{ch4}$ showed growing CI for MLE when biogas production settled after feedings (days 16-19), whereas CI for BI remained constant. Conversely, BI's CI of AC scaled with the peak height but remained about constant for MLE. \fi

Output CI revealed another perspective. 
Low output sensitivities $S_\mathrm{y}$ for IN, e.g., equalized the different magnitudes of parameter uncertainties obtained with MLE and BI (Eq.~\eqref{eq:meth:uncertainty_propagation} and Table~\ref{tab:res:mu_sigma_r3_core}). 
This applied especially to $K_\mathrm{I,nh3}, k_\mathrm{dec}, \mu_\mathrm{m,ac}$, and $K_\mathrm{S,ac}$, which all affect the dynamics of IN, cf. Eq.~\sieqref{eq:si:r3-frac:f:IN} in \siref{sec:si:model_equations}. Despite their highly different parameter uncertainties for MLE and BI, CI of IN were almost equally wide for both MLE and BI, Fig.~\ref{fig:res:conf_intervals}. 

The similar CI of MLE and BI despite clearly different parameter uncertainties suggest that the second-order approximation in the Cramér-Rao bound of MLE, Eq.~\eqref{eq:meth:cramer_rao}, may be inaccurate for the ADM1-R3 as a strongly nonlinear model. Similar observations were reported by Batstone et al., who considered a linearized approximation of uncertainty as inaccurate for the full ADM1~\cite{batstone2009}. 
SD resulting from BI, by contrast, appeared to represent the true parameter uncertainties more realistically. Notably, CI of IN agreed well with Delory et al.~\cite{delory2025}, who applied bootstrapping to quantify uncertainty. They reported SD in line with BI results, particularly for $K_\mathrm{I,nh3}$ and $k_\mathrm{dec}$ (Table~\ref{tab:res:mu_sigma_r3_core}). 

For the given dataset, these observations imply a gradual relevance of ADM1-R3 outputs given the present, high-resolution dataset: while IN adds little value to PI, pH and AC do so more clearly, and biogas production most evidently. 


%
\subsection{Convergence analysis}
\label{sec:res:convergence}
Solver convergence is discussed by means of (i) global presearch, (ii) local refinement starting from the best global-only result, and (iii) local-only searches, 
Fig.~\ref{fig:meth:metaheuristics}. 
The first two analyses are exemplified by MLE, benefitting more clearly from global presearch than BI, and the third compares MLE with BI. 

\subsubsection{Global search}
\begin{figure}
\centering
\includegraphics[width=\linewidth]{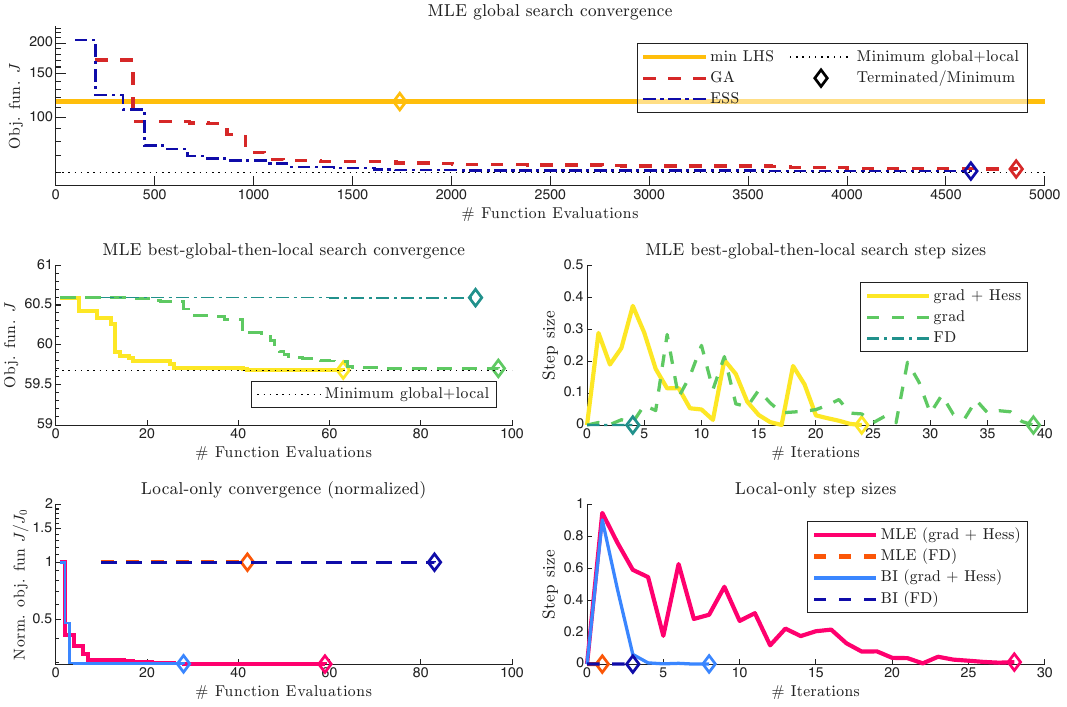}
\caption{Convergence of global searches (MLE, top row), as well as convergence and step sizes of local searches (MLE, second row) and local-only searches (MLE and BI, third row). \textsc{Matlab} starts counting iterations at zero, hence the offset with respect to Tables~\ref{tab:res:kpi_mle_global_local} and \ref{tab:res:kpi_local_only}.}
\label{fig:res:convergence_joint}
\end{figure}

Among global search algorithms, ESS decreased quicker than GA (except between function calls 390--449) and reached slightly lower terminal values, Fig.~\ref{fig:res:convergence_joint} (top, blue and red lines) and Table~\ref{tab:res:kpi_mle_global_local} (global search columns). \ifarxiv{}Neither ESS nor GA exceeded their limit of 5000 function calls\fi. 

LHS, by contrast, delivered nearly twice the terminal OF value (yellow line in top subplot of Fig.~\ref{fig:res:convergence_joint}, Table~\ref{tab:res:kpi_mle_global_local}). Many of the LHS samples resulted in very high OF values, and only the minimum value is shown in Fig.~\ref{fig:res:convergence_joint} (top). \ifarxiv{}The location of the LHS terminal marker is of course a random result caused by the random permutation of samples. \fi{}The corresponding model fit after LHS was clearly worse than for GA and ESS, cf. Fig.~\xref{fig:si:fits_global_only_cross} in \siref{sec:si:res:plots_tables}. 

Although running on 10 parallel cores, GA still required about half the computation time of ESS, which was executed serially on a single core (Table~\ref{tab:res:kpi_mle_global_local}). LHS, also run in parallel, took the least computation time. 
Yet, the relative advantage of parallelization improved with more function calls and cores, thus involving a smaller relative overhead for setting up and managing the parallel pool~\cite{niculescu2022}. 

ESS and GA nearly reached the same terminal OF value, and their associated model simulations compared well, cf. Fig.~\xref{fig:si:fits_global_only_cross} in \siref{sec:si:res:plots_tables}. However, they delivered different parameters, notably for $k_\mathrm{pr}$, $\mu_\mathrm{m,ac}$ and $K_\mathrm{S,ac}$, cf. Table~\xref{tab:si:mu_sigma_r3_core} in \siref{sec:si:res:plots_tables}. Moreover, subsequent local searches converged to different optima, reflected in different terminal OF values, Table~\ref{tab:res:kpi_mle_global_local}. 

GA and ESS share evolutionary principles. Additionally, ESS includes gradient-based search, 
giving an advantage to ESS over gradient-free GA. This explains why the local refinement of ESS eventually delivered the best of all estimates, which likely marks the global optimum, indicated by a dotted line in the top subplot of Fig.~\ref{fig:res:convergence_joint}. 

%
\begin{table}
\begin{threeparttable}
\centering
\renewcommand{\arraystretch}{1.1} 
\caption{Global and local search comparison of MLE.$^{\mathrm{a}}$}
\label{tab:res:kpi_mle_global_local}
\begin{tabularx}{\linewidth}{X*{3}{r}p{2mm}*{3}{r}}
\toprule
& \multicolumn{3}{c}{Global search} & & \multicolumn{3}{c}{Best-global-then-local search}\\
\cmidrule(lr){2-4} \cmidrule(lr){6-8}
Criterion & LHS & GA & ESS & & FD$^{\mathrm{b}}$ & grad$^{\mathrm{b}}$ & grad + Hess$^{\mathrm{b}}$ \\
\midrule
Min. obj. function value & 115.87 & 61.75 & 60.59 & & 60.59 & 59.70 & 59.68 \\
No. of obj. function calls & 5000 & 4855 & 4626 & & 105 & 117 & 90 \\
No. of iterations & $-$ & 50 & 39 & & 5 & 40 & 25 \\
No. of calls per iteration$^{\mathrm{c}}$ & $-$ & 97 & 119 & & 21 & 3 & 4 \\
Computation time [s] & 5736 & 6977 & 15261 & & 423 & 653 & 363 \\
\botrule
\end{tabularx}
\begin{tablenotes}
\footnotesize
\item[a] Best-global-then-local searches were initialized at the global solution with minimal objective function value (ESS, cf. Table~\ref{tab:res:kpi_mle_vs_bi}).

\item[b] FD: finite differences, grad: gradient, grad + Hess: gradient and Hessian.

\item[c] Average number of objective function calls per iteration (rounded).
\end{tablenotes}
\end{threeparttable}
\end{table}
\begin{table}
\begin{threeparttable}
\centering
\renewcommand{\arraystretch}{1.1} 
\caption{Local-only search of MLE and BI comparing finite differences with analytical gradient and Hessian approximation, all initialized at literature prior.}
\label{tab:res:kpi_local_only}
\begin{tabularx}{\linewidth}{X*{4}{r}}
\toprule
& \multicolumn{2}{c}{MLE} & \multicolumn{2}{c}{BI} \\ 
\cmidrule(lr){2-3} \cmidrule(lr){4-5} 

Criterion & grad + Hess$^{\mathrm{a}}$ & FD$^{\mathrm{a}}$ & grad + Hess$^{\mathrm{a}}$ & FD$^{\mathrm{a}}$ \\
\midrule
Min. obj. function value & 60.33 & 206.06 & 65.00 & 221.27 \\
Min. log-likelihood  & 60.33 & 206.06 & 62.16 & 221.27 \\
No. of obj. function calls & 92 & 64 & 57 & 94 \\
No. of iterations & 29 & 2 & 9 & 4 \\
No. of calls per iteration$^{\mathrm{b}}$ & 3 & 32 & 6 & 24 \\
\botrule
\end{tabularx}
\begin{tablenotes}
\footnotesize
\item[a] FD: finite differences, grad + Hess: gradient and Hessian.

\item[b] Average number of objective function calls per iteration (rounded).
%
\end{tablenotes}
\end{threeparttable}
\end{table}
%

\subsubsection{Local refinement}
Local refinement after the global presearch delivered additional improvement, albeit small, Fig.~\ref{fig:res:convergence_joint} (second and third subplot). All three local search variants were initialized at the best global-only result (ESS), Fig.~\ref{fig:meth:metaheuristics}. Clearly, the most sophisticated variant of analytical gradient and Hessian approximation (grad + Hess) converged the fastest, as shown in Fig.~\ref{fig:res:convergence_joint} (second subplot, yellow). It took the largest steps especially in the beginning, shown in Fig.~\ref{fig:res:convergence_joint} (third subplot). Usage of gradient and Hessian also delivered the lowest OF values (Table~\ref{tab:res:kpi_mle_global_local}, local search columns). However, analytical gradient (grad) alone achieved nearly the same terminal value, after progressing slower 
in the beginning, indicated by smaller step sizes in Fig.~\ref{fig:res:convergence_joint} (third subplot, green). 

Finite differences (FD), by contrast, delivered no improvement over the global presearch at all, reflected by a horizontal line in Fig.~\ref{fig:res:convergence_joint} (teal), negligibly small step sizes and 
termination after only 5 iterations, Fig.~\ref{fig:res:convergence_joint}. FD, which represents the default setting of \textsc{Matlab}'s standard gradient-based nonlinear optimizer \texttt{fmincon}~\cite{matlab2024a}, required about 20 function evaluations per iteration compared to 3 or 4 for analytical gradient or analytical gradient and Hessian, respectively (Tables~\ref{tab:res:kpi_mle_global_local} and \ref{tab:res:kpi_local_only}). This is due to forward and backward perturbations of individual parameter entries required to approximate the gradient~\cite{frohlich2017a}. Contrarily, FD spares the need to integrate sensitivity equations. However, the additional effort of implementing sensitivities was evidently rewarded by better approximation of the topology of the OF~\cite{raue2013}, and thus superior search directions. 
While this partly confirmed Villaverde's best practices~\cite{villaverde2019}, the Gauss-Newton algorithm delivered a Hessian approximation that only relies on first-order derivatives, cf. Eqs.~\sieqref{eq:meth:hessian_mle} and~\sieqref{eq:meth:hessian_map} in \siref{sec:si_meth:algorithms}. It thus poses only a marginal extra effort over the implementation of analytical gradients, and is therefore endorsed for comparable PI problems. 

In fact, an additional local search in linear-scaled coordinates, initialized at the optimum found via global search and local refinement (in log10-scale), delivered no further improvement of the OF, even when supplied with analytical gradient and Hessian information. This reinforces that the global optimum was indeed found in log10-scale.

Global and local searches illustrated that the model behaves similarly for ambiguous parameter values of high uncertainties (especially $K_\mathrm{I,nh3}$), caused by low output sensitivities (particularly of IN). 
This underlines a certain degree of sloppiness~\cite{chis2016,mannakee2016}. Furthermore, this coincides with flat topologies around the local optima, which cause long iterations of little improvement before termination, Fig.~\ref{fig:res:convergence_joint}. Lastly, this likely causes the different global solvers to terminate in different local optima, cf. \siref{sec:si:res:plots_tables}, Fig.~\xref{fig:si:fits_global_only_cross} and Table~\xref{tab:si:kpi_bi_global_local} (global search columns). 

\subsubsection{Local-only search}
The relative improvement by the local refinement over the global presearch may seem small. Yet, these improvements were only enabled by appropriate gradient and curvature information. 
To this end, local-only searches using analytical gradients and Hessian approximation were initialized at literature values (Fig.~\ref{fig:meth:metaheuristics_local_only}) and yielded rapid convergence for both MLE and BI, Fig.~\ref{fig:res:convergence_joint} (bottom two subplots, solid lines in magenta and blue).  
\ifarxiv{}This applied especially just after only a few function evaluations. \fi 
OF values in the second last subplot of Fig.~\ref{fig:res:convergence_joint} were normalized by the initial OF, $J_{\mathrm{MLE},0}$ and $J_{\mathrm{MAP},0}$, to fairly compare MLE and BI results in spite of different underlying OFs (Eqs.~\eqref{eq:meth:mle_obj_fun} and \eqref{eq:meth:map_obj_fun}). 

By contrast, local searches using FD (dashed lines) failed to progress entirely. Despite numerous (futile) function evaluations, solvers took negligibly small steps before terminating after only a few iterations. Consequently, the solvers remained stuck at prior values, shown by model simulations that do not differ from the prior, cf. Fig.~\xref{fig:si:fits_local_only_mle_vs_bi} in \siref{sec:si:res:plots_tables}. 


Moreover, the solid lines in the last two subplots of Fig.~\ref{fig:res:convergence_joint} show that BI converged much faster than MLE, which underlines the guiding effect of the prior in reshaping the OF topology. Where BI took few large steps in the first iterations before plateauing and then terminating rapidly, MLE continued searching nearly twice as long, and steps decreased in size only slowly. Naturally, the bias of the prior in BI's OF caused local-only searches of BI to terminate at different estimates than MLE, cf. Table~\xref{tab:si:mu_sigma_r3_core} in \siref{sec:si:res:plots_tables}.

\ifarxiv
\subsection{Limitations and perspectives}
\label{sec:res:limitations_perspectives}
Results presented were based on the single dataset considered and the ADM1-R3 model variants applied. Performance in other scenarios (dataset, AD model, different biological process etc.) remains to be investigated in more detail.  

Moreover, a second-order approximation quantified parameter uncertainties. This involved first-order sensitivities which are exact only for linear models~\cite{lopezc.2015}. The ADM1-R3, however, is strongly nonlinear, and thus uncertainty estimates could surely be improved. Alternative approaches to estimate parameter uncertainty should be investigated, e.g., bootstrapping~\cite{delory2025}, the unscented transformation~\cite{heine2008}, Monte-Carlo sampling~\cite{batstone2009}, or by applying the Bayesian approach to jointly estimate parameters and uncertainties~\cite{picard-weibel2024}. 

Identifiability issues could be mitigated by (i) more informative datasets, and (ii) carefully selecting parameters involved, thus easing the challenging calibration of AD models especially in full-scale. On the one hand, more sophisticated sampling of offline measurements could provide more valuable information~\cite{wilms2026,balsa-canto2008,durr2021}. In the present setup, however, merely changing the timing of sampling is unlikely to reduce the uncertainty, as shown by the constant amplitudes of confidence intervals for IN and AC, Fig.~\ref{fig:res:conf_intervals}. 

Optimal experimental design maximizes the expected information content of future datasets by, e.g., determining optimal feeding trajectories~\cite{heine2008}. This could help to reveal short-term AC peaks and pH drops in full scale AD plants, where neither are commonly observed or even sampled~\cite{segura2025,gaida2012,jimenez2015}. 
Furthermore, spectroscopy-guided sample subset selection~\cite{wilms2026} identifies the most relevant offline samples, avoiding the need to process and consider all of them for PI. 

On the other hand, parameter subset selection (PSS) 
determines the most influential model parameters by excluding those with low influence and high redundancy from PI, increasing practical identifiability for a fixed dataset~\cite{catenacci2021,brun2001,lopezc.2015}. 

\fi
\section{Conclusion}
MLE and BI were demonstrated as powerful frameworks in finding optimal model parameters of the ADM1-R3 as a reduced-order ADM1. Both MLE and BI delivered instructive fits for experimental data of a lab-scale AD reactor. While parameter uncertainties estimated with BI were much smaller than those of MLE, output sensitivities compensated their difference such that output uncertainties of both BI and MLE agreed well. 

The effort of implementing forward sensitivities for analytical gradients and approximated Hessians was rewarded by much better local optimization results and faster convergence. 
BI leveraged an appropriately quantified prior guess of model parameters, which helped avoid the need for global optimization. MLE, by contrast, benefitted from a global presearch, where ESS delivered the best estimate, which likely marks the global optimum. 

The ADM1-R3 was shown to be well suitable for describing dynamic AD process conditions. The most important measurements for its successful calibration were biogas production and AC, while \chfour content, pH, and IN were less important.

Confident parameter estimates were found for an effective ion concentration and hydrolysis constants, whereas coupled kinetic parameters of acetoclastic methanogenesis showed wide uncertainty bands. 
The free ammonia inhibition constant was shown to be practically non-identifiable with MLE. 

The results appear promising for deployment at a larger scale, and lay a rigorous methodical base for model-based monitoring and control applications of AD plants 
and other biological multirate systems.

\phantomsection 
\section*{Data availability statement}
\addcontentsline{toc}{section}{Data availability statement}
The processed measurement data used for model calibration and validation are openly available at \mbox{\href{https://zenodo.org/records/23730272?token=eyJhbGciOiJIUzUxMiJ9.eyJpZCI6IjkxNWZhYjE2LTc4YmEtNGIxOS1hMmUxLTQ1YzM3NzRiYWY2YyIsImRhdGEiOnt9LCJyYW5kb20iOiIxMDc5ZDBiOThlYmVkNjljMjYzZTMyZTAwMjRmODUzMiJ9.hBUl2ZkL9NiwTcKSjUFQLv_XLplfsmOZMzeaInUSMRqi_WTTlwvFfrZnvncosflXTY3ms3x1CaaT26cC_rraXQ}{zenodo.org}}~\cite{hellmann2026c}. 
Raw data as well as underlying \textsc{matlab} and Python code are available from the corresponding author upon reasonable request.

\ifarxiv{}\fi
 
\phantomsection 
\section*{Acknowledgements}
\addcontentsline{toc}{section}{Acknowledgements}
The authors express their gratitude to Fachagentur Nachwachsende Rohstoffe (FNR) and the Federal Ministry of Agriculture, Food and Regional Identity for funding of the junior research group at DBFZ (grant no 2219NR333). SH acknowledges André Benndorf for diligent support in the lab, Michael Goldstein for fruitful discussions on influent concentrations, and Rudibert King for his excellent teaching materials. 

\phantomsection 
\section*{CRediT authorship contribution statement}
\addcontentsline{toc}{section}{CRediT authorship contribution statement}
\textbf{Simon Hellmann:} Conceptualization, Data Curation, Investigation, Methodology, Software, Validation, Visualization, Writing - Original Draft. 
\textbf{Leander Lerch:} Conceptualization, Data Curation, Investigation, Methodology, Software, Validation, Visualization, Writing - Original Draft. 
\textbf{Terrance Wilms:} Conceptualization, Methodology, Validation, Supervision, Writing - Review \& Editing.  
\textbf{Moritz Reumschüssel:} Conceptualization, Methodology, Visualization, Writing - Review \& Editing. 
\textbf{Alberto Meola:} Data Curation, Methodology, Software, Validation, Writing - Review \& Editing. 
\textbf{Félix Delory:} Investigation, Data Curation, Writing - Review \& Editing. 
\textbf{Steffi Knorn:} Conceptualization, Methodology, Supervision, Writing - Review \& Editing. 
\textbf{Stefan Streif:} Conceptualization, Methodology, Supervision, Writing - Review \& Editing. 
\textbf{Sören Weinrich:} Conceptualization, Methodology, Visualization, Resources, Supervision, Funding acquisition, Project administration, Writing - Review \& Editing. 

\section*{Declaration of competing interest}
The authors declare that they have no known competing financial interests or personal relationships that could have appeared to influence the work reported in this paper.

\section*{Declaration of generative AI and AI-assisted technologies in the manuscript preparation process}
During the preparation of this work the authors used Anthropic Claude Opus 5.5 in order to draft individual code snippets, identify bugs and typos, aggregate simulation results into tables, and draft the first version of the graphical abstract. After using these tools/service, the authors reviewed and edited the content as needed and take full responsibility for the content of the published article.

\newpage
\pdfbookmark[0]{Appendix}{appendix}
\section*{Appendix}
\appendix

\setcounter{table}{0}
\setcounter{figure}{0}
\setcounter{equation}{0}
\renewcommand{\thetable}{\thesection.\arabic{table}}
\renewcommand{\thefigure}{\thesection.\arabic{figure}}
\renewcommand{\theequation}{\thesection.\arabic{equation}}
\renewcommand{\theHequation}{\thesection.\arabic{equation}} 

\section{Supplementary results}
\label{sec:si:results}
Two additional case studies are presented before supporting plots and tables provide further details on results discussed in Sec.~\mainref{sec:results}. 

\subsection{Case studies of models and data}
\label{sec:res:aspects}

The following explores selected extensions of model variants (1 vs. 2 carbohydrate fractions) 
and considered model outputs on resulting fits\ifarxiv{}\fi.
\subsubsection{Carbohydrates fractionation}
\label{sec:res:ch_fractionation}

\begin{figure}[!t]
\centering
\includegraphics[width=0.46\linewidth]{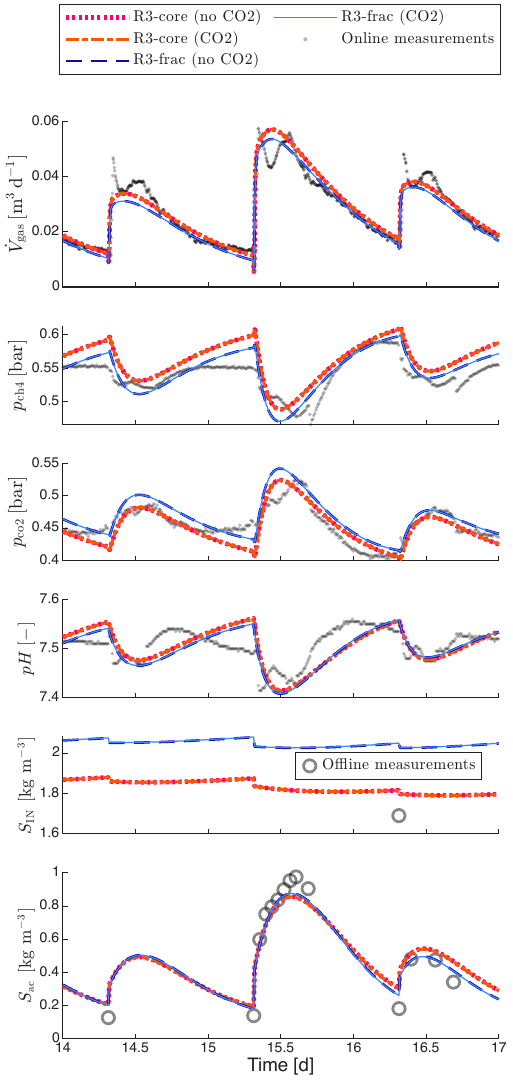}
\caption{Effect of considering \cotwo measurement data during calibration of ADM1-R3 model with one (ADM1-R3-core) and two carbohydrate fractions (ADM1-R3-core-frac): Measurements (auto validation dataset, excerpt of days 14-17) and dynamic models based on MLE local-only search. Fits against the full length auto validation dataset are shown in Fig.~\ref{fig:si:carbs_frac_co2}.}
\label{fig:res:carbs_frac_co2}
\end{figure}

Splitting the carbohydrates (CH) into a slow and a fast fraction (ADM1-R3-core-frac) provided two additional degrees of freedom, cf. Sec.~\mainref{sec:meth:ad_models}. Compared with just a single CH fraction (ADM1-R3-core), local-only search delivered a slightly improved fit w.r.t. biogas composition ($p_\mathrm{ch4}$ and $p_\mathrm{co2}$), Fig.~\ref{fig:res:carbs_frac_co2}. 
However, fits improved only marginally for pH and \ifarxiv{}AC\fi{}, and clearly deteriorated for \ifarxiv{}IN\fi{}. There was practically no difference w.r.t. biogas production. 

Notably, even the ADM1-R3-core-frac failed to capture the fast dynamics of biogas production just after feeding events (dent and subsequent rebound) induced by the interplay of \cotwo and pH~\cite{mcCarty1964}. These observations agree with Delory et al.~\cite{delory2025}. 

\begin{figure}
\centering
\includegraphics[width=\linewidth]{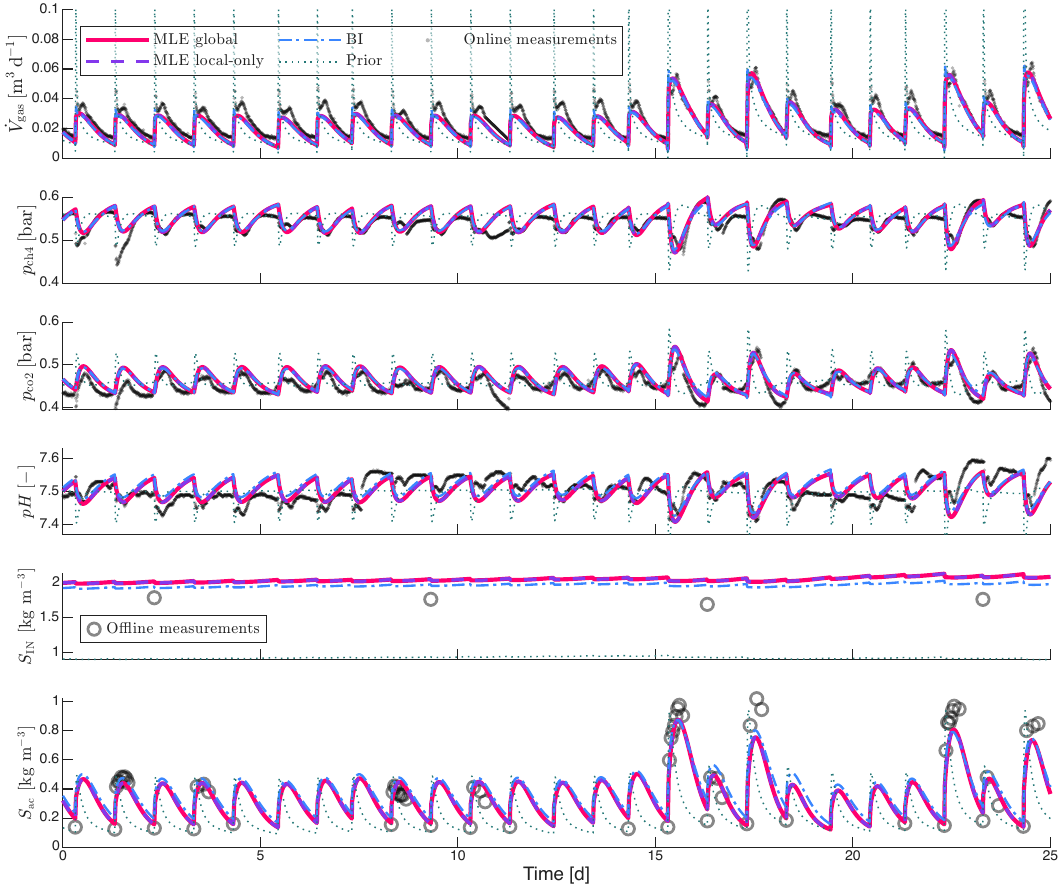}
\caption{Measurements (auto validation dataset) and dynamic model ADM1-R3-core-\textbf{frac} for MLE, BI, and prior. Global searches with ESS, local refinement with analytical gradient and Hessian information.}
\label{fig:si:fit_mle_bi_r3_frac_cross}
\end{figure}

A global presearch with local refinement did not improve the fit substantially, neither did a \ifarxiv{}BI OF\fi{}, Fig.~\ref{fig:si:fit_mle_bi_r3_frac_cross}. 

Resulting estimates are provided in Table~\ref{tab:si:mu_sigma_r3_frac}. Compared with the ADM1-R3-core, a diminished practical identifiability was implied by considerably increased uncertainties of $\mu_\mathrm{m,ac}$ and of all first-order rate constants except $k_\mathrm{dec}$ (Table~\ref{tab:si:mu_sigma_r3_frac}). Therefore, the use of a single CH fraction is recommended, as implemented in the conventional ADM1~\cite{rosen_jeppsson.2006}. 

%
\subsubsection{Effects of calibration data}
It was clearly insignificant whether \cotwo measurements were considered in the OF. Fits of both model variants overlapped regardless of \cotwo data (Fig.~\ref{fig:res:carbs_frac_co2}). This reinforces that the gas phase equations are fully determined by two out of three available outputs $\dot V_\mathrm{gas}, p_\mathrm{ch4}$, and $p_\mathrm{co2}$.



\begin{figure}
\centering
\includegraphics[width=\linewidth]{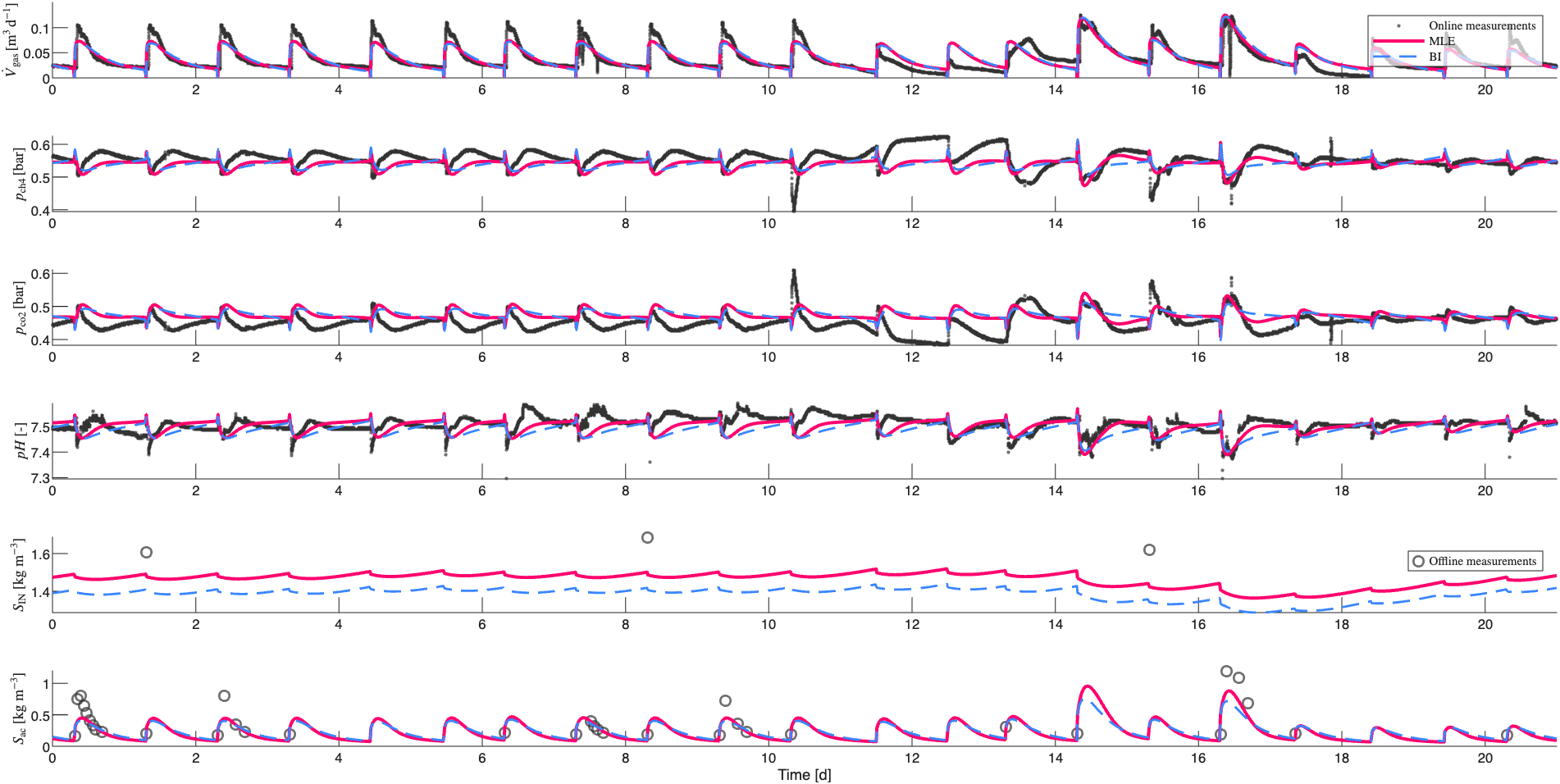}
\caption{Measurements (auto validation dataset of 2023/02/28 -- 2023/03/21) and dynamic model ADM1-R3-core-frac. Fits are shown for BI using local-only search and considering measurement data of all outputs. High AC peaks cannot be captured adequately due to low and short pH drops.}
\label{fig:si:fit_bi_low_ph_drops}
\end{figure}

When applying the same PI framework to a different excerpt of data from the same experiment (
average OLR of 6.5\,$\mathrm{kg\,VS \,m}^{-3} \,\mathrm{d}^{-1}$), and considering all model outputs of Table~\mainref{tab:meth:measurements} (cf. Eqs.~\sieqref{eq:si:r3-core:y} and \sieqref{eq:si:r3-frac:y} in \siref{sec:si:model_equations}), clearly worse model fits were obtained for AC and pH. This is exemplified by BI local-only results in Fig.~\ref{fig:si:fit_bi_low_ph_drops}. Evidently, the solver failed to find a compromise between both low pH drops and high AC peaks, which cannot be captured adequately by the model, given a weighted least squares objective and measurement covariances as per Table~\mainref{tab:meth:measurements}. 

\begin{figure}
\centering
\includegraphics[width=\linewidth]{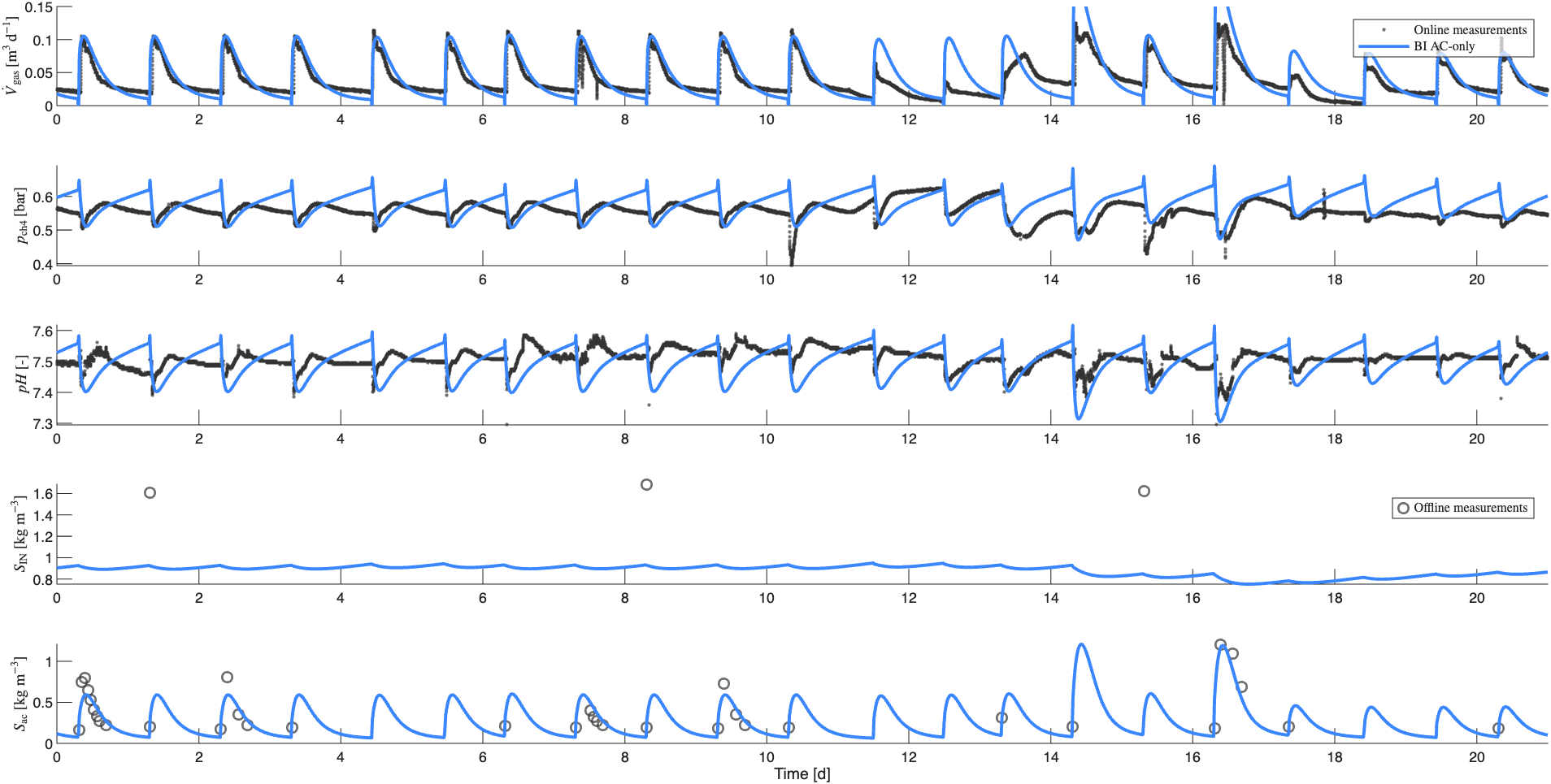}
\caption{Measurements (auto validation dataset of 2023/02/28 -- 2023/03/21) and dynamic model ADM1-R3-core. Fits are shown for BI using local-only search and considering only AC measurement data. MLE estimates (not shown) result in a fit of the same quality.}
\label{fig:si:fit_bi_ac_only}
\end{figure}

One way to address this is by assuming less accurate pH measurements, i.e., by increasing \ifarxiv{}\fi{}$\sigma_\mathrm{y}$ for pH. An extreme variant thereof is to disregard all measurements except AC, as proposed by Batstone and Steyer~\cite{batstone2007}. A local-only search with BI achieved very suitable fits, Fig.~\ref{fig:si:fit_bi_ac_only}. This agrees with the findings of Sec.~\mainref{sec:res:output_uncertainty}, which showed that only few outputs (especially biogas production and AC) are of high significance for model calibration, whereas others (especially \cotwo and IN) can be ignored without major information loss. 
\clearpage
\subsection{Supporting plots and tables}
\label{sec:si:res:plots_tables}
\begin{figure}
\centering
\includegraphics[width=\linewidth]{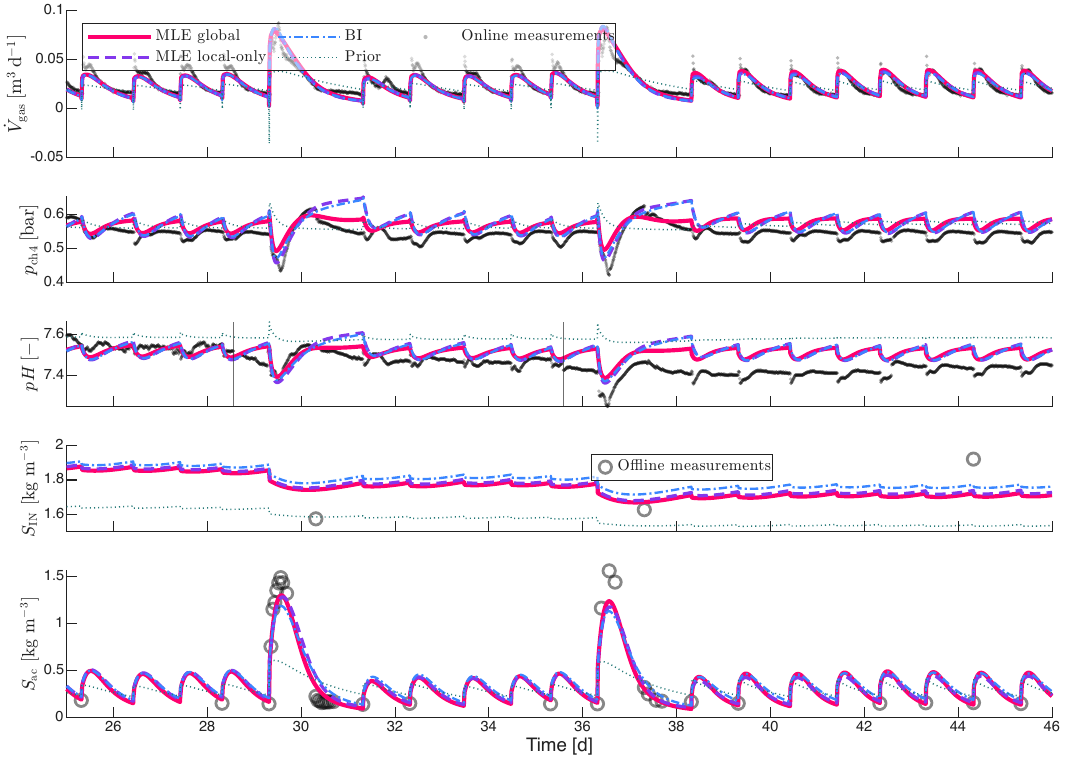}
\caption{Measurement data (\textbf{cross} validation dataset) and dynamic model ADM1-R3-core. Fits are shown for MLE and BI optima after global search with local refinement, after local-only search, and for literature prior. BI graphs for global + local and local-only overlap, hence only one set of graphs is shown. Vertical grey lines in pH tile mark re-calibration times of the pH electrode.}
\label{fig:si:fit_mle_bi_cross}
\end{figure}
\begin{figure}
\centering
\includegraphics[width=\linewidth]{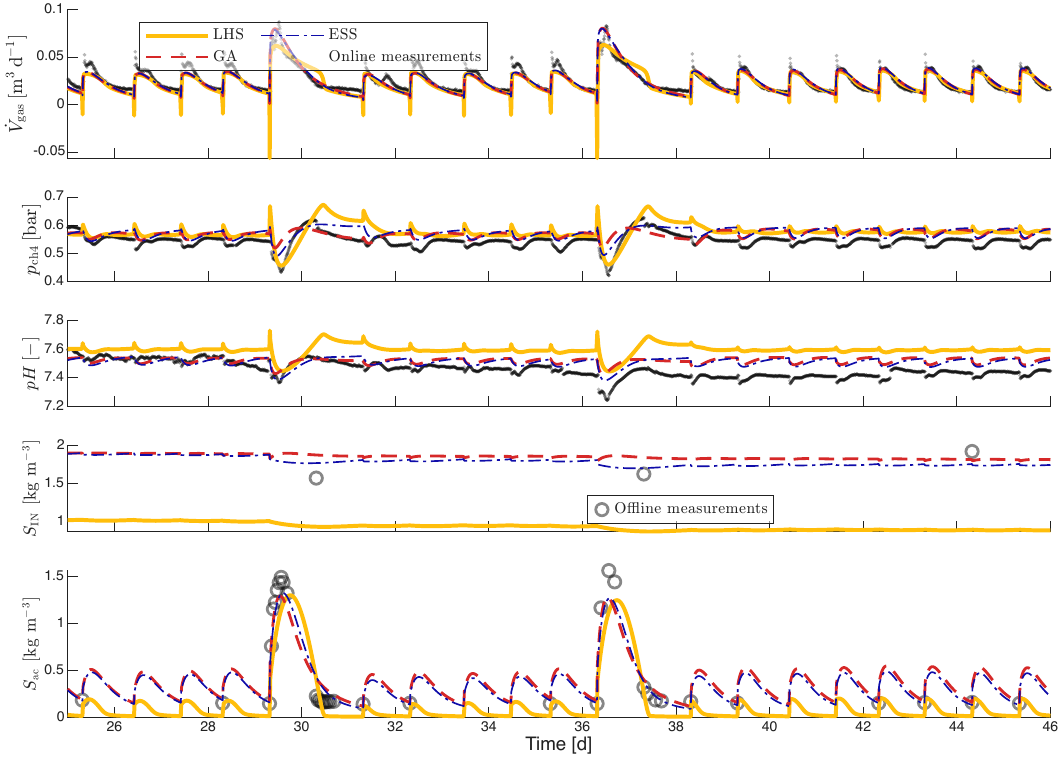}
\caption{Measurements (cross validation dataset) and model dynamics of the ADM1-R3-core after global MLE searches with LHS, GA, and ESS.}
\label{fig:si:fits_global_only_cross}
\end{figure}
\begin{figure}
\centering
\includegraphics[width=\linewidth]{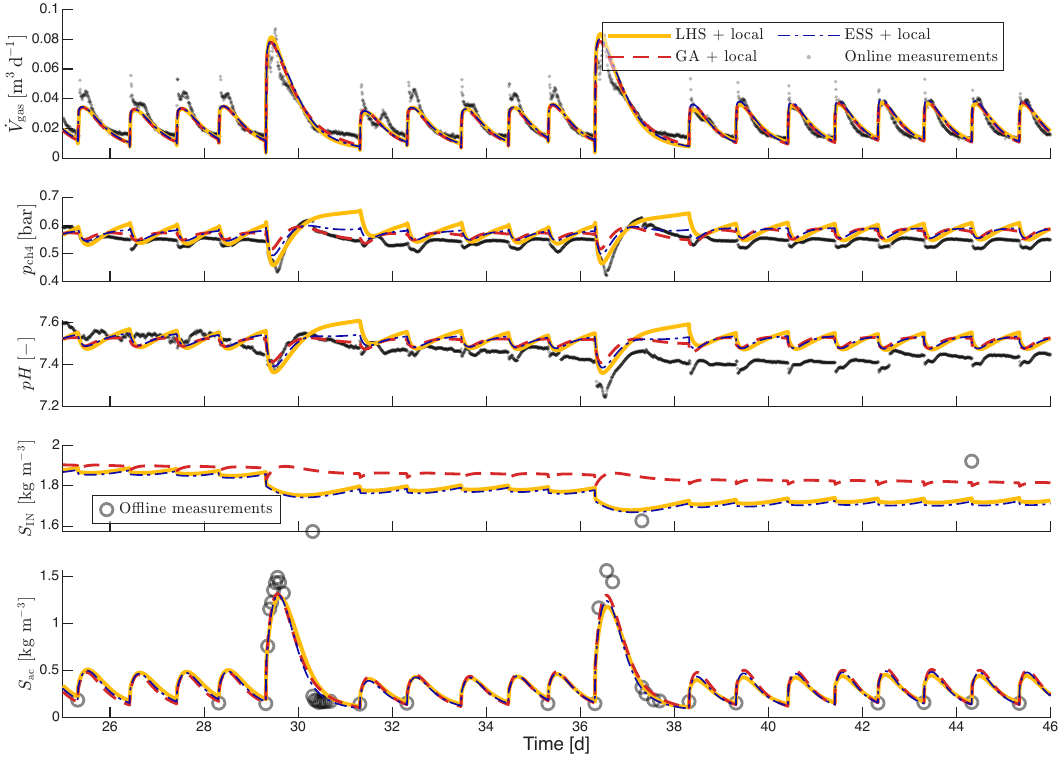}
\caption{Measurements (cross validation dataset) and model dynamics of the ADM1-R3-core after global MLE searches (LHS, GA, and ESS) and subsequent local refinements.}
\label{fig:si:fits_mle_3global_then_local}
\end{figure}
\begin{figure}
\centering
\includegraphics[width=\linewidth]{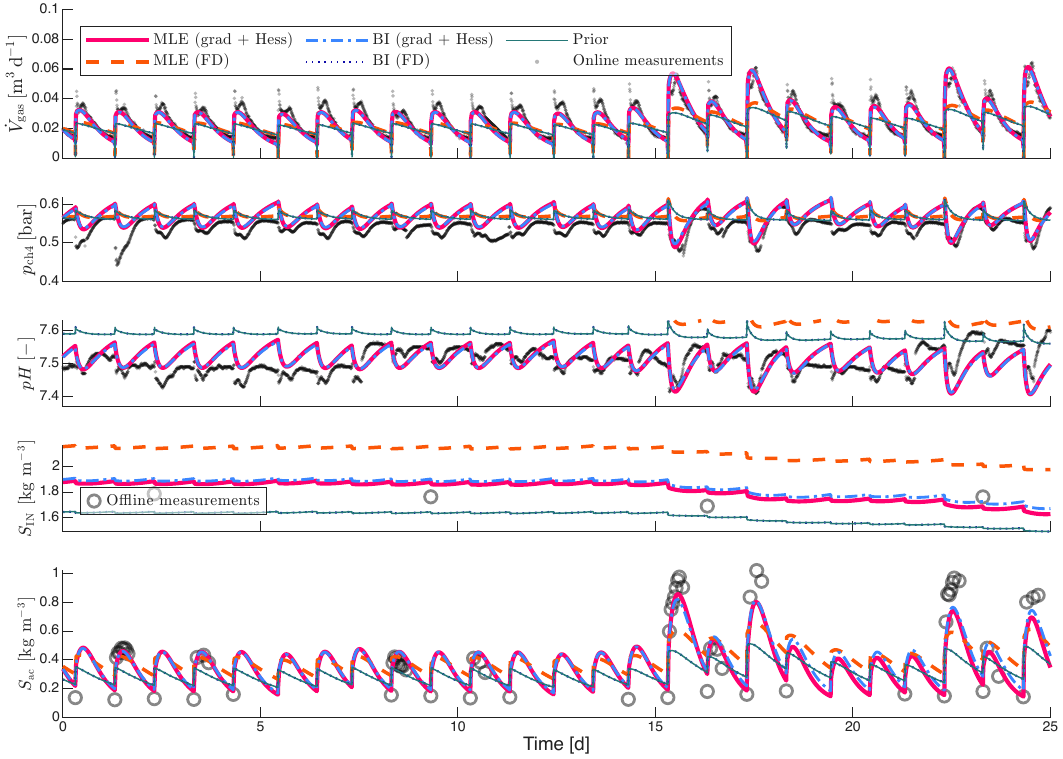}
\caption{Measurements (auto validation dataset) and model dynamics of the ADM1-R3-core. Fits are shown for \textbf{local-only} searches with MLE and BI objective functions, comparing finite differences (FD) with analytical gradient and Hessian information (grad + Hess).}
\label{fig:si:fits_local_only_mle_vs_bi}
\end{figure}
\begin{figure}
\centering
\includegraphics[width=\linewidth]{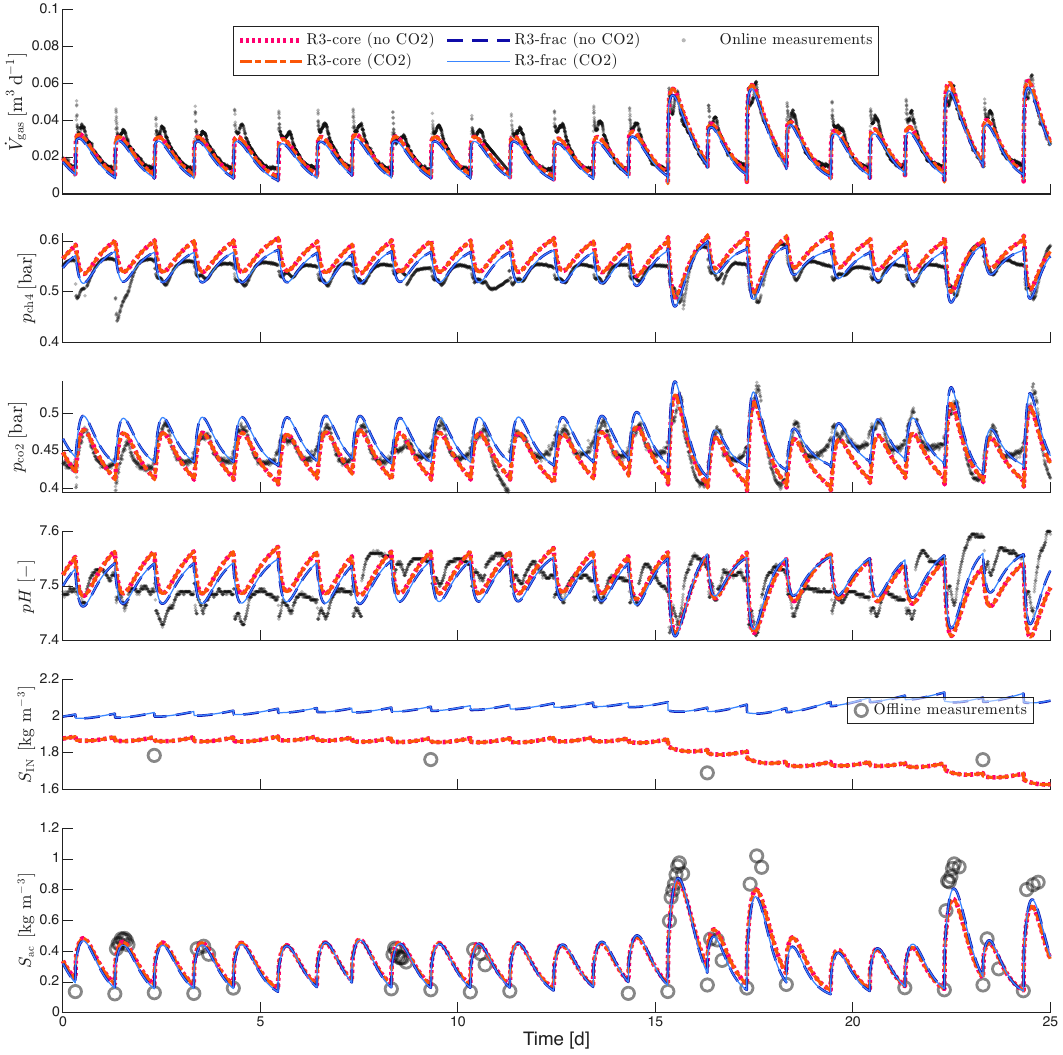}
\caption{Effect of considering \cotwo measurement data during calibration of ADM1-R3 model with one (ADM1-R3-core) and two carbohydrate fractions (ADM1-R3-core-frac): Measurements (full auto validation dataset) and dynamic models based on MLE local-only search.}
\label{fig:si:carbs_frac_co2}
\end{figure}
\begin{figure}
    \centering
    \begin{subfigure}[b]{0.46\linewidth}
        \includegraphics[width=\linewidth]{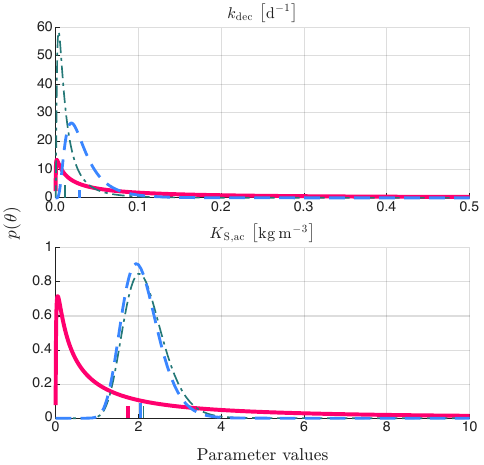}
        \caption{Log10-scaled parameters $k_\mathrm{dec}$ (top) and $K_\mathrm{S,ac}$ (bottom) on linear x-axis}
        \label{fig:si:gaussians_poor_log_abs}
    \end{subfigure}
    \par \vspace{0.5cm}
    \begin{subfigure}[b]{0.46\linewidth}
        \includegraphics[width=\linewidth]{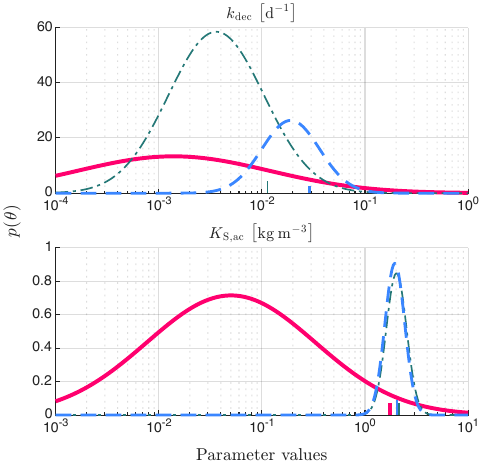}
        \caption{Log10-scaled parameters $k_\mathrm{dec}$ (top) and $K_\mathrm{S,ac}$ (bottom) on log10 x-axis}
        \label{fig:si:gaussians_poor_log_log}   
    \end{subfigure}
    \caption{Approximated prior and posterior distributions of MLE and BI estimates for selected abs- and log10-scaled model parameters of ADM1-R3-core with high uncertainty.}
    \label{fig:si:gaussians_poor}
\end{figure}
\begin{table}
\begin{threeparttable}
\centering
\renewcommand{\arraystretch}{1.1} 
\caption{Parameter estimates of \textbf{ADM1-R3-core} model according to MLE and BI after local-only search, and after global-only searches with MLE and BI. Standard deviations are omitted for global-only searches, as these do not mark terminal points of any complete search.}
\label{tab:si:mu_sigma_r3_core}
\begin{tabularx}{\linewidth}{YXd{3}d{2}d{3}d{2}d{3}d{3}d{3}d{3}d{3}d{3}}
\toprule
& & \multicolumn{4}{c}{Local-only$^{\mathrm{a,b}}$} & \multicolumn{6}{c}{Global-only$^{\mathrm{a}}$} \\
\cmidrule(lr){3-6} \cmidrule(lr){7-12}
& & \multicolumn{2}{c}{MLE} & \multicolumn{2}{c}{BI} & \multicolumn{3}{c}{MLE} & \multicolumn{3}{c}{BI} \\
\cmidrule(lr){3-4} \cmidrule(lr){5-6} \cmidrule(lr){7-9} \cmidrule(lr){10-12}
Index$^{\mathrm{c}}$ & Symbol & \multicolumn{1}{c}{$\mu$} & \multicolumn{1}{c}{$\sigma$} & \multicolumn{1}{c}{$\mu$} & \multicolumn{1}{c}{$\sigma$} & \multicolumn{1}{c}{LHS$^{\mathrm{c}}$} & \multicolumn{1}{c}{GA$^{\mathrm{c}}$} & \multicolumn{1}{c}{ESS$^{\mathrm{c}}$} & \multicolumn{1}{c}{LHS$^{\mathrm{c}}$} & \multicolumn{1}{c}{GA$^{\mathrm{c}}$} & \multicolumn{1}{c}{ESS$^{\mathrm{c}}$} \\
\midrule
1 & $k_\mathrm{ch}$ & 2.279 & 1.10 & 2.206 & 1.07 & 1.699 & 1.091 & 1.577 & 2.010 & 2.219 & 2.208 \\
2 & $k_\mathrm{pr}$ & 0.044 & 2.10 & 0.067 & 1.60 & 0.017 & 5.029 & 0.076 & 0.146 & 0.079 & 0.069 \\
3 & $k_\mathrm{li}$ & 0.407 & 4.03 & 0.565 & 1.52 & 0.647 & 3.407 & 4.421 & 4.354 & 0.565 & 0.558 \\
4 & $k_\mathrm{dec}$ & 0.112 & 9.21 & 0.029 & 1.91 & 0.600 & 0.008 & 0.032 & 0.000 & 0.023 & 0.028 \\
5 & $\mu_\mathrm{m,ac}$ & 1.237 & 4.71 & 0.729 & 1.52 & 1.404 & 0.564 & 0.390 & 0.131 & 0.686 & 0.696 \\
6 & $K_{\mathrm{S,ac}}$ & 0.930 & 6.01 & 2.044 & 1.25 & 0.029 & 6.399 & 1.157 & 3.349 & 2.158 & 2.011 \\
7 & $K_{\mathrm{I,nh3}}$ & 0.045 & 117.70 & 0.055 & 1.59 & 0.136 & 2.441 & 0.113 & 0.170 & 0.054 & 0.057 \\
8 & $\Delta S_\mathrm{ion,eff}$ & 0.085 & 0.01 & 0.083 & 0.01 & 0.165 & 0.086 & 0.083 & 0.146 & 0.083 & 0.082 \\
9 & $\varphi_\mathrm{IN,in}$ & 0.756 & 0.53 & 0.911 & 0.12 & 0.948 & 0.773 & 0.751 & 1.189 & 0.926 & 0.911 \\
\botrule
\end{tabularx}
\begin{tablenotes}
\footnotesize
\item[a] For log10-scaled parameters, $\mu$ refers to geometric mean (median), and $\sigma$ denotes the geometric (multiplicative) standard deviation derived from the inverse Hessian. Cf. Table~\mainref{tab:meth:parameters} for parameters' scales and units. 

\item[b] Analytical gradient and Hessian approximation were provided. 

\item[c] Idx: Index, LHS: Latin hypercube sampling, GA: genetic algorithm, ESS: enhanced scatter search
\end{tablenotes}
\end{threeparttable}
\end{table}
\begin{table}
\begin{threeparttable}
\centering
\renewcommand{\arraystretch}{1.1} 
\caption{KPIs of BI global search and local refinement with analytical gradient and Hessian approximation.}
\label{tab:si:kpi_bi_global_local}
\begin{tabularx}{\linewidth}{X*{3}{r}p{2mm}*{3}{r}}
\toprule
& \multicolumn{3}{c}{Global search} & & \multicolumn{3}{c}{Local refinement$^{\mathrm{a}}$}\\
\cmidrule(lr){2-4} \cmidrule(lr){6-8}
Criterion & LHS & GA & ESS & & LHS $+$ g\&H & GA $+$ g\&H & ESS $+$ g\&H \\
\midrule
Min. obj. function value & 179.43 & 65.12 & 65.01 & & 65.00 & 65.00 & 65.00 \\
Min. log-likelihood & 152.25 & 62.55 & 62.19 & & 62.16 & 62.15 & 62.15 \\
No. of obj. function calls & 5000 & 4855 & 4568 & & 47 & 43 & 39 \\
No. of iterations & $-$ & 50 & 40 & & 12 & 7 & 4 \\
No. of calls per iteration$^{\mathrm{b}}$ & $-$ & 97 & 114 & & 4 & 6 & 10 \\
\botrule
\end{tabularx}
\begin{tablenotes}
\footnotesize
\item[a] g\&H: gradient and Hessian
\item[b] Average number of objective function calls per iteration (rounded).
\end{tablenotes}
\end{threeparttable}
\end{table}
\begin{table}
\begin{threeparttable}
\centering
\renewcommand{\arraystretch}{1.1} 
\caption{Parameter estimates and standard deviations of \textbf{ADM1-R3-core-frac} model according to MLE and BI after global search with local refinement (best final estimates according to objective function value).}
\label{tab:si:mu_sigma_r3_frac}
\begin{tabularx}{\linewidth}{clYYd{3}d{2}d{3}d{2}d{3}d{2}}
\toprule
& & & & \multicolumn{2}{c}{MLE$^{\mathrm{a}}$} & \multicolumn{2}{c}{BI$^{\mathrm{a}}$} & \multicolumn{2}{c}{Prior$^{\mathrm{a,b}}$} \\
\cmidrule(lr){5-6} \cmidrule(lr){7-8} \cmidrule(lr){9-10}
Index & Symbol & Scale & Unit & \multicolumn{1}{c}{$\mu$} & \multicolumn{1}{c}{$\sigma$} & \multicolumn{1}{c}{$\mu$} & \multicolumn{1}{c}{$\sigma$} & \multicolumn{1}{c}{$\mu$} & \multicolumn{1}{c}{$\sigma$} \\
\midrule
1 & $k_\mathrm{ch,f}$ & log10 & \si{\per\day} & 2.655 & 4.69 & 64.726 & 1.05 & 67.570 & 1.05 \\
2 & $k_\mathrm{ch,s}$ & log10 & \si{\per\day} & 1.332 & 305.78 & 2.081 & 1.08 & 0.810 & 1.72 \\
3 & $k_\mathrm{pr}$ & log10 & \si{\per\day} & 0.843 & 38.25 & 0.687 & 2.41 & 0.120 & 3.06 \\
4 & $k_\mathrm{li}$ & log10 & \si{\per\day} & 8.505 & \multicolumn{1}{c}{\num{5.15e251}} & 0.660 & 1.62 & 0.630 & 1.62 \\
5 & $k_\mathrm{dec}$ & log10 & \si{\per\day} & 0.421 & 2.98 & 0.202 & 1.71 & 0.020 & 4.31 \\
6 & $\mu_\mathrm{m,ac}$ & log10 & \si{\per\day} & 4.990 & 70.62 & 2.711 & 1.63 & 0.710 & 2.08 \\
7 & $K_{\mathrm{S,ac}}$ & log10 & \si{\kilogram\per\cubic\meter} & 0.770 & 5.28 & 1.925 & 1.25 & 2.180 & 1.26 \\
8 & $K_{\mathrm{I,nh3}}$ & log10 & \si{\kilogram\per\cubic\meter} & 0.030 & 940.36 & 0.074 & 1.65 & 0.060 & 1.68 \\
9 & $\Delta S_\mathrm{ion,eff}$ & abs. & \si{\kilo\mole\per\cubic\meter} & 0.076 & 0.01 & 0.086 & 0.01 & 0.130 & 0.09 \\
10 & $\varphi_\mathrm{IN,in}$ & abs. & $-$ & 0.750 & 0.40 & 0.803 & 0.11 & 1.000 & 0.13 \\
11 & $\varphi_\mathrm{ch,in,f}$ & abs. & $-$ & 0.707 & 3.38 & 0.038 & 0.01 & 0.320 & 0.02 \\
\botrule
\end{tabularx}
\begin{tablenotes}
\footnotesize
\item[a] For log10-scaled parameters, values $\mu$ refer to geometric mean (median), and standard deviations $\sigma$ denote geometric (multiplicative) values. 

\item[b] Prior distribution values of log10-scaled parameters apply \textit{after} conversion with delta method, and hence differ from those in Table~\mainref{tab:meth:parameters}.    
\end{tablenotes}
\end{threeparttable}
\end{table}

\newpage 

\clearpage
\section{Supplementary materials and methods}
The following section highlights details on the algorithms applied and their numerical implementation. Next, conversion to log10 scale is explained before data preprocessing steps are detailed. Finally, the full sets of model equations of the ADM1-R3 model variants are summarized.

\subsection{Algorithms}
\label{sec:si_meth:algorithms}
\subsubsection{Multirate MLE objective function}
\label{sec:si:obj_funcs}
Eq.~\eqref{eq:si:obj_fun_mle_mr} shows the maximum likelihood estimation (MLE) objective function (OF) as a sum over $n_\mathrm{y}$ individual (multirate) model outputs
\begin{equation} \label{eq:si:obj_fun_mle_mr}
    J_\mathrm{MLE}(\hat \theta) = \frac{1}{2} \sum_{l=1}^{n_\mathrm{y}} \sum_{k=1}^{N_l} \left(y_{l,k} - g_{l,k}\right)^T \, \left(\sigma_{l,k}^2 \, N_l\right)^{-1} \, \left(y_{l,k} - g_{l,k} \right), 
\end{equation}
which provides an alternative to Eq.~\maineqref{eq:meth:mle_obj_fun}, which sums over individual sample times. Usually, the standard deviations $\sigma$ are assumed to be time-invariant, hence $\sigma_{l,k}$ simplifies to $\sigma_l$ per model output $l$.
\subsubsection{Objective function gradients and Hessian}
\label{sec:meth_si:grad_hess_derivation}
\paragraph{Maximum likelihood estimation}
The analytical gradient of the OF of MLE follows by applying the chain rule to Eq.~\maineqref{eq:meth:mle_obj_fun}~\cite{frohlich2017a}, giving
\begin{equation} 
\nabla_{\theta} J_\mathrm{MLE} = - \sum_{k=1}^{N} \left(y_k - g_k\right)^T \, \tilde C_k^{-1} \, \left.\frac{dg}{d\theta} \right|_k, 
\label{eq:meth:gradient_mle}
\end{equation}
where $\frac{dg}{d\theta} \coloneq S_\mathrm{y}$ denotes the output sensitivity w.r.t. model parameters $\theta$, cf. Sec.~\ref{sec:meth_si:sensitivities}.

Close to the optimum $\hat \theta_\mathrm{MLE}$, second-order output sensitivities $\frac{d^2 g}{d\theta^2}$ can be neglected, resulting in the Gauss-Newton approximation of the Hessian~\cite{nocedal2006} 
\begin{equation} 
\nabla^2_{\theta} J_\mathrm{MLE} \approx \sum_{k=1}^{N} \left(\left.\frac{dg}{d\theta} \right|_k\right)^T \, \tilde C_k^{-1} \, \left.\frac{dg}{d\theta} \right|_k.  
\label{eq:meth:hessian_mle}
\end{equation}

\paragraph{Bayesian inference}
Since the OF of Bayesian inference (BI), Eq.~\maineqref{eq:meth:map_obj_fun}, describes an extension to the MLE OF, Eq.~\maineqref{eq:meth:mle_obj_fun}, analogous expressions of the analytical gradient and the Hessian approximation are obtained as 
\begin{subequations}
\begin{align}
    \nabla_{\theta} J_\mathrm{MAP} &= - \sum_{k=1}^{N} \left(y_k - g_k\right)^T \, \tilde C_k^{-1} \, \left.\frac{dg}{d\theta} \right|_k + C_\mathrm{\theta\theta,0}^{-1} \left( \theta - \theta_0 \right), \label{eq:meth:gradient_map} \\
    \nabla^2_{\theta} J_\mathrm{MAP} &\approx \sum_{k=1}^{N} \left(\left.\frac{dg}{d\theta} \right|_k\right)^T \, \tilde C_k^{-1} \, \left.\frac{dg}{d\theta} \right|_k + C_\mathrm{\theta\theta,0}^{-1}. \label{eq:meth:hessian_map}
\end{align}
\end{subequations}
Again, second-order output sensitivities were neglected for Eq.~\eqref{eq:meth:hessian_map}~\cite{nocedal2006}. The prior values of model parameters are given in Table~\mainref{tab:meth:parameters}. 
\subsubsection{Forward sensitivities}
\label{sec:meth_si:sensitivities}
Fröhlich et al.~\cite{frohlich2017a} provide an overview of sensitivity computation schemes and endorse forward sensitivities as accurate and efficient for low and medium-dimensional models. Forward sensitivities describe analytical first-order sensitivities of the states $x$ w.r.t. parameters $\theta$, $S_{\mathrm{x}} = \frac{dx}{d\theta}$. They are obtained by integrating the following matrix-valued ordinary differential equations (ODEs) alongside the model ODEs (Eq.~\maineqref{eq:meth:ode_x})~\cite{barz2018,frohlich2017a}
\begin{equation}
\dot{S}_\mathrm{x} = \frac{\partial f}{\partial x} \, S_\mathrm{x} + \frac{\partial f}{\partial \theta}, \quad S_\mathrm{x}(t_{0}) = \frac{dx_0}{d\theta}.
\label{eq:meth:sens_ode}
\end{equation}
Output sensitivities $S_\mathrm{y}$ then result from state sensitivities $S_\mathrm{x}$ via the chain rule
\begin{equation}
S_\mathrm{y} = \frac{dg}{d\theta} = \frac{\partial g}{\partial x} \, S_\mathrm{x} + \frac{\partial g}{\partial \theta}.
\label{eq:meth:sens_output}
\end{equation}

\subsection{Implementation}
\subsubsection{Increased numerical efficiency}
\label{sec:meth_si:imp:numerical_tricks}
A range of numerical techniques improved the efficiency of the optimization algorithms, enabling reasonable simulation results and ensuring reliable solver convergence.

\paragraph{Dynamic initialization}
Simulations started in a dynamic initial state $x_0$. It was obtained by starting at a steady state (SS) $x_\mathrm{ss}$ and applying the dynamic feed pattern during the initialization dataset, Fig.~\mainref{fig:meth:feeding}. $x_\mathrm{ss}$ in turn resulted from a constant feed load, applied over a simulative period of \SI{500}{d} and starting at fixed values $x_{0,\mathrm{ss}}$ as per Table~\ref{tab:si:influent_init}. Influent concentrations remained fixed at average values during the initialization period. As of the auto validation period, influent concentrations varied with each new substrate batch.
\paragraph{Accelerated simulations}
\textsc{Matlab}'s \texttt{ode15s} function integrated the stiff model equations using relative and absolute tolerances of \mbox{1E-5} and \mbox{1E-6}, found as sufficiently accurate during initial tests. Furthermore, use of symbolically derived Jacobians eased synchronous integration of states and sensitivities~\cite{frohlich2017}. Additionally, non-negativity was enforced on all states. 

Latin hypercube sampling (LHS) and genetic algorithm (GA) ran with parallelized optimizations. Parallelization is not natively possible for the enhanced scatter search (ESS) implementation provided by the \textsc{MEIGO} toolbox~\cite{egea2014}\ifarxiv\footnote{Instead, parallelization can be achieved by running multiple \textsc{Matlab} sessions in parallel and coordinating them via an overarching main session~\cite{egea2014}, which was not pursued here.}\fi. 

Certain parameter combinations within the search space caused stalling simulations due to the ODE solver, where usually SS values were unphysical or implausible yet~\cite{delory2025}. Plausibility checks of SS values during global optimization prevented unnecessary computational effort on infeasible candidate solutions. Unphysical SS values showing negative concentrations and those with extremely high ratios of AC over alkalinities (washout of acetoclastic methanogens~\cite{bernard2001,bade2026}) were discarded by assigning them a very high OF value of 1E14. The next paragraph 
provides details on filtering criteria. 

Finally, a cutoff time of \SI{15}{\second} per simulation over the entire dataset during global searches and \SI{30}{\second} during local searches ensured effective optimizer progress. Initial tests identified these as sufficiently long given usual simulation times of less than \SI{1}{\second}. 
\paragraph{Thresholds for discarding steady states}
\label{sec:si:thresholds_ss}
Negative SSs were discarded if any state entry was lower than \SI{-0.005}{\kilogram\per\cubic\meter}. 

The washout of acetoclastic methanogens was examined by means of the ratio of volatile fatty acids (VFA) and alkalinities (total inorganic carbonates, TIC)~\cite{drosg2013,ahmed2020,koch2010}. For the ADM1-R3-core, VFAs were approximated as $\textit{VFA} \sim S_\mathrm{ac}$, and $\textit{TIC} \sim S_\mathrm{nh4} + S_\mathrm{hco3}$, where $S_\mathrm{nh4} = S_\mathrm{IN} - S_\mathrm{nh3}$. A threshold of $\textit{VFA}/\textit{TIC} = 10$ identified the biomass washout sufficiently accurate. 
\subsection{Conversion between linear and log10 scale}
\label{sec:si:log10_conversion}
Literature values of parameters are usually given in terms of linear scale mean and standard deviation (SD).\footnote{If instead plausible bounds are available, these can be converted to SDs by interpreting them as confidence intervals of $\pm 2$ or 3 SDs~\cite{yoko2026}, covering 95.5 or \SI{99.7}{\percent} of a Gaussian distribution, respectively.} 
When assuming Gaussian priors and applying log10 conversion on parameters, the first two moments (mean and SD) of the prior must be converted to log10-scale. 
Two methods to achieve this are briefly reviewed: moment conservation, and the delta method. Note that neither transforms the distributions exactly, but only approximates the first two moments. 

In this context, three scales are considered: absolute (abs) scale ($x$), log-scale (based on the natural logarithm $\ln$, $y$), and log10-scale ($z$). The three scales are linked by the coordinate transformation $\phi: x \mapsto z$ via
\begin{equation}
    z = \log_{10}(x) = \frac{\ln(x)}{\ln (10)} = \frac{y}{\ln (10)}.
    \label{eq:si:trafo_log10}
\end{equation}
Recall that a random variable $\chi$ is log-normally distributed, if its natural logarithm $\psi=\ln(\chi)$ follows a normal distribution $\psi \sim \mathcal N (\mu_\mathrm{y}, \sigma_\mathrm{y}^2)$. This is denoted as $\chi \sim \mathcal{LN} (\mu_\mathrm{y}, \sigma_\mathrm{y}^2)$, where $\mu_\mathrm{y}$ and $\sigma_\mathrm{y}$ describe mean (expected value) and standard deviation (SD) of the log-transformed variable $\psi$. While for normal distributions, mean, median and mode overlap, they all differ for log-normal (and log10-normal) distributions. This is due to the log-normal's positive skewness (tail heaviness), and it holds that the mode $<$ median $<$ mean. 

Conversion between log- and log10-scale follows a simple linear transformation given in Eq.~\eqref{eq:si:trafo_log10}, which applies for both conversions discussed below
\begin{subequations}
\begin{align}
    \mu_\mathrm{z} &= \frac{\mu_\mathrm{y}}{\ln (10)} \\
    \sigma_\mathrm{z} &= \frac{\sigma_\mathrm{y}}{\ln (10)}.
\end{align}
\end{subequations}
The following describes the conversion of mean and SD from linear to log scale (instead of log10 scale), because the natural logarithm represents the most widely used parametrization, e.g., in \textsc{Matlab}'s \texttt{lognpdf}. Conversions refer to scalar distributions first. Multivariate distributions and re-transformation to linear scale are discussed thereafter. 

Notably, individual values such as bounds can be converted by simply applying the coordinate transformation in Eq.~\eqref{eq:si:trafo_log10}. 

\subsubsection{Moment conservation}
\label{sec:si:moment_conservation}
Converting a linear-scale normal distribution with given mean $\mu_\mathrm{x}$ and SD $\sigma_\mathrm{x}$ to log scale while maintaining the first two moments is achieved by chossing the following log-normal parameters
\begin{subequations}
\label{eq:si:moment_conservation}
\begin{align}
    \mu_\mathrm{y} &= \ln(\mu_\mathrm{x}) - \frac{\sigma_\mathrm{y}^2}{2} = \ln \left( \frac{\mu_\mathrm{x}^2}{\sqrt{\mu_\mathrm{x}^2 + \sigma_\mathrm{x}^2}} \right), \\
    \sigma_\mathrm{y}^2 &= \ln \left( 1 + \left(\frac{\sigma_\mathrm{x}}{\mu_\mathrm{x}}\right)^2 \right).  
\end{align}
\end{subequations}
This follows directly from the definition of mean and variance of the log-normal. 
\subsubsection{Delta method}
For small ratios of linear-scale SD and mean, $\frac{\sigma_\mathrm{x}}{\mu_\mathrm{x}} \ll 1$, the coordinate transformation in Eq.~\eqref{eq:si:trafo_log10} can be linearized around $\mu_\mathrm{x}$, resulting in a simplified conversion 
\begin{subequations}
\label{eq:si:delta_method}
\begin{align}
    \mu_\mathrm{y} &\approx \ln (\mu_\mathrm{x}), \label{eq:si:delta_method_mean} \\
    \sigma_\mathrm{y} &\approx \frac{\sigma_\mathrm{x}}{\mu_\mathrm{x}}. 
\end{align}
\end{subequations}
In the context of variance conversion, this is frequently referred to as the delta method. Note that Eq.~\eqref{eq:si:delta_method_mean} in fact describes the median (not the mean) of a log-normal distribution. 
\subsubsection{Matrix-valued conversion}
Above conversions of variances and SD apply for scalar variables. The PI problems in this contribution involve parameter \textit{vectors} (of dimension $n$) with associated covariance \textit{matrices}.  Matrix-valued operations shall be illustrated by means of converting a vector-valued, linear-scale prior mean $\mu_\mathrm{x} \in \mathbb{R}^n$ with covariance $C_\mathrm{x} \in \mathbb{R}^{n \times n}$ to log-scale, where $C_\mathrm{x}$ is usually chosen as diagonal
\begin{equation}
    C_\mathrm{x} = diag(\sigma_{\mathrm{x},1}^2, \ldots , \sigma_{\mathrm{x},n}^2)
\end{equation}
but does not need to \cite{yoko2026}. First, the delta method delivers
\begin{equation}
    C_\mathrm{z} = C_\mathrm{z}^\mathrm{\delta} = \left. \frac{\partial \phi}{\partial x} \right|_{\mu_\mathrm{x}} \, C_\mathrm{x} \, \left( \left. \frac{\partial \phi}{\partial x} \right|_{\mu_\mathrm{x}} \right)^T,  \label{eq:si:delta_method:vectorial}
\end{equation}
where $\left. \frac{\partial \phi}{\partial x} \right|_{\mu_\mathrm{x}}$ denotes the first derivative of the coordinate transformation of Eq.~\eqref{eq:si:trafo_log10}, evaluated at the linear-scale value $\mu_\mathrm{x}$. In the scalar case, this first derivative reads
\begin{equation}
    \frac{\partial \phi}{\partial x} = \frac{1}{\ln (10)} \frac{1}{x}. 
\end{equation}
In matrix form, this gives a diagonal matrix
\begin{equation}   
     \left. \frac{\partial \phi}{\partial x} \right|_{\mu_\mathrm{x}} = diag \left( \left(\ln(10) \, \mu_{\mathrm{x},1}\right)^{-1}, \ldots ,\left(\ln(10) \, \mu_{\mathrm{x},n}\right)^{-1} \right), 
\end{equation}
where $a_i$ denotes the $i$\textsuperscript{th} entry of any vector $a$. The delta method affects diagonal and off-diagonal entries alike, provided $C_\mathrm{x}$ contains non-zero off-diagonal values.

Second (optional), moment conservation substitutes diagonal elements of the covariance matrix resulting from Eq.~\eqref{eq:si:delta_method:vectorial} with
\begin{equation}
    C_{\mathrm{z} [k,k]} = \frac{1}{\left( \ln (10) \right)^2} \ln \left( 1 + \frac{C_{\mathrm{x} [k,k]}}{\mu_{\mathrm{x},k}^2} \right), \quad k = 1, \ldots, n.
\end{equation}
Herein, $A_{[i,j]}$ denotes the $i$\textsuperscript{th} row and $j$\textsuperscript{th} column of any matrix $A$. The second step is omitted, if the delta method is considered sufficiently accurate. The converted SDs follow from
\begin{equation}
    \sigma_{\mathrm{z},k} = \sqrt{C_{\mathrm{z} [k,k]}}.
\end{equation}
\subsubsection{Interpreting log10-scaled parameter estimates}
\label{sec:si:log10_params_interpretation}
Both MLE and BI deliver parameter estimates and approximate parameter uncertainties in the shape of covariance matrices (Sec.~\mainref{sec:meth:algorithms}). BI involves a posterior parameter distribution which, in this work, is approximated by a Gaussian. Contrarily, MLE does not make an assumption on the parameters' underlying distribution. Nonetheless, MLE estimates are also approximated by a Gaussian, following from the Cramér-Rao bound.

The above equations could be inverted to obtain approximate expressions converting log10- to linear-scaled estimates. Alternatively, log10-scaled values $\mu_\mathrm{z}$ and $\sigma_\mathrm{z}$ can be expressed as \textit{geometric} mean and SD, indicated by a superscript star ($*$)
\begin{subequations}
\begin{align}
    \mu^*_\mathrm{x} &= \exp(\mu_\mathrm{y}) = 10^{\mu_\mathrm{z}}, \\ 
    \sigma_\mathrm{x}^* &= \exp(\sigma_\mathrm{y}) = 10^{\sigma_\mathrm{z}}.
\end{align}
\end{subequations}
These result in a multiplicative (instead of additive) interpretation of a $\pm 1 \sigma$ confidence intervals as 
\begin{equation}
    \left[ \frac{\mu_\mathrm{x}^*}{\sigma^*_\mathrm{x}} , \, \mu_\mathrm{x}^*\cdot \sigma^*_\mathrm{x} \right].    
\end{equation}
Note that these geometric values hold for BI posteriors and MLE estimates alike.

\subsection{Data preprocessing}
\label{sec:si:data_preprocessing}
In addition to applying the data preprocessing pipeline of Meola et al.~\cite{meola2023}, biogas production data was clipped to an interval of \SIrange{0.001}{0.14}{\cubic\meter\per\day}. Next, a maximum time derivative of \SI{1}{\cubic\meter\per\square\day} filtered outliers well, which were mostly caused by faulty gas measurements during digestate sampling. Finally, offline measurements of digestate and substrate characterization were replaced by default values where the data at hand was deficient. Resulting mass-weighted average values of influent concentrations are provided in Table~\ref{tab:si:influent_init}. The underlying computation of influent concentrations and digestate samples is described in the following. 

%

%
\subsubsection{Influent concentrations}
\label{sec:si:influent}
Influent concentrations were computed according to standard in-house procedures described in Liebetrau and Pfeiffer~\cite{liebetrau2020}, and are summarized in Table~\ref{tab:si:influent_init}. 
\begin{table}
\begin{threeparttable}
\renewcommand{\arraystretch}{1.1}
\centering
\caption{Average influent concentrations$^{\mathrm{a}}$ $\xi$ and initial values $x_{0,\mathrm{ss}}$ for dynamic steady-state (SS) computation of ADM1-R3 variants. All values in [\si{\kilogram\per\cubic\meter}].} 
\label{tab:si:influent_init}
\begin{tabularx}{\linewidth}{ccYd{2}d{2}d{2}d{3}d{3}}
    \toprule
    \multicolumn{2}{c}{State index} & & \multicolumn{3}{c}{$\xi_i$} & \multicolumn{2}{c}{$x_{0,\mathrm{ss}}$} \\
    \cmidrule(lr){1-2} \cmidrule(lr){4-6} \cmidrule(lr){7-8}
    R3-core$^{\mathrm{b}}$ & R3-frac$^{\mathrm{b}}$ & Component$^{\mathrm{c}}$ & \multicolumn{1}{c}{MS$^{\mathrm{b, d}}$} & \multicolumn{1}{c}{CM$^{\mathrm{b, d}}$} & \multicolumn{1}{c}{mix$^{\mathrm{e}}$} & \multicolumn{1}{c}{R3-core$^{\mathrm{b}}$} & \multicolumn{1}{c}{R3-frac$^{\mathrm{b,f}}$} \\
    \midrule
    1   & 1   & $S_\mathrm{ac}$ & 8.41   & 6.73  & 7.49 & 0.049 & 0.049 \\
    2   & 2   & $S_\mathrm{ch4}$ & 0 & 0 & 0 & 0.012 & 0.012 \\
    3   & 3   & $S_\mathrm{IC}$ & 0 & 0 & 0 & 4.975 & 4.975 \\
    4   & 4   & $S_\mathrm{IN}$ & 0.36    & 1.47  & 0.97 & 0.964 & 0.964 \\
    5   & 5   & $X_\mathrm{ch}$/$X_\mathrm{ch,f}$ & 212.69  & 16.46 & 104.91 & 2.962 & 0.948 \\
    $-$ & 6   & $X_\mathrm{ch,s}$ & 0 & 0 & 0 & & 2.014 \\
    6   & 7   & $X_\mathrm{pr}$ & 21.10   & 12.63 & 16.45 & 0.949 & 0.949 \\
    7   & 8   & $X_\mathrm{li}$ & 18.15   & 5.23  & 11.06 & 0.412 & 0.412 \\
    8   & 9   & $X_\mathrm{bac}$ & 0 & 0 & 0 & 1.926 & 1.926 \\
    9   & 10  & $X_\mathrm{ac}$ & 0 & 0 & 0 & 0.522 & 0.522 \\
    10  & 11  & $S_\mathrm{ac^-}$ & 0 & 0 & 0 & 0.049 & 0.049 \\
    11  & 12  & $S_\mathrm{hco3^-}$ & 0 & 0 & 0 & 4.546 & 4.546 \\
    12  & 13  & $S_\mathrm{nh3}$ & 0 & 0 & 0 & 0.022 & 0.022 \\
    13  & 14  & $S_\mathrm{ch4,gas}$ & 0 & 0 & 0 & 0.358 & 0.358 \\
    14  & 15  & $S_\mathrm{co2,gas}$ & 0 & 0 & 0 & 0.660 & 0.660 \\
    \bottomrule
\end{tabularx}
\begin{tablenotes}
\footnotesize
    \item[a] Computed from raw substrate characterization data with in-house procedures~\cite{liebetrau2020}.
    
    \item[b] R3-core: ADM1-R3-core, R3-frac: ADM1-R3-core-frac. MS: maize silage. CM: cattle manure.
    
    \item[c] For the ADM1-R3-core-frac, influent carbohydrates (CH) are entirely attributed to the fast CH fraction, and split between the fractions with the fraction parameter $\varphi_\mathrm{ch,in,f}$.
    
    \item[d] Values were averaged from $n$ batches of the same substrate type fed, weighted with individual feed masses applied; $n_\mathrm{MS} = 3, n_\mathrm{CM} = 5$. 
    
    \item[e] Mix value averaged from individual substrates' average values, weighted with total substrate masses applied.
    
    \item[f] For ADM1-R3-core-frac, total initial CH are attributed to fast and slow fraction according to the default fraction parameter $\varphi_\mathrm{ch,in,f}$ (Table~\mainref{tab:meth:parameters}).
\end{tablenotes}
\end{threeparttable}
\end{table}

Computation of influent concentrations are described for silages in the following. For influent IN and AC ($S_\mathrm{IN}$ and $S_\mathrm{ac}$) of cattle manure, measurements were corrected by their raw water (RW) content as described in Section~\ref{sec:si:influent:rq_correction}. 

Determining substrate measurements involves to elute substrate samples for \SI{24}{\hour}. After this period, relevant constituents considered were assumed to be dissolved. Samples of manure and digestate samples do not require this step. Values directly measured are denoted by a superscript l (laboratory) to distinguish them from model variables. 
\paragraph{Total inorganic nitrogen (IN)}
Concentrations of total inorganic nitrogen were measured at neutral pH~\cite{strach2020}. Here, the acid-base equilibrium was assumed to be entirely shifted towards ammonium (\mbox{\nhfour-N}), thus accounting for all of the inorganic nitrogen
\begin{equation}
    S_\mathrm{IN} = S_\mathrm{NH_4-N}^\mathrm{l}. \label{eq:si:influent:in}
\end{equation}
Any dilutions due to eluting the sample were assumed to be incorporated in the measurement $S_\mathrm{NH_4-N}^\mathrm{l}$.


%
\paragraph{Acetic acid (AC)}
The ADM1-R3 summarizes all volatile fatty acids of the ADM1 to an effective AC concentration. This was considered by transforming mole equivalents to AC mass concentrations as
\begin{equation}
    S_{ac} = S^\mathrm{l}_\mathrm{ac} + S^\mathrm{l}_\mathrm{pro}\frac{\bar M_\mathrm{ac}}{\bar M_\mathrm{pro}} + S^\mathrm{l}_\mathrm{bu} \frac{\bar M_\mathrm{ac}}{\bar M_\mathrm{bu}} + S^\mathrm{l}_\mathrm{va}\frac{\bar M_\mathrm{ac}}{\bar M_\mathrm{va}}. \label{eq:si:influent:ac}
\end{equation}
Likewise, any dilution factors due to eluting the sample were assumed to be incorporated in $S^\mathrm{l}_i$. 

%
\paragraph{Macro nutrients}
Influent carbohydrates (CH), proteins (PR), and lipids (LI) were computed based on substrate characterization according to the established procedure of Weender and van Soest analysis. In particular, this delivered crude proteins $XP$, crude lipids $XL$, and crude ash $XA$. These crude measurements are given in kg $\textrm{kg}^{-1}\textrm{TS}$, where TS denotes total solids and is given in \%\,FM (fresh matter). They were converted to mass concentrations via TS and the FM density of the samples, which was assumed as \SI{1000}{\kilogram\per\cubic\meter}, giving 
\begin{subequations}
\begin{align}
    X_\mathrm{pr} &= XP \, \frac{TS}{100} \, \rho_\mathrm{FM}, \\ 
    X_\mathrm{li} &= XL \, \frac{TS}{100} \, \rho_\mathrm{FM}.  
\end{align}
\end{subequations}
CH, by contrast, were derived via a substrate-specific fermentability ratio of CH, $FQ_\mathrm{ch}$, which was linked to the ratio between the typically measured and the stoichiometric biochemical methane potential (BMP)~\cite{hellmann2026a} as
\begin{subequations}
\begin{align}
    FQ_\mathrm{ch} &= \frac{FQ \, (1 - XA) - XP - XL}{XC}, \quad \mathrm{where} \\ 
    FQ &= \frac{BMP_\mathrm{meas}}{BMP_\mathrm{stoic}}, \quad \mathrm{and} \label{eq:si:influent:fq} \\
    XC &= 1 - XA - XP - XL.
\end{align}
\end{subequations}
This assumes that TS is comprised by the sum of crude ash and all three crude macro nutrients including crude CH $XC$.  
The stoichiometric BMP in Eq.~\eqref{eq:si:influent:fq} was assumed to be 420\,$\mathrm{L\, kg}^{-1}\mathrm{VS}$ for agricultural substrates\cite{weissbach2009}, where VS denotes volatile solids. Measured BMP values of 357 and 230\,$\mathrm{L\, kg}^{-1}\mathrm{VS}$ were assumed for maize silage and cattle manure, respectively~\cite{KTBL2021}. This delivers the mass concentration of CH as 
\begin{equation}
    X_\mathrm{ch} = FQ_\mathrm{ch} \, XC \, \frac{TS}{100} \, \rho_\mathrm{FM}. 
\end{equation}
For the ADM1-R3-core-frac, all influent CH were attributed to the fast CH fraction (Table~\ref{tab:si:influent_init} and Section~\ref{sec:si:model_equations:r3-frac}). 
\subsubsection{Raw water correction}
\label{sec:si:influent:rq_correction}
Computations of influent IN and AC were thus far described for solid substrates, which presuppose an eluate. For manures, by contrast, computations additionally require a RW correction, i.e., values resulting from Eqs.~\eqref{eq:si:influent:in} and \eqref{eq:si:influent:ac} must be multiplied with $\phi_\mathrm{RW}$ where
\begin{subequations} \label{eq:si:rw_correction}
\begin{align}
    \phi_\mathrm{RW} &= \frac{RW}{100} , \quad \mathrm{and} \\
    RW &= 100 - TS.
\end{align}
\end{subequations}
This accounts for the processing of raw samples, which involves centrifuging them prior to analysis and measuring only the supernatant. Furthermore, it accounts for the assumption that all measured constituents are dissolved in the liquid supernatant. 

Offline digestate samples of IN and AC follow exactly the same computations as substrate samples of manure, i.e., Eqs.~\eqref{eq:si:influent:in}, \eqref{eq:si:influent:ac}, and \eqref{eq:si:rw_correction}.

\subsection{Model equations}
\label{sec:si:model_equations}

\subsubsection{ADM1-R3-core}
\label{sec:si:model_equations:r3-core}
The ADM1-R3-core neglects individual states of ash and water. Additionally, the states of free cations and anions are replaced by a tunable parameter $\Delta S_\mathrm{ion,eff}$, describing the effective cation concentration. This enables fitting of the pH independent of feeding events. Moreover, influent concentration of inorganic nitrogen (IN) is scaled by a factor $\varphi_\mathrm{IN,in}$ to improve fitting of pH and IN. The state vector, thus, reads: 
\begin{equation}
x = \left[S_\mathrm{ac}, S_\mathrm{ch4}, S_\mathrm{IC}, S_\mathrm{IN}, X_\mathrm{ch}, X_\mathrm{pr}, X_\mathrm{li}, X_\mathrm{bac}, X_\mathrm{ac}, S_\mathrm{ac^-}, S_\mathrm{hco3^-}, S_\mathrm{nh3}, S_\mathrm{ch4,gas}, S_\mathrm{co2,gas}\right]^T. \label{eq:si:r3-core:x}
\end{equation}
The system of ordinary differential equations (ODEs) then reads
\begin{subequations} \label{eq:si:r3-core:f}
	\small
	\begin{align}
	\dot x_1 &= c_1 \left(\xi_1 - x_1\right) u + a_{11} \theta_1 x_5 + a_{12} \theta_2 x_6 + a_{13} \theta_3 x_7 + a_{14} \theta_5 \frac{x_1 \, x_9}{\theta_6 + x_1} I_\mathrm{ac}, \\
	\dot x_2 &= c_1 \left(\xi_2 - x_2\right) u + a_{21} \theta_1 x_5 + a_{22} \theta_2 x_6 + a_{23} \theta_3 x_7 - c_5 x_2 + c_6 x_{13} + a_{24} \theta_5  \frac{x_1 \, x_9}{\theta_6 + x_1} I_\mathrm{ac}, \\
	\dot x_3 &= c_1 \left(\xi_3 - x_3\right) u + a_{31} \theta_1 x_5 + a_{32} \theta_2 x_6 + a_{33} \theta_3 x_7 - c_5 x_3 + c_5 x_{11} + c_7 x_{14} + a_{34} \theta_5 \frac{x_1 \, x_9}{\theta_6 + x_1} I_\mathrm{ac}, \\
	\dot x_4 &= c_1 \left(\xi_4 \, \theta_9 - x_4\right) u + a_{41} \theta_1 x_5 + a_{42} \theta_2 x_6 + a_{43} \theta_3 x_7 + a_{44} \theta_5 \frac{x_1 \, x_9}{\theta_6 + x_1} I_\mathrm{ac}, \\
	\dot x_5 &= c_1 \left(\xi_5 - x_5\right) u - \theta_1 x_5 + a_{55} \theta_4 x_8 + a_{56} \theta_4 x_9, \\
	\dot x_6 &= c_1 \left(\xi_6 - x_6\right) u - \theta_2 x_6 + a_{65} \theta_4 x_8 + a_{66} \theta_4 x_9, \\ 
	\dot x_7 &= c_1 \left(\xi_7 - x_7\right) u - \theta_3 x_7 + a_{75} \theta_4 x_8 + a_{76} \theta_4 x_9, \\ 
	\dot x_8 &= c_1 \left(\xi_8 - x_8\right) u + a_{81} \theta_1 x_5 + a_{82} \theta_2 x_6 + a_{83} \theta_3 x_7 - \theta_4 x_8, \\
	\dot x_9 &= c_1 \left(\xi_9 - x_9\right) u + \theta_5 \frac{x_1 \, x_9}{\theta_6 + x_1} I_\mathrm{ac} - \theta_4 x_9, \\
	%
	%
	\dot x_{10} &= c_{28} \left(x_1 - x_{10}\right) - c_9 x_{10} S_\mathrm{H^+}, \\
	\dot x_{11} &= c_{29} \left(x_3 - x_{11}\right) - c_{10} x_{11} S_\mathrm{H^+}, \\
	\dot x_{12} &= c_{30} \left(x_4 - x_{12}\right) - c_{11} x_{12} S_\mathrm{H^+}, \\
	\dot x_{13} &= c_{21} x_{13}^3 + c_{22} x_{13}^2 x_{14} + c_{23} x_{13} x_{14}^2 + c_{24} x_{13}^2 + c_{25} x_{13} x_{14} + c_{12} x_2 + c_{26} x_{13}, \\
	\dot x_{14} &= c_{23} x_{14}^3 + c_{22} x_{13} x_{14}^2 + c_{21} x_{13}^2 x_{14} + c_{25} x_{14}^2 + c_{24} x_{13} x_{14} + c_{12} x_3 - c_{12} x_{11} + c_{27} x_{14}.
	\end{align}
\end{subequations}
$I_\mathrm{ac}$ and $S_\mathrm{H^+}$ are defined as  
\begin{align}
I_\mathrm{ac} &= \underbrace{\frac{c_3}{c_3 + S_\mathrm{H^+}^{c_2}}}_{I_\mathrm{pH}} \underbrace{\frac{x_4}{x_4 + c_8}}_{I_\mathrm{IN,lim}} \underbrace{\frac{\theta_7}{\theta_7 + x_{12}}}_{I_\mathrm{nh3}}, \label{eq:si:R3-inhibit} \\
S_\mathrm{H^+} &= -\frac{\Phi}{2} + \frac{1}{2} \sqrt{\Phi^2 + c_4} \text{, \quad where} \\
\Phi &= \theta_8 + \frac{x_4 - x_{12}}{17} - \frac{x_{11}}{44} - \frac{x_{10}}{60}.
\end{align}
Output equations read
\begin{subequations}
	\label{eq:si:r3-core:y}
	\begin{align}
	y_1 &= \dot V_\mathrm{gas} = c_{13} x_{13}^2 + c_{14} x_{13} x_{14} + c_{15} x_{14}^2 + c_{16} x_{13} + c_{17} x_{14} + c_{18}, \\
	y_2 &= p_\mathrm{ch4} = c_{19} x_{13}, \\
	y_3 &= p_\mathrm{co2} = c_{20} x_{14}, \\
	y_4 &= pH = -\log_{10} S_\mathrm{H^+}, \\
	y_5 &= S_\mathrm{IN} = x_4, \\
	y_6 &= S_\mathrm{ac} = x_1.
	\end{align}
\end{subequations}
Stoichiometric coefficients $a_{ij}$ are given in the Petersen matrix, Table~\ref{tab:si:petersen:r3-core}. Parameters $c$ are considered time-invariant and aggregate variables used in Weinrich and Nelles \cite{weinrich2021} (Table~\ref{tab:si:adm1-r3:c}). Tunable model parameters $\theta$ are found in Table~\mainref{tab:meth:parameters}. 
\begin{table}
\centering
\begin{threeparttable}
\footnotesize
\caption{Aggregation of time-invariant ADM1-R3 model parameters.} 
\label{tab:si:adm1-r3:c}
\begin{tabular}{clrl}
    \toprule
    Parameter & Computation & Value$^{\mathrm{b}}$ & Unit \\
    \midrule
    $c_1$ & $V_\mathrm{liq}^{-1}$ & 8.333\,E1 & \si{\per\cubic\meter} \\ 
    $c_2$ & $n_\mathrm{ac} = 3 \, \left(pH_{\mathrm{UL,ac}} - pH_{\mathrm{LL,ac}}\right)^{-1}$ & 3 & $-$ \\
    $c_3$ & $10^{- \frac{3}{2} \frac{pH_{\mathrm{UL,ac}} + pH_{\mathrm{LL,ac}}}{pH_{\mathrm{UL,ac}} - pH_{\mathrm{LL,ac}}}}$ & 3.162\,E-20 & $-$ \\
    $c_4$ & $4 K_\mathrm{W}$ & 8.315\,E-14 & \si{\kilo\mole\per\cubic\meter} \\
    $c_5$ & $k_\mathrm{L} a$ & 2\,E2 & \si{\per\day} \\
    $c_6$ & $k_\mathrm{L} a \, K_{\mathrm{H,ch4}} \bar{R} T$ & 4.997 & \si{\per\day} \\ 
    $c_7$ & $k_\mathrm{L} a \, K_{\mathrm{H,co2}} \bar{R} T$ & 1.136\,E2 & \si{\per\day} \\	
    $c_8$ & $K_{\mathrm{S,IN}}$ & 1.7\,E-3 & \si{\kilogram\per\cubic\meter} \\  
    $c_9$ & $k_{\mathrm{AB,ac}}$ & 1\,E10 & \si{\cubic\meter\per\kilo\mole\day} \\ 		
    $c_{10}$ & $k_{\mathrm{AB,co2}}$ & 1\,E10 & \si{\cubic\meter\per\kilo\mole\day} \\
    $c_{11}$ & $k_{\mathrm{AB,IN}}$ & 1\,E10 & \si{\cubic\meter\per\kilo\mole\day} \\		
    $c_{12}$ & $k_\mathrm{L} a \, V_\mathrm{liq} V_\mathrm{gas}^{-1}$ & 8\,E2 & \si{\per\day} \\
    $c_{13}$ & $k_\mathrm{p} p_0^{-1} \left(\bar R T \bar{M}_\mathrm{ch4}^{-1}\right)^2$ & 9.944\,E4 & $\textrm{m}^9\, \textrm{kg}^{-2} \textrm{d}^{-1}$ \\
    $c_{14}$ & $2 k_\mathrm{p} p_0^{-1} \left(\bar R T\right)^2 \bar{M}_\mathrm{ch4}^{-1} \bar{M}_\mathrm{co2}^{-1}$ & 7.232\,E4 & $\textrm{m}^9\, \textrm{kg}^{-2} \textrm{d}^{-1}$ \\	
    $c_{15}$ & $k_\mathrm{p} p_0^{-1} \left(\bar R T \bar{M}_\mathrm{co2}^{-1}\right)^2$ & 1.315\,E4 & $\textrm{m}^9\, \textrm{kg}^{-2} \textrm{d}^{-1}$ \\
    $c_{16}$ & $k_\mathrm{p} p_0^{-1} \bar R T \bar{M}_\mathrm{ch4}^{-1} \left(2 p_\mathrm{h2o} - p_0\right)$ & -7.098\,E4 & $\textrm{m}^6\, \textrm{kg}^{-1} \textrm{d}^{-1}$ \\
    $c_{17}$ & $k_\mathrm{p} p_0^{-1} \bar R T \bar{M}_\mathrm{co2}^{-1} \left(2 p_\mathrm{h2o} - p_0\right)$ & -2.581\,E4 & $\textrm{m}^6\, \textrm{kg}^{-1} \textrm{d}^{-1}$ \\
    $c_{18}$ & $k_\mathrm{p} p_0^{-1} \left(p_\mathrm{h2o} - p_0\right) p_\mathrm{h2o}$ & 0 & \si{\cubic\meter\per\day}{} \\
    $c_{19}$ & $\bar R T \bar{M}_\mathrm{ch4}^{-1}$ & 1.420 & \si{\bar\cubic\meter\per\kilogram} \\
    $c_{20}$ & $\bar R T \bar{M}_\mathrm{co2}^{-1}$ & 0.516 & \si{\bar\cubic\meter\per\kilogram} \\
    $c_{21}$ & $-k_\mathrm{p} V_\mathrm{gas}^{-1} p_0^{-1} \left(\bar R T \bar{M}_\mathrm{ch4}^{-1}\right)^2 $ & -3.315\,E7 & $\textrm{m}^6\, \textrm{kg}^{-2} \textrm{d}^{-1}$ \\
    $c_{22}$ & $-2 k_\mathrm{p} V_\mathrm{gas}^{-1} p_0^{-1} \left(\bar R T\right)^2 \bar{M}_\mathrm{ch4}^{-1} \bar{M}_\mathrm{co2}^{-1}$ & -2.411\,E7 & $\textrm{m}^6\, \textrm{kg}^{-2} \textrm{d}^{-1}$ \\  
    $c_{23}$ & $-k_\mathrm{p} V_\mathrm{gas}^{-1} p_0^{-1} \left(\bar R T \bar{M}_\mathrm{co2}^{-1}\right)^2$ & -4.383\,E6 & $\textrm{m}^6\, \textrm{kg}^{-2} \textrm{d}^{-1}$ \\ 
    $c_{24}$ & $-k_\mathrm{p} V_\mathrm{gas}^{-1} p_0^{-1} \bar R T \bar{M}_\mathrm{ch4}^{-1} \left(2 p_\mathrm{h2o} - p_0\right)$ & 2.366\,E7 & \si{\cubic\meter\per\kilogram\day} \\ 		
    $c_{25}$ & $-k_\mathrm{p} V_\mathrm{gas}^{-1} p_0^{-1} \bar R T \bar{M}_\mathrm{co2}^{-1} \left(2 p_\mathrm{h2o} - p_0\right)$ & 8.603\,E6 & \si{\cubic\meter\per\kilogram\day} \\ 
    $c_{26}$ & $- k_\mathrm{L} a \, V_\mathrm{liq} V_\mathrm{gas}^{-1} K_{\mathrm{H,ch4}} \bar{R} T - k_\mathrm{p} V_\mathrm{gas}^{-1} p_0^{-1} \left(p_\mathrm{h2o} - p_0\right) p_\mathrm{h2o}$ & -1.999\,E1 & \si{\per\day} \\ 
    $c_{27}$ & $- k_\mathrm{L} a \, V_\mathrm{liq} V_\mathrm{gas}^{-1} K_{\mathrm{H,co2}} \bar{R} T - k_\mathrm{p} V_\mathrm{gas}^{-1} p_0^{-1} \left(p_\mathrm{h2o} - p_0\right) p_\mathrm{h2o}$ & -4.542\,E2 & \si{\per\day} \\
    $c_{28}$ & $k_{\mathrm{AB,ac}} K_{\mathrm{a,ac}}$ & 1.738\,E5 & \si{\per\day} \\ 
    $c_{29}$ & $k_{\mathrm{AB,co2}} K_{\mathrm{a,co2}}$ & 4.937\,E3 & \si{\per\day} \\  
    $c_{30}$ & $k_{\mathrm{AB,IN}} K_{\mathrm{a,IN}}$ & 1.110\,E1 & \si{\per\day} \\ 
    $c_{31}$ & $V_\mathrm{liq} V_\mathrm{gas}^{-1}$ & 4 & $-$ \\ 
    \bottomrule
\end{tabular}
\begin{tablenotes}
\footnotesize
	\item[a] Liquid and gas volumes of the reactor were 12 and \SI{3}{\liter}, otherwise default values of Weinrich and Nelles~\cite{weinrich2021} were used.
	
	\item[b] Scientific E notation applies, i.e., E$\pm$n means $\times 10^{\pm n}$.
\end{tablenotes}
\end{threeparttable}
\end{table}   
%
\begin{landscape}
	\scriptsize
	\renewcommand{\arraystretch}{1.4}
	\setlength{\tabcolsep}{1pt}
	\centering
	%
	%
	\begin{longtable}{ll*{9}{C{1.2cm}}l} 
	\caption{Petersen matrix of ADM1-R3-core, derived from Weinrich and Nelles (2021).} \label{tab:si:petersen:r3-core} \\
	\toprule 
	\multicolumn{2}{l}{\textbf{Component i $\rightarrow$}} & 1 & 2 & 3 & 4 & 5 & 6 & 7 & 8 & 9 \\ [-0.7mm]
	\textbf{j} & \textbf{Process j $\downarrow$} & $S_\mathrm{ac}$ & $S_\mathrm{ch4}$ & $S_\mathrm{IC}$ & $S_\mathrm{IN}$ & $X_\mathrm{ch}$ & $X_\mathrm{pr}$ & $X_\mathrm{li}$ & $X_\mathrm{bac}$ & $X_\mathrm{ac}$ & \textbf{Process rate} $r_j$\\ [-0.7mm]
	\midrule 
	1 & Fermentation $X_\mathrm{ch}$ & 0.6555 & 0.0818 & 0.2245 & -0.0169 & -1 & & & 0.1125 & & $\theta_1 \,  x_5$ \\ 
	2 & Fermentation $X_\mathrm{pr}$ & 0.9947 & 0.0696 & 0.1029 & 0.1746 & & -1 & & 0.1349 & & $\theta_2 \,  x_6$ \\ 
	3 & Fermentation $X_\mathrm{li}$ & 1.7651 & 0.1913 & -0.6472 & -0.0244 & & & -1 & 0.1621 & & $\theta_3 \,  x_7$ \\ 
	\hdashline[0.6pt/1.8pt]
	4 & Methanogenesis $S_\mathrm{ac}$ & -26.5447 & 6.7367 & 18.4808 & -0.1506 & & & & & 1 & $\theta_5 \, \frac{x_1}{\theta_6 + x_1} \, x_9 \, I_{ac}$\\ 
	\hdashline[0.6pt/1.8pt]
	5 & Decay $X_\mathrm{bac}$ & & & & & 0.18 & 0.77 & 0.05 & -1 & & $\theta_4 \,  x_8$ \\ 
	6 & Decay $X_\mathrm{ac}$ & & & & & 0.18 & 0.77 & 0.05 & & -1 & $\theta_4 \,  x_9$ \\ 
	\midrule  
	\midrule   
	\multicolumn{2}{l}{\textbf{Component $i \rightarrow$}} & & 2 & 3 & $\ldots$ & 10 & 11 & 12 & 13 & 14 & \\ [-0.7mm]
	\textbf{j} & \textbf{Process j $\downarrow$} & & $S_\mathrm{ch4}$ & $S_\mathrm{IC}$ & & $S_\mathrm{ac^-}$ & $S_\mathrm{hco3^-}$ & $S_\mathrm{nh3}$ & $S_\mathrm{ch4,gas}$ & $S_\mathrm{co2,gas}$ & \\ [-0.7mm]		 
	\midrule 
	7 & Dissociation $S_\mathrm{ac}$ & & & & & -1 & & & & & $c_{28} \, (x_{10} - x_1) + c_9 \, x_{10} \, S_\mathrm{H^+}$\\ 
	8 & Dissociation $S_\mathrm{IC}$ & & & & & & -1 & & & & $c_{29} \, (x_{11} - x_3) + c_{10} \, x_{11} \, S_\mathrm{H^+}$ \\ 
	9 & Dissociation $S_\mathrm{IN}$ & & & & & & & -1 & & & $c_{30} \, (x_{12} - x_4) + c_{11} \, x_{12} \, S_\mathrm{H^+}$ \\
	\hdashline[0.6pt/1.8pt] 
	10 & Phase transition $S_\mathrm{ch4}$& & -1 & & & & & & $c_{31}$ & & $c_5 \, x_2 - c_6 \, x_{13}$ \\ 
	11 & Phase transition $S_\mathrm{co2}$& & & -1 & & & & & & $c_{31}$ & $c_5 \, (x_3 - x_{11}) - c_7 \, x_{14}$ \\ 
	\bottomrule
	\end{longtable}
	%
	%
	\begin{longtable}{ll*{10}{C{1.2cm}}l} 
	\caption{Petersen matrix of ADM1-R3-core-frac, derived from Weinrich and Nelles (2021).} \label{tab:si:petersen:r3-frac} \\
	\toprule 
	\multicolumn{2}{l}{\textbf{Component i $\rightarrow$}} & 1 & 2 & 3 & 4 & 5 & 6 & 7 & 8 & 9 & 10 \\ [-0.7mm]
	\textbf{j} & \textbf{Process j $\downarrow$} & $S_\mathrm{ac}$ & $S_\mathrm{ch4}$ & $S_\mathrm{IC}$ & $S_\mathrm{IN}$ & $X_\mathrm{ch,f}$ & $X_\mathrm{ch,s}$ & $X_\mathrm{pr}$ & $X_\mathrm{li}$ & $X_\mathrm{bac}$ & $X_\mathrm{ac}$ & \textbf{Process rate} $r_j$ \\ [-0.7mm] 
    \midrule 
    1 & Fermentation $X_\mathrm{ch,f}$ & 0.6555 & 0.0818 & 0.2245 & -0.0169 & -1 & & & & 0.1125 & & $\theta_1 \,  x_5$\\ 
    2 & Fermentation $X_\mathrm{ch,s}$ & 0.6555 & 0.0818 & 0.2245 & -0.0169 & & -1 & & & 0.1125 & & $\theta_2 \,  x_6$\\ 
    3 & Fermentation $X_\mathrm{pr}$ & 0.9947 & 0.0696 & 0.1029 & 0.1746 & & & -1 & & 0.1349 & & $\theta_3 \,  x_7$\\
    4 & Fermentation $X_\mathrm{li}$ & 1.7651 & 0.1913 & -0.6472 & -0.0244 & & & & -1 & 0.1621 & & $\theta_4 \,  x_8$\\
    \hdashline[0.6pt/1.8pt] 
    5 & Methanogenesis $S_\mathrm{ac}$ & -26.5447 & 6.7367 & 18.4808 & -0.1506 & & & & & & 1 & $\theta_6 \, \frac{x_1}{\theta_7 + x_1} \, x_{10} \, I_{ac}$\\ 
    \hdashline[0.6pt/1.8pt] 
    6 & Decay $X_\mathrm{bac}$ & & & & & 0.18 & & 0.77 & 0.05 & -1 & & $\theta_5 \, x_9$\\
    7 & Decay $X_\mathrm{ac}$ & & & & & 0.18 & & 0.77 & 0.05 & & -1 & $\theta_5 \, x_{10}$\\
    \midrule  
    \midrule 
    \multicolumn{2}{l}{\textbf{Component $i \rightarrow$}} & & 2 & 3 & $\ldots$ & & 11 & 12 & 13 & 14 & 15 & \\ [-0.7mm]
    \textbf{j} & \textbf{Process j $\downarrow$} & & $S_\mathrm{ch4}$ & $S_\mathrm{IC}$ & & & $S_\mathrm{ac^-}$ & $S_\mathrm{hco3^-}$ & $S_\mathrm{nh3}$ & $S_\mathrm{ch4,gas}$ & $S_\mathrm{co2,gas}$ & \\ [-0.7mm]
    \midrule 
    8 & Dissociation $S_\mathrm{ac}$ & & & & & & -1 & & & & & $c_{28} \, (x_{11} - x_1) + c_9 \, x_{11} \, S_\mathrm{H^+}$\\
    9 & Dissociation $S_\mathrm{IC}$ & & & & & & & -1 & & & & $c_{29} \, (x_{12} - x_3) + c_{10} \, x_{12} \, S_\mathrm{H^+}$\\ 
    10 & Dissociation $S_\mathrm{IN}$ & & & & & & & & -1 & & & $c_{30} \, (x_{13} - x_4) + c_{11} \, x_{13} \, S_\mathrm{H^+}$\\
    \hdashline[0.6pt/1.8pt]
    11 & Phase transition $S_\mathrm{ch4}$& & -1 & & & & & & & $c_{31}$ & & $c_5 \, x_2 - c_6 \, x_{14}$\\
    12 & Phase transition $S_\mathrm{co2}$& & & -1 & & & & & & & $c_{31}$ & $c_5 \, (x_3 - x_{12}) - c_7 \, x_{15}$\\ 
	\bottomrule
	\end{longtable}
\end{landscape}

\subsubsection{ADM1-R3-core-frac}
\label{sec:si:model_equations:r3-frac}
Compared to the ADM1-R3-core, total carbohydrates (CH) are split into two fractions, a fast and a slowly degradable one, $X_\mathrm{ch,f}$ and $X_\mathrm{ch,s}$. Both are assigned individual hydrolysis constants $k_\mathrm{ch,f}$ and $k_\mathrm{ch,s}$. Total influent CH are allocated to the fast fraction, and then attributed by means of a fraction parameter $\varphi_\mathrm{ch,in,f}$. Stoichiometric degradation pathways of both carbohydrate fractions are considered identical. Hence, two identical rows of stoichiometric coefficients result for the Petersen matrix (Table~\ref{tab:si:petersen:r3-frac}). Lastly, the stoichiometric decay of biomass is assumed to only add to the formation of fast CH.

The state vector thus reads
\begin{equation}
x = \left[S_\mathrm{ac}, S_\mathrm{ch4}, S_\mathrm{IC}, S_\mathrm{IN}, X_\mathrm{ch,f}, X_\mathrm{ch,s}, X_\mathrm{pr}, X_\mathrm{li}, X_\mathrm{bac}, X_\mathrm{ac}, S_\mathrm{ac^-}, S_\mathrm{hco3^-}, S_\mathrm{nh3}, S_\mathrm{ch4,gas}, S_\mathrm{co2,gas}\right]^T. \label{eq:si:r3-frac:x}
\end{equation}
Model equations change only slightly compared to the ADM1-R3-core. The system of ODEs reads
\begin{subequations}  \label{eq:si:r3-frac:f}
	\small
	\begin{align}
	\dot x_1 &= c_1 \left(\xi_1 - x_1\right) u + a_{11} \theta_1 x_5 + a_{12} \theta_2 x_6 + a_{13} \theta_3 x_7 + a_{14} \theta_4 x_8 + a_{15} \theta_6 \frac{x_1 \, x_{10}}{\theta_7 + x_1} I_\mathrm{ac}, \\
	\dot x_2 &= c_1 \left(\xi_2 - x_2\right) u + a_{21} \theta_1 x_5 + a_{22} \theta_2 x_6 + a_{23} \theta_3 x_7 + a_{24} \theta_4 x_8 - c_5 x_2 + c_6 x_{14} + a_{25} \theta_6 \frac{x_1 \, x_{10}}{\theta_7 + x_1} I_\mathrm{ac}, \\
	\dot x_3 &= c_1 \left(\xi_3 - x_3\right) u + a_{31} \theta_1 x_5 + a_{32} \theta_2 x_6 + a_{33} \theta_3 x_7 + a_{34} \theta_4 x_8 - c_5 x_3 + c_5 x_{12} + c_7 x_{15} + a_{35} \theta_6 \frac{x_1 \, x_{10}}{\theta_7 + x_1} I_\mathrm{ac}, \\
	\dot x_4 &= c_1 \left(\xi_4 \, \theta_{10} - x_4\right) u + a_{41} \theta_1 x_5 + a_{42} \theta_2 x_6 + a_{43} \theta_3 x_7 + a_{44} \theta_4 x_8 + a_{45} \theta_6 \frac{x_1 \, x_{10}}{\theta_7 + x_1} I_\mathrm{ac}, \label{eq:si:r3-frac:f:IN} \\
	\dot x_5 &= c_1 \left(\xi_5 \, \theta_{11} - x_5\right) u - \theta_1 x_5 + a_{56} \theta_5 x_9 + a_{57} \theta_5 x_{10}, \\
    \dot x_6 &= c_1 \left(\xi_5 \, (1 - \theta_{11}) - x_6\right) u - \theta_2 x_6, \\
	\dot x_7 &= c_1 \left(\xi_7 - x_7\right) u - \theta_3 x_7 + a_{76} \theta_5 x_9 + a_{77} \theta_5 x_{10}, \\ 
	\dot x_8 &= c_1 \left(\xi_8 - x_8\right) u - \theta_4 x_8 + a_{86} \theta_5 x_9 + a_{87} \theta_5 x_{10}, \\ 
	\dot x_9 &= c_1 \left(\xi_9 - x_9\right) u + a_{91} \theta_1 x_5 + a_{92} \theta_2 x_6 + a_{93} \theta_3 x_7 + a_{94} \theta_4 x_8 - \theta_5 x_9, \\
	\dot x_{10} &= c_1 \left(\xi_{10} - x_{10}\right) u + \theta_6 \frac{x_1 \, x_{10}}{\theta_7 + x_1} I_\mathrm{ac} - \theta_5 x_{10}, \\
	\dot x_{11} &= c_{28} \left(x_1 - x_{11}\right) - c_9 x_{11} S_\mathrm{H^+}, \\
	\dot x_{12} &= c_{29} \left(x_3 - x_{12}\right) - c_{10} x_{12} S_\mathrm{H^+}, \\
	\dot x_{13} &= c_{30} \left(x_4 - x_{13}\right) - c_{11} x_{13} S_\mathrm{H^+}, \\
	\dot x_{14} &= c_{21} x_{14}^3 + c_{22} x_{14}^2 x_{15} + c_{23} x_{14} x_{15}^2 + c_{24} x_{14}^2 + c_{25} x_{14} x_{15} + c_{12} x_2 + c_{26} x_{14}, \\
	\dot x_{15} &= c_{23} x_{15}^3 + c_{22} x_{14} x_{15}^2 + c_{21} x_{14}^2 x_{15} + c_{25} x_{15}^2 + c_{24} x_{14} x_{15} + c_{12} x_3 - c_{12} x_{12} + c_{27} x_{15},
	\end{align}
\end{subequations}
with $I_\mathrm{ac}$ and $S_\mathrm{H^+}$ defined as  
\begin{subequations}
\label{eq:si:r3-frac:Iac_SHPlus_Phi}
\begin{align}
I_\mathrm{ac} &= \underbrace{\frac{c_3}{c_3 + S_\mathrm{H^+}^{c_2}}}_{I_\mathrm{pH}} \underbrace{\frac{x_4}{x_4 + c_8}}_{I_\mathrm{IN,lim}} \underbrace{\frac{\theta_8}{\theta_8 + x_{13}}}_{I_\mathrm{nh3}}, \\ 
S_\mathrm{H^+} &= -\frac{\Phi}{2} + \frac{1}{2} \sqrt{\Phi^2 + c_4}  \text{, \quad where} \\
\Phi &= \theta_9 + \frac{x_4 - x_{13}}{17} - \frac{x_{12}}{44} - \frac{x_{11}}{60}.
\end{align}
\end{subequations}
Output equations read
\begin{subequations}
	\label{eq:si:r3-frac:y}
	\begin{align}
	y_1 &= \dot V_\mathrm{gas} = c_{13} x_{14}^2 + c_{14} x_{14} x_{15} + c_{15} x_{15}^2 + c_{16} x_{14} + c_{17} x_{15} + c_{18}, \\
	y_2 &= p_\mathrm{ch4} = c_{19} x_{14}, \\
	y_3 &= p_\mathrm{co2} = c_{20} x_{15}, \\
	y_4 &= pH = -\log_{10} S_\mathrm{H^+}, \\
	y_5 &= S_\mathrm{IN} = x_4, \\
	y_6 &= S_\mathrm{ac} = x_1.
	\end{align}
\end{subequations}
Stoichiometric coefficients $a_{ij}$ are provided in the Petersen matrix of the ADM1-R3-core-frac (Table~\ref{tab:si:petersen:r3-frac}). Time-invariant parameters $c$ are summarized in Table~\ref{tab:si:adm1-r3:c} and apply for both ADM1-R3-core and -frac. For tunable model parameters $\theta$, cf. Table~\mainref{tab:meth:parameters}. 
%

\bookmarksetup{startatroot} 
\phantomsection 
\addcontentsline{toc}{section}{References}
\sloppy
\printbibliography

\end{document}